\documentclass[11pt,letterpaper]{article}

\usepackage[utf8]{inputenc}
\usepackage[T1]{fontenc}
\usepackage{microtype}

\usepackage[letterpaper,top=1in,bottom=1in,left=1in,right=1in]{geometry}

\usepackage[skip=5pt plus 3pt minus 1pt,parfill=0pt]{parskip}
\makeatletter
\def\thm@space@setup{%
  \thm@preskip=\parskip \thm@postskip=0pt
}
\renewcommand\paragraph{\@startsection{paragraph}{4}{\z@}{\parskip}{-1em}{\normalfont\normalsize\bfseries}}
\makeatother

\usepackage{amsmath}
\usepackage{amsthm}
\usepackage{thm-restate}
\makeatletter
\@ifundefined{newcounteralias}{}{%
  \renewcommand\thmt@autorefsetup{%
    \@xa\def\csname\thmt@envname autorefname\@xa\endcsname\@xa{\thmt@thmname}}}
\makeatother
\usepackage{mathtools}

\usepackage{amssymb}
\usepackage{bm,dsfont}
\usepackage{csquotes}
\usepackage{nicefrac}
\usepackage{soul}

\usepackage{xcolor}

\definecolor{mymagenta}{RGB}{200,60,140}
\definecolor{myblue}{RGB}{50,100,200}
\definecolor{myred}{RGB}{210,60,60}
\definecolor{myorange}{RGB}{235,140,40}
\definecolor{mygreen}{RGB}{60,170,90}
\definecolor{mypurple}{RGB}{120,80,180}
\definecolor{myteal}{RGB}{0,128,128}
\definecolor{mybrown}{RGB}{140,86,75}

\usepackage{paralist}

\usepackage[small]{caption}
\usepackage{subcaption}
\usepackage{tikz}
\usepackage{pgfplots}
\pgfplotsset{compat=newest}
\usepackage{wrapfig}

\usepackage{booktabs,tabularew}
\usepackage[para]{threeparttable}

\usepackage[ruled,linesnumbered,noline]{algorithm2e}

\usepackage{array}

\usepackage[shortlabels]{enumitem}
\newlist{obstacles}{enumerate}{1}
\setlist[obstacles]{label=(O\arabic*),ref=O\arabic*}

\usepackage[toc,page]{appendix}

\usepackage[
 backend=biber,
 style=alphabetic,
 citetracker,
 hyperref=auto,
 maxcitenames=10, 
 minalphanames=3,
 maxalphanames=4,
 sortcites,      
 sorting=nyt,
 maxbibnames=12, 
 date=year,      
 isbn=false,     
 url=false,      
 doi=false,      
 eprint=false,   
]{biblatex}
\newbibmacro{string+doiurl}[1]{%
  \iffieldundef{doi}{%
    \iffieldundef{url}{%
      #1
    }{%
      \href{\thefield{url}}{#1}%
    }%
  }{%
    \href{http://dx.doi.org/\thefield{doi}}{#1}%
  }%
}

\renewbibmacro*{volume+number+eid}{%
  \printfield{volume}%
  \setunit*{\adddot}%
  \printfield{number}}

\newbibmacro*{eid-or-pages}{%
  \iffieldundef{eid}
    {\printfield{pages}}%
    {\printfield{eid}}}

\renewbibmacro*{note+pages}{%
  \printfield{note}%
  \setunit{\bibpagespunct}%
  \usebibmacro{eid-or-pages}%
  \newunit}

\renewbibmacro*{chapter+pages}{%
  \printfield{chapter}%
  \setunit{\bibpagespunct}%
  \usebibmacro{eid-or-pages}%
  \newunit}

\AtEveryCitekey{%
  \ifciteseen
    {\defcounter{maxnames}{2}}
    {}%
}

\DeclareFieldFormat%
[article,inbook,incollection,inproceedings,patent,thesis,unpublished]
  {title}{\usebibmacro{string+doiurl}{#1}}

\DeclareFieldFormat{eid}{art.~#1}

\usepackage[colorlinks]{hyperref}
\hypersetup{citecolor=green!60!black,linkcolor=blue!70!black,urlcolor=blue!70!black}
\usepackage[capitalize, noabbrev, nameinlink]{cleveref}
\crefname{obstaclesi}{obstruction}{obstructions}
\Crefname{obstaclesi}{Obstruction}{Obstructions}
\usepackage{xurl}
\hypersetup{breaklinks=true}

\usepackage{xspace}

\allowdisplaybreaks

\newtheorem{theorem}{Theorem}[section]
\newtheorem{lemma}[theorem]{Lemma}

\newtheorem{corollary}[theorem]{Corollary}

\newtheorem{proposition}[theorem]{Proposition}

\theoremstyle{definition}
\newtheorem{definition}[theorem]{Definition}

\theoremstyle{remark}
\newtheorem{remark}[theorem]{Remark}

\DeclarePairedDelimiter{\abs}{\lvert}{\rvert}
\DeclarePairedDelimiter{\scp}{\langle}{\rangle}
\DeclareMathOperator*{\argmax}{arg\,max}

\newcommand{\symmdiff}{\mathbin{\triangle}}
\newcommand{\supp}{\mathrm{supp}}

\newcommand{\RR}{\mathbb{R}}

\newcommand{\NN}{\mathbb{N}}

\newcommand{\pr}{\mathrm{Pr}}
\newcommand{\EX}{\mathbb{E}}

\newcommand{\ind}{\ensuremath{\mathds{1}}}
\newcommand{\dd}{\,\mathrm{d}}

\newcommand{\psp}{\relax\ifmmode{\mathrm{PSP}}\else PSP\xspace\fi}

\newcommand{\alg}{{\mathrm{ALG}}}

\newcommand{\pf}{\mathrm{PF}\xspace}
\newcommand{\PF}{\mathrm{PF}\xspace}

\newcommand{\opt}{{\mathrm{OPT}}}
\newcommand{\OPT}{{\mathrm{OPT}}}
\newcommand{\alphagr}{{\mathrm{GR}_\alpha}}

\newcommand{\LP}{\mathrm{LP}}

\newcommand{\DLP}{\mathrm{DLP}}

\newcommand{\cP}{\mathcal{P}}

\newcommand{\cQ}{\mathcal{Q}}
\newcommand{\cS}{\mathcal{S}}

\newcommand{\dualSa}{\bm\bar\alpha}
\newcommand{\dualSb}{\bm\bar\beta}

\newcommand{\eps}{\varepsilon}

\definecolor{bluegreen}{HTML}{33615E}
\definecolor{lime}{HTML}{56A778}
\definecolor{yellow}{HTML}{E5CB6D}
\definecolor{orange}{HTML}{FB8318}
\definecolor{lightred}{HTML}{DD6143}

\definecolor{job1}{HTML}{A6CEE3}
\definecolor{job2}{HTML}{1F78B4}
\definecolor{job5}{HTML}{B2DF8A}
\definecolor{job6}{HTML}{33A02C}
\definecolor{job3}{HTML}{FB9A99}
\definecolor{job4}{HTML}{E31A1C}

\definecolor{col1}{HTML}{56b3e9}
\definecolor{col2}{HTML}{e69f00}
\definecolor{col3}{HTML}{009e74}
\definecolor{col4}{HTML}{cc79a7}
\definecolor{col5}{HTML}{d55e00}
\definecolor{col6}{HTML}{0071b2}

\usepackage[colorinlistoftodos,prependcaption,textsize=scriptsize]{todonotes}

\title{Single-Machine Guarantees on Unrelated Machines and Beyond: Proportional Fairness for Nonclairvoyant Polytope Scheduling}

\author{%
  Sven Jäger\thanks{Institut für Mathematik, Technische Universität Berlin, Germany (\texttt{jaeger@tu-berlin.de}). }
  \and
  Alexander Lindermayr\thanks{Institut für Mathematik, Technische Universität Berlin, Germany (\texttt{alexander.lindermayr@tu-berlin.de}). }
  \and
  Bart Zondervan\thanks{Faculty of Mathematics and Computer Science, University of Bremen, Germany (\texttt{bart@uni-bremen.de}).}
}
\date{}

\begin{document}
\pagenumbering{roman}

\maketitle

\begin{abstract}
The Polytope Scheduling Problem (PSP) is a general framework for preemptive scheduling in which feasible instantaneous job allocations form a packing polytope.
We study PSP in the nonclairvoyant online setting, where job processing requirements are unknown until completion.
In this setting, Proportional Fairness (PF), a practical allocation rule with deep connections to market theory, is the leading general-purpose algorithm.
On a single machine, PF coincides with weighted Round-Robin, which is $2$-competitive for weighted completion time with uniform release dates (optimal for nonclairvoyant algorithms) and $(2+\varepsilon)$-speed $O(1/\varepsilon)$-competitive for weighted flow time with arbitrary release dates.
Beyond a single machine, PF was known to be $3.62$-competitive on unrelated machines and $4$-competitive on polymatroids for weighted completion time with uniform release dates (SICOMP 2026), and $(e+\varepsilon)$-speed $O(1/\varepsilon^2)$-competitive on polymatroids for weighted flow time (JACM 2018).
This raises a central question:
\emph{Are these heterogeneous environments intrinsically harder than a single machine under nonclairvoyance?}

We show that they are not: PF retains both single-machine guarantees on the novel class of \emph{support-submodular} polytopes, which includes unrelated machines and polymatroids.
On unrelated machines, this refutes a belief (JACM 2018, SODA 2025) that PF is not $O(1)$-speed $O(1)$-competitive for weighted flow time.
Our main technical contribution is an equivalent characterization: a polytope is support submodular if and only if, for all job weights, the optimal weighted completion time is concave in the processing requirements. This structural insight underlies both guarantees.

For general PSP, the picture changes: we give the first lower bound above~$2$, showing that every randomized nonclairvoyant algorithm has a competitive ratio of at least~$2+10^{-5}$ for weighted completion time, even with unit weights and uniform release dates.
On the positive side, we substantially improve the best known guarantee for weighted completion time from~$27$ (SICOMP 2026) to~$4$ via a new dual-fitting analysis of PF.
For weighted flow time, we show that PF achieves the first $O(1)$-speed $O(1)$-competitive nonclairvoyant guarantee for graph scheduling, where jobs are the edges of a multigraph and feasible allocations form its matching polytope.
More generally, this extends to every packing polytope between a binary packing polytope with constant column sparsity and its integer hull.
\end{abstract}

\thispagestyle{empty}
\newpage
\setcounter{tocdepth}{2}
{\small
\tableofcontents
}
\thispagestyle{empty}
\newpage
\pagenumbering{arabic}

\section{Introduction}

Schedulers must often allocate scarce resources without knowing which jobs will arrive next or how much work current jobs require.
Such uncertainty is central to operating systems, cloud computing, and network resource allocation.
Its two canonical forms are unknown arrival times~$r_j$ (online arrivals) and unknown processing requirements~$p_j$ (nonclairvoyance)~\cite{MotwaniPT94,ShmoysWW95,HallSSW97}.

The Polytope Scheduling Problem (PSP), introduced by Im, Kulkarni, and Munagala~\cite{ImKM18}, provides a common language for preemptive scheduling. 
It views scheduling as instantaneous resource allocation over time, where active jobs compete for limited resources.
Formally, at each time $t$, a scheduler irrevocably selects a \emph{rate vector}~$y(t)$ from a downward-closed polytope~$\cP \subseteq \RR_{\ge 0}^n$ that encodes all feasible allocations.
Job~$j\in[n]$ completes at the first time~$C_j$ such that $\int_{r_j}^{C_j} y_j(t)\dd t = p_j$. A feasible schedule must complete all jobs, so we assume that every job~$j$ has~$y_j>0$ for some~$y\in\cP$.

Despite its compact formulation, PSP captures a wide range of independently studied models, including multidimensional scheduling~\cite{GhodsiZHKSS10}, multicommodity flow scheduling~\cite{KellyMT98}, fractional broadcast scheduling~\cite{BansalKN14,EdmondsP03}, graph scheduling~\cite{DinitzM20,RohwedderS25,LindermayrLM26,RohwedderS26}, and scheduling with concave speedup curves~\cite{Edmonds00}; see~\cite{ImKM18,ChenIP25,JLM2026} for an overview.
Thus, results for PSP apply across many diverse scheduling environments.
A prominent example is \emph{unrelated machines}~\cite{HallSSW97,LenstraST90,BSS21,AnandGK12,Skutella01}, where machine~$i$ processes job~$j$ at a (known) speed~$s_{ij} \geq 0$. Its rate polytope arises from fractional bipartite matchings between jobs and machines.

Most work on PSP focuses on two important quality-of-service metrics: weighted completion time and weighted flow time.
Before this work, the state of the art on nonclairvoyant online algorithms for major subclasses of PSP can be summarized as follows:
\begin{itemize}
	\item \textbf{Weighted Completion Time $\sum_j w_jC_j$:}
	For this objective, we assume that release dates are uniform unless stated otherwise.
	The best previous competitive ratios were $27$ for PSP,
	$4$ when $\cP$ is a polymatroid,\footnote{\label{fn:monotone-psp}These results hold even for PF-monotone PSP, a larger class defined via Proportional Fairness (see \cref{def:monPSP}).} and $3.62$ for unrelated machines~\cite{JLM2026}.
	Yet no randomized nonclairvoyant algorithm can achieve a ratio below~$2$, even on a single machine~\cite{MotwaniPT94}.
	\item \textbf{Weighted Flow Time $\sum_j w_j (C_j-r_j)$:}
	Without speed augmentation, no nonclairvoyant algorithm is $O(1)$-competitive, even on a single machine~\cite{MotwaniPT94}.
	Moreover, for PSP, no deterministic nonclairvoyant algorithm is $o(\sqrt{\log n})$-speed $O(1)$-competitive~\cite{ImKM18}.
	For polymatroids, an $O(1/\eps^2)$-competitive algorithm with speed augmentation~$e+\eps$ is known~\cite{ImKM18}.\textsuperscript{\ref{fn:monotone-psp}}
\end{itemize}
\emph{All} of these positive results are achieved by the \emph{Proportional Fairness} (PF) algorithm,
a prominent practical scheduling algorithm~\cite{ImKM18,GhodsiZHKSS10,KellyMT98} with deep connections to market theory and fair allocation~\cite{nash1950bargaining,eisenberg1959consensus,JainV10}.
At each time~$t$, PF chooses the rate vector
$y^{\textsc{pf}}(t) = \argmax_{y \in \cP} \sum_{j \in A(t)} w_j \log(y_j)$
for the active jobs~$A(t)$.
On a single machine, PF is precisely (weighted) Round-Robin, which assigns rates proportional to job weights, and is $2$-competitive for weighted completion time and $(2+\eps)$-speed $O(1/\eps)$-competitive for weighted flow time~\cite{KimC03,Edmonds00,MotwaniPT94}.

For general PSP, it has remained open whether the factor of~$2$ is achievable~\cite{JLM2026} and which polytopes admit an $O(1)$-speed $O(1)$-competitive algorithm~\cite{ImKM18,ChenIP25}.
For unrelated machines, Jäger, Lindermayr, and Megow conjectured that PF is $2$-competitive for weighted completion time~\cite{JLM2026}, whereas for weighted flow time, PF is ``believed not to be $O(1)$-competitive'' even with $O(1)$ speed augmentation~\cite{ImKM18,ChenIP25}.
This raises the following natural questions:
\begin{quote}
\emph{For which polytopes does PF retain the (optimal) single-machine guarantees of Round-Robin?
Are rich polyhedral allocation environments such as unrelated machines and polymatroids intrinsically harder than a single machine under nonclairvoyance?}
\end{quote}

\subsection{Our Results}

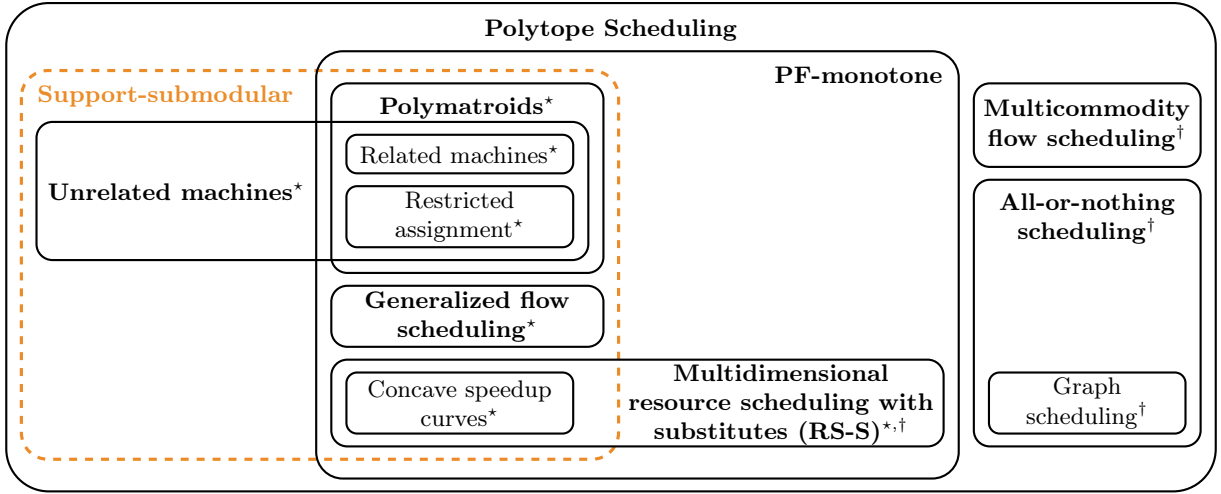
\begin{figure}[tb]
\centering
\begin{tikzpicture}[yscale=0.85]

    \colorlet{pspcolor}{black}
    \colorlet{supercolor}{black} 
    \colorlet{sscolor}{myorange} 
    \colorlet{moncolor}{black} 
    \colorlet{unrelcolor}{black} 
    \colorlet{polycolor}{black} 
    \colorlet{multicolor}{black} 
    \colorlet{multiflow}{black} 


    \draw[rounded corners=16pt, thick, pspcolor]
        (0,0) rectangle (16.0,-7.6);
    \node[anchor=north, font=\footnotesize, color=pspcolor] at (8.0,-0.1) {\textbf{Polytope Scheduling}};


    \draw[rounded corners=6pt, thick, multiflow]
        (12.8,-1.25) rectangle (15.8,-2.55);
    \node[font=\footnotesize, align=center, color=multiflow]
        at (14.3,-1.9) {\textbf{Multicommodity}\\\textbf{flow scheduling}$^\dagger$};

    \draw[rounded corners=4pt, thick]
        (13.0,-5.75) rectangle (15.6,-6.7);
    \node[font=\footnotesize, align=center]
        at (14.3,-6.225) {Graph\\ scheduling$^\dagger$};

    \draw[rounded corners=6pt, thick, supercolor]
        (12.8,-2.75) rectangle (15.8,-6.9);
    \node[anchor=north, font=\footnotesize, align=center, color=supercolor]
        at (14.3,-2.8) {\textbf{All-or-nothing}\\\textbf{scheduling}$^\dagger$};

    \draw[rounded corners=10pt, very thick, sscolor, dashed]
        (0.2,-1.05) rectangle (8.1,-7.1);
    \node[anchor=north west, font=\footnotesize, color=sscolor]
        at (0.28,-1.13) {\textbf{Support-submodular}};

    \draw[rounded corners=10pt, thick, moncolor]
        (4.1,-0.75) rectangle (12.6,-7.4);
    \node[anchor=north east, font=\footnotesize, color=moncolor]
        at (12.52,-0.83) {\textbf{PF-monotone}};

    \draw[rounded corners=6pt, thick, polycolor]
        (4.3,-1.25) rectangle (7.9,-4.2);
    \node[anchor=north, font=\footnotesize, align=center, color=polycolor]
        at (6.1,-1.3) {\textbf{Polymatroids}$^\star$};

    \draw[rounded corners=6pt, thick, unrelcolor]
        (0.4,-1.85) rectangle (7.7,-4.0);
    \node[font=\footnotesize, align=center, color=unrelcolor]
        at (2.25,-2.925) {\textbf{Unrelated machines}$^\star$};

    \draw[rounded corners=4pt, thick]
        (4.5,-2.05) rectangle (7.5,-2.65);
    \node[font=\footnotesize, align=center]
        at (6.0,-2.35) {Related machines$^\star$};

    \draw[rounded corners=4pt, thick]
        (4.5,-2.85) rectangle (7.5,-3.8);
    \node[font=\footnotesize, align=center]
        at (6.0,-3.325) {Restricted\\ assignment$^\star$};

    \draw[rounded corners=6pt, thick]
        (4.3,-4.4) rectangle (7.9,-5.35);
    \node[font=\footnotesize, align=center]
        at (6.1,-4.875) {\textbf{Generalized flow}\\ \textbf{scheduling}$^\star$};

    \draw[rounded corners=4pt, thick]
        (4.5,-5.75) rectangle (7.5,-6.7);
    \node[font=\footnotesize, align=center]
        at (6.0,-6.225) {Concave speedup\\ curves$^\star$};

    \draw[rounded corners=6pt, thick, multicolor]
        (4.3,-5.55) rectangle (12.4,-6.9);
    \node[font=\footnotesize, align=center, color=multicolor]
        at (10.25,-6.25) {\textbf{Multidimensional}\\ \textbf{resource scheduling with}\\ \textbf{substitutes (RS-S)}$^{\star,\dagger}$};
\end{tikzpicture}
\caption{
    Schematic relationships among the principal subclasses of PSP that we consider.
    Nesting indicates containment, but disjoint or partially overlapping boxes do not indicate disjoint classes. E.g., every class shown contains the single machine.
    The marks~$\star$ and~$\dagger$ refer to the support-submodularity results in \cref{sec:ssPSP} and the counterexamples in \cref{sec:non-support-submodular-PSP}, respectively. Detailed descriptions are also given in~\cite{ImKM18,ChenIP25}.}
\label{fig:PSP-Venn-diagram}
\end{figure}

\paragraph{Support-submodular PSP.}
We identify a novel class of \emph{support-submodular} polytopes
on which PF is $2$-competitive for weighted completion time and $(2+\eps)$-speed $O(1/\eps)$-competitive for weighted flow time.
This class contains \emph{both} unrelated machines and polymatroids. Moreover, it captures
generalized flow scheduling~\cite{ChenIP25} and
resource scheduling with substitutes (RS-S) with separable utilities~\cite{ImKM18}, including concave speedup curves (\Cref{fig:PSP-Venn-diagram,sec:ssPSP}).
Informally, support submodularity requires that, for every nonnegative linear objective over rate allocations, the maximum value attainable by a set of jobs exhibits diminishing returns as jobs are added.

\begin{definition}[Support-submodular PSP]
    \label{def:ssPSP-intro}
    A downward-closed polytope $\cP \subseteq \RR^n_{\ge 0}$ is \emph{support submodular} if for every $\alpha \in \RR^n_{\ge 0}$ the support function $H_\alpha \colon 2^{[n]} \to \RR_{\geq0}$, $U \mapsto \max\{\langle \alpha, y\rangle \mid y \in \cP_U\}$, is submodular, where $\cP_U \coloneqq \{y \in \cP \mid y_j = 0 \ \forall j \notin U\}$.
    A \emph{support-submodular PSP} is a polytope scheduling problem whose rate polytope is support submodular.
\end{definition}

Our main algorithmic results rest on a characterization of support submodularity via the optimal weighted completion time,
viewed as the function $p \mapsto \opt(p)$ of processing requirements $p \in \RR^n_{\ge 0}$.

\begin{theorem}[Characterization]\label{thm:main-concave}
    A downward-closed polytope $\cP \subseteq \RR^n_{\ge 0}$ is support submodular if and only if, for every weight vector $w \in \RR^n_{>0}$, the optimal weighted completion time $p \mapsto \opt(p)$ with uniform release dates is concave on $\RR^n_{\ge 0}$.
\end{theorem}

This characterization allows us to apply a simplified variant of the superadditivity framework of Jäger, Lindermayr, and Megow~\cite{JLM2026} to obtain the following result.

\begin{restatable}[]{theorem}{maintheoremcompletiontime}
    \label{thm:main-completion}
    Proportional Fairness is $2$-competitive for minimizing weighted completion time in support-submodular PSP and $3$-competitive with arbitrary release dates.
\end{restatable}

Since the single-machine polytope is support submodular, no nonclairvoyant algorithm can achieve a ratio below~$2$ for the entire class~\cite{MotwaniPT94}.
Thus, \Cref{thm:main-completion} determines the optimal nonclairvoyant competitive ratio for weighted completion time with uniform release dates on unrelated machines and on all of support-submodular PSP.
We next put this result into perspective from three angles.
\begin{itemize}[nosep]
    \item \textbf{Unrelated machines vs.\ a single machine.}
    Offline, Smith's rule solves the problem efficiently on a single machine~\cite{smith1956various}.
    On unrelated machines, the problem remains solvable in strongly polynomial time for unit weights without preemption~\cite{Horn73,BrunoCS74}, but becomes APX-hard with preemption and migration~\cite{Sitters17} or with arbitrary weights~\cite{HoogeveenSW01}.
    Nevertheless, this increase in computational complexity causes essentially no loss in nonclairvoyant competitiveness, even among polynomial-time algorithms.
    \item \textbf{Nonclairvoyant vs.\ clairvoyant.}
    With arbitrary release dates, our nonclairvoyant ratio of~$3$ matches the best known guarantee of deterministic \emph{clairvoyant} online algorithms on unrelated machines with preemption and migration~\cite{BienkowskiKL21}, which learn the exact processing requirements upon job arrival.
    \item \textbf{Nonclairvoyant vs.\ offline.}
    Our results imply that PF is a polynomial-time $(2+\eps)$-approximation for support-submodular PSP,\footnote{We lose a factor of~$1+\eps$ to approximate the instantaneous convex optimization problem in polynomial time. We give details in \Cref{sec:pf-polynomial-time}.} which precisely matches the best known \emph{offline} approximation ratio for weighted completion time in PSP~\cite{JLM2025,LindermayrLM26}.
\end{itemize}

The structural insights underlying \Cref{thm:main-concave}
are also the key to our main result for flow time.

\begin{restatable}[]{theorem}{maintheoremflowtime}
    \label{thm:main-flow}
    For every $\eps>0$, Proportional Fairness is $O(1/\eps)$-competitive for minimizing weighted flow time in support-submodular PSP with speed augmentation~$2+\eps$.
\end{restatable}

This guarantee is tight for PF in both speed and competitive ratio, even on a single machine~\cite{Edmonds00}.
Im, Kulkarni, and Munagala~\cite{ImKM18} showed that PF is $(e+\eps)$-speed $O(1/\eps^2)$-competitive on PF-monotone polytopes, where removing active jobs never decreases the PF rates of the remaining jobs.
We improve both the required speed and the dependence on~$\eps$ relative to~\cite{ImKM18} for polytopes that are both PF-monotone and support submodular, such as polymatroids, generalized flow scheduling, and RS-S with separable utilities.
The two classes are incomparable (\Cref{fig:PSP-Venn-diagram}), as \mbox{RS-S}~\cite{ImKM18} is PF-monotone but in general not support submodular (\Cref{subsec:multidim-ssPSP,subsec:rho-leq-1}), whereas unrelated machines are support submodular (\Cref{subsec:unrel-ssPSP}) but not PF-monotone~\cite{JLM2026}.
Hence, our theorem \emph{strictly enlarges} the set of polytopes for which PF is known to be $O(1)$-speed $O(1)$-competitive for weighted flow time.
In particular, we show that PF is $(2+\eps)$-speed $O(1/\eps)$-competitive on unrelated machines.
This refutes the belief of Chen, Im, and Petety~\cite{ChenIP25} that PF is not $O(1)$-competitive on unrelated machines with $O(1)$-speed augmentation.
The belief is in line with the result of Gupta, Im, Krishnaswamy, Moseley, and Pruhs~\cite{GuptaIKMP12}, who showed that the most natural extension of weighted Round-Robin to heterogeneous machines is not $O(1)$-speed $O(1)$-competitive for weighted flow time, concluding that scheduling heterogeneous processors ``isn't as easy as you think''.
In this paper, we show that with Proportional Fairness, it \emph{is} as easy as you think.
A more complex nonclairvoyant algorithm is known to be $(1+\eps)$-speed $O(1/\eps^2)$-competitive on unrelated machines~\cite{ImKMP14}, but it is tailored to this setting.

\paragraph{General PSP.}
We show that no (randomized) nonclairvoyant algorithm can be $2$-competitive for total completion time with uniform release dates.
To our knowledge, this is the \emph{first} lower bound for PSP beyond the single-machine lower bound of~$2$ from 1994~\cite{MotwaniPT94}.
\begin{restatable}[]{theorem}{mainrandomizedLB}
    \label{thm:randomized-LB}
    The competitive ratio of every randomized nonclairvoyant algorithm for total completion time in PSP is at least~$2+10^{-5}$, even for graph scheduling with uniform release dates.
\end{restatable}
Since this bound holds for every nonclairvoyant algorithm, the separation between support-submodular PSP and general PSP is intrinsic to the problem rather than an artifact of PF or of our analysis.
Together, \Cref{thm:main-completion,thm:randomized-LB} resolve the open question discussed above in both directions: unrelated machines are no harder than a single machine for nonclairvoyant weighted completion time, whereas PSP in general is strictly harder.

On the technical side, our concavity-based approach does not yield any competitiveness guarantee for general PSP:
by \Cref{thm:main-concave}, whenever the polytope is not support submodular, the optimal weighted completion time is not concave for some weights, and in general it does not even admit a constant-factor concave approximation (\Cref{sec:unbounded-concavity}).
Instead, we give a novel dual-fitting analysis of PF, which significantly improves the previous bounds of~$27$~\cite{JLM2026} and~$128$~\cite{ImKM18}.

\begin{restatable}[]{theorem}{maintheoremgeneralPSP}
    \label{thm:general-psp-completion}
    Proportional Fairness is $4$-competitive for minimizing weighted completion time in PSP with arbitrary release dates.
\end{restatable}

Finally, for weighted flow time, we give the \emph{first} $O(1)$-speed $O(1)$-competitive nonclairvoyant algorithm for \emph{graph scheduling}~\cite{DinitzM20,RohwedderS25,LindermayrLM26,RohwedderS26}, where jobs are the edges of an undirected multigraph and rate vectors must lie in the graph's matching polytope.
Graph scheduling is in general neither support submodular nor PF-monotone, its optimal weighted completion time need not be concave, even with unit weights (cf.~\Cref{subsec:graph-scheduling}), and by \Cref{thm:randomized-LB} it is strictly harder than a single machine for completion time.
Nevertheless, we establish a relaxed structural property that yields the following result.
\begin{theorem}\label{thm:main-flow-graph-scheduling}
	For every $\eps > 0$, Proportional Fairness is $(4+\eps)$-speed $O(1/\eps)$-competitive for minimizing weighted flow time in graph scheduling.
\end{theorem}
In fact, \Cref{thm:main-flow-graph-scheduling} is a special case of a more general result.
The matching polytope is the integer hull of the fractional matching polytope (the degree-constraint relaxation), whose constraint matrix is the incidence matrix of the graph and has column sparsity~$2$.
We show that, for every~$c \in \NN$, PF is $(2c+\eps)$-speed $O(c/\eps)$-competitive on every downward-closed polytope between a binary packing polytope with at most~$c$ ones per column and its integer hull (see \Cref{cor:flow-time-c-sparse-binary}).

\subsection{Further Related Work}

Kashiwabara and Takabatake~\cite{KT03} studied downward-closed polyhedra whose support function is submodular on the entire nonnegative lattice.
Support submodularity only requires submodularity among the restrictions~$\alpha\odot\bm{1}_U$ of a fixed nonnegative vector~$\alpha$, where~$\bm{1}_U$ denotes the indicator vector of a job set~$U$ and~$\odot$ the coordinate-wise product, and to our knowledge it has not been studied previously.
We show that the condition of~\cite{KT03} implies support submodularity and PF-monotonicity and holds for polymatroids, generalized flow scheduling, and separable RS-S (\Cref{sec:ssPSP}).
Unrelated machines are support submodular but not PF-monotone~\cite{JLM2026}, so they violate the condition of~\cite{KT03}, and hence support-submodular PSP is strictly broader.

\paragraph{Nonclairvoyant weighted completion time.}
PF is known to be $2$-competitive on a single machine~\cite{MotwaniPT94,KimC03}, on identical parallel machines~\cite{DengGBL00,BBEM12}, and,
with unit weights, on related machines and machines with restricted assignment~\cite{JLM2026}.
Nonclairvoyant completion-time scheduling with precedence constraints on a single machine has also been studied~\cite{GargGKS19,JagerW24,LassotaLMS23}.
With arbitrary release dates, PF has a competitive ratio of at least~$2.19$ even on a single machine~\cite{JLM2026}, although a different algorithm is $2$-competitive in this setting~\cite{JaegerSSW24}.
For minimizing makespan, PF is $O(\log n)$-competitive, which is best possible~\cite{LindermayrLM26,ShmoysWW95}.

\paragraph{Clairvoyant weighted completion time.}
In clairvoyant online scheduling, processing requirements are revealed when jobs arrive.
On unrelated machines with preemption and migration, Bienkowski, Kraska, and Liu~\cite{BienkowskiKL21} refined the doubling framework of~\cite{HallSSW97}, improving the competitive ratio from~$4$~\cite{HallSSW97} to~$3$ for deterministic algorithms and from~$2.886$~\cite{ChakrabartiPSSSW96} to~$2.443$ for randomized algorithms.
Thus, PF matches their deterministic clairvoyant bound of~$3$, and, in contrast to the algorithm of~\cite{BienkowskiKL21},
PF can be implemented in polynomial time while losing only a factor of~$1+\eps$ (see \Cref{sec:pf-polynomial-time}).
The previously best known polynomial-time guarantee for clairvoyant weighted completion time was~$4.62$~\cite{JLM2026}.

\paragraph{Nonclairvoyant weighted flow time.}
Every randomized nonclairvoyant algorithm has a competitive ratio in $\Omega(\log n)$, even with unit weights~\cite{MotwaniPT94}, and this bound is achieved~\cite{BecchettiL04}.
In their seminal paper, Kalyanasundaram and Pruhs~\cite{KalyanasundaramP00} showed that a simple nonclairvoyant algorithm for unit weights is $O(1)$-competitive with speed~$1+\eps$; such an algorithm is called \emph{scalable}. This was later extended to weighted jobs~\cite{KimC03,BansalD07} and identical parallel machines~\cite{ChekuriGKK04}.
On unrelated machines, SelfishMigrate is $(1+\eps)$-speed $O(1/\eps^2)$-competitive for weighted flow time~\cite{ImKMP14}.
Several related models have received problem-specific treatment. For example, scalable nonclairvoyant algorithms are known for unweighted flow time with sublinear speedup curves~\cite{EdmondsP12}, broadcast scheduling~\cite{BansalKN14}, and more general flow-time norms~\cite{EdmondsIM11}.
For related machines and unit weights, Gupta, Krishnaswamy, and Pruhs~\cite{GuptaKP11} showed that PF is $(2+\eps)$-speed $O(1/\eps^3)$-competitive, and a scalable algorithm is known~\cite{GuptaIKMP12}.
Beyond these models, Im, Kulkarni, Moseley, and Munagala~\cite{ImKMM16MultiClusterPSP} gave an $(e+\eps)$-speed $O(1/\eps^2)$-competitive algorithm for weighted flow time on monotone multicluster PSP that uses PF on each cluster.
Constant guarantees in the unit-weight setting on a single machine can also be achieved via processing time predictions~\cite{AzarLT21,AzarLT22,GuptaKPW26}, $\eps$-clairvoyance~\cite{GuptaKLSY25}, or both~\cite{GuptaKLSY27}.

\paragraph{Clairvoyant weighted flow time.}
Anand, Garg, and Kumar~\cite{AnandGK12} showed that a greedy algorithm is scalable and $O(1/\eps)$-competitive on unrelated machines. Chen, Im, and Petety~\cite{ChenIP25} use gradient descent on residual scheduling objectives to obtain scalable clairvoyant algorithms under substitutes conditions. We analyze nonclairvoyant PF through an exact characterization of concavity. Our support-submodularity condition is strictly weaker than their general rate-polytope substitutes hypothesis, which implies the lattice submodularity studied in \Cref{subsec:lattice-ssPSP}. They treat unrelated machines separately. We give more details on this comparison in \Cref{sec:comparison-chen-im-petety}.
For graph scheduling, Dinitz and Moseley gave a $(2+\eps)$-speed $O(1/\eps^2)$-competitive algorithm~\cite{DinitzM20}.
Without speed augmentation on a single machine, total flow time can be solved optimally with Shortest-Remaining-Processing-Time first (SRPT)~\cite{Schrage68}; for weighted flow time, no deterministic online algorithm is $O(1)$-competitive~\cite{BansalC09}, but parameterized upper bounds are known~\cite{AzarT18,BansalD07,ChekuriKZ01}.

\paragraph{Offline approximations.}
In the offline setting, where all jobs and their processing requirements are known in advance, there is a $(2+\eps)$-approximation for weighted completion time in general PSP~\cite{JLM2026,LindermayrLM26}.
For the special case of unrelated machines, a stronger $1.698$-approximation is known~\cite{Sitters17}, and for related machines, even a PTAS is possible~\cite{ChekuriK01}.
There is also a PTAS for minimizing weighted flow time on a single machine~\cite{ArmbrusterRW26}.

\subsection{Paper Organization}
In \Cref{sec:tech-overview}, we give an overview of our techniques and proofs of our main results.
We then give detailed preliminaries on PF and PSP in \Cref{sec:preliminaries} and prove support submodularity for various problems in \Cref{sec:ssPSP}. We give the full proofs of \Cref{thm:main-concave,thm:main-completion} in \Cref{sec:completion_time_results} and prove \Cref{thm:main-flow,thm:main-flow-graph-scheduling} in \Cref{sec:flow-time}. Finally, in \Cref{sec:general-psp}, we prove our results for general PSP.

\section{Technical Overview}\label{sec:tech-overview}

\subsection{Weighted Completion Time for Support-Submodular PSP}
\label{subsec:overview-completion}

We sketch the proofs of \Cref{thm:main-concave} and of the $2$-competitiveness in \Cref{thm:main-completion}.
First, a simplified variant of the superadditivity framework of Jäger, Lindermayr, and Megow~\cite{JLM2026} shows that PF is $2$-competitive whenever the optimal weighted completion time is a concave function of the processing requirements (\Cref{subsubsec:reduction-concavity}).
Second, and this is our main technical contribution, we show that support submodularity is precisely the property of the rate polytope that makes this function concave for all weights.

Fix a PSP instance with jobs~$J\coloneqq[n]$, weights~$w\in\RR^n_{>0}$, rate polytope~$\cP$, and all release dates equal to~$0$, and let~$w(U)\coloneqq\sum_{j\in U}w_j$ for~$U\subseteq J$.
For processing requirements $p\in\RR^n_{\geq0}$, let~$\pf(p)$ denote the weighted completion time of PF and~$\opt(p)$ the optimal weighted completion time.

\subsubsection{Reduction to Concavity}
\label{subsubsec:reduction-concavity}

The framework of~\cite{JLM2026} decomposes the PF schedule at completion times into \emph{structured} subinstances, in which PF completes all jobs simultaneously.
PF is~$2$-competitive on these structured subinstances and PF's total objective is the sum of these subinstance objectives, because PF is nonclairvoyant.
Hence, PF is $2c$-competitive whenever the optimal objective is \emph{$c$-superadditive}, that is, whenever for every decomposition of the processing requirements, the optimal values of the parts sum to at most~$c$ times the optimal value of the whole instance.
For unrelated machines,~\cite{JLM2026} obtain~$c=1.81$ from the $1$-superadditivity of the optimal \emph{nonpreemptive} objective and Sitters' bound of~$1.81$ on the power of preemption~\cite{Sitters17}, which gives the ratio~$3.62$.
We show that~$c=1$ for every support-submodular polytope, using the following pointwise form of the reduction.

Let~$A(t)$ be the set of active jobs and~$y^{\textsc{pf}}(t)\in\cP$ the rate vector of PF at time~$t$, so that~$\pf(p)=\int_0^\infty w(A(t))\dd t$.
The first-order optimality conditions of PF's convex program yield the bound~$\opt(y^{\textsc{pf}}(t))\geq\frac12w(A(t))$ at every time $t$ (\cref{lem:subinstance-of-structured-instance}), which~\cite{JLM2026} established via dual fitting.
The function~$p\mapsto\opt(p)$ is positively homogeneous by time scaling, so it is concave if and only if it is superadditive.
Since~$\int_0^\infty y^{\textsc{pf}}(t)\dd t=p$, concavity thus implies~$\int_0^\infty\opt(y^{\textsc{pf}}(t))\dd t\leq\opt(p)$, and hence
\begin{align*}
    \pf(p)
    =\int_0^\infty w(A(t))\dd t
    \leq 2\int_0^\infty \opt(y^{\textsc{pf}}(t))\dd t
    \leq 2\cdot\opt(p). 
\end{align*}
Consequently, the online analysis reduces to an offline geometric question: \emph{when is the optimal weighted completion time concave in the processing requirements~$p$?}

\subsubsection{Warm-up: A Single Machine}
\label{subsubsec:single-machine-warm-up}

Let~$\mathcal S_n$ be the set of permutations of~$J$, which we call \emph{orders}.
For~$\pi\in\mathcal S_n$ and~$b\in[n]$, let~$S_b^\pi\coloneqq\{\pi(b),\ldots,\pi(n)\}$ be the suffix of~$\pi$ beginning at position~$b$, and let~$S_{n+1}^\pi\coloneqq\emptyset$.
On a single machine, consider the schedule that processes the jobs one after another in the order~$\pi$.
While it processes job~$\pi(b)$, the unfinished jobs are contained in~$S_b^\pi$, so its weighted completion time is at most
\begin{equation*}
    \sum_{b=1}^n w(S_b^\pi)\,p_{\pi(b)}=\scp{\alpha^\pi,p},
    \qquad\text{where}\quad \alpha^\pi_{\pi(b)}\coloneqq w(S_b^\pi).
\end{equation*}
Since an optimal schedule achieves this for its completion time order, $\opt(p)=\min_{\pi\in\mathcal S_n}\scp{\alpha^\pi,p}$ is a minimum of finitely many linear functions and hence concave.
The same holds for nonpreemptive schedules on unrelated machines, because fixing an assignment of the jobs to the machines and an order on each machine again gives a weighted completion time that is linear in~$p$~\cite{Horn73,BrunoCS74}.
This yields the superadditivity of the nonpreemptive optimum used in~\cite{JLM2026}.

Preemption and migration destroy this representation, because a job may receive processing from several machines while other jobs progress simultaneously.
Nevertheless, for every rate polytope, we construct price vectors~$\alpha^\pi\in\RR^n_{\geq0}$ that depend on the order~$\pi$, the weights, and the polytope, but not on~$p$, and we show that~$\opt(p)=\min_{\pi\in\mathcal S_n}\scp{\alpha^\pi,p}$ whenever the polytope is support submodular.

\subsubsection{Fixed-Order Bounds}
\label{subsubsec:fixed-order}

For~$\pi\in\mathcal S_n$, let~$\opt_\pi(p)$ be the infimum of the weighted completion times of the schedules for~$p$ that complete the jobs in the order~$\pi$, that is, with~$C_{\pi(1)}\leq\cdots\leq C_{\pi(n)}$, where~$\inf\emptyset\coloneqq\infty$.
Since every schedule completes the jobs in some order, we have~$\opt(p)=\min_{\pi\in\mathcal S_n}\opt_\pi(p)$.
With preemption,~$\opt_\pi$ need not be concave.
For example, for two unit-weight jobs on two identical parallel machines, we have $\opt_{\operatorname{id}}(p)=p_1+\max\{p_1,p_2\}$ for~$p_2>0$.
Instead of~$\opt_\pi$, we use the linear function~$p\mapsto\scp{\alpha^\pi,p}$ for the price vector~$\alpha^\pi$ constructed below, and we show that, for a support-submodular polytope and every~$p\in\RR^n_{\geq0}$, it holds that
\begin{equation}
    \opt(p)\leq\scp{\alpha^\pi,p}\leq\opt_\pi(p).
    \label{ineq:fixed-order-bound}
\end{equation}
Taking the minimum over all orders then gives~$\opt(p)\leq\min_{\pi}\scp{\alpha^\pi,p}\leq\min_{\pi}\opt_\pi(p)=\opt(p)$, so equality holds throughout.

\paragraph{Schedule budgets.}
For both inequalities in~\eqref{ineq:fixed-order-bound}, we use the same technique to compare prices to weighted completion time.
Consider a feasible schedule for~$p\in\RR^n_{\ge0}$ with rate vectors~$y(t)$, sets~$U(t)$ of unfinished jobs, completion times $C_1,\ldots,C_n$, and remaining processing requirements~$p(t)$, and let~$\alpha\in\RR^n_{\geq0}$.
The weighted completion time of the schedule accrues at rate~$w(U(t))$, whereas its \emph{budget}~$\scp{\alpha,p(t)}$ decreases at rate~$\scp{\alpha,y(t)}$.
Since rates are zero after completion, we have~$y(t)\in\cP_{U(t)}$, so this rate is at most the value of the support function~$H_\alpha(U(t))=\max_{y\in\cP_{U(t)}}\scp{\alpha,y}$.
Integrating over time gives
\begin{equation}
    \sum_{j\in J}w_jC_j=\int_0^\infty w(U(t))\dd t
    \qquad\text{and}\qquad
    \scp{\alpha,p}=\int_0^\infty\scp{\alpha,y(t)}\dd t\leq\int_0^\infty H_\alpha(U(t))\dd t.
    \label{eq:overview-budget}
\end{equation}
Hence, the weighted completion time of the schedule is at least~$\scp{\alpha,p}$ if~$H_\alpha(U(t))\leq w(U(t))$ at every time~$t$, and at most~$\scp{\alpha,p}$ if~$\scp{\alpha,y(t)}\geq w(U(t))$ at every time~$t$.

\paragraph{Prices with tight suffixes.}
A schedule that completes the jobs in the order~$\pi$ has only suffixes of~$\pi$ as sets of unfinished jobs, i.e., for all times $t$ there is a $b \in [n+1]$ such that $U(t) = S_b^\pi$.
For the second inequality in~\eqref{ineq:fixed-order-bound}, it therefore suffices that~$H_{\alpha^\pi}(S_b^\pi)\leq w(S_b^\pi)$ for every~$b \in [n]$. 
We call a set~$S\subseteq J$ \emph{tight} for~$\alpha$ if~$H_\alpha(S)=w(S)$, and we construct~$\alpha^\pi$ such that every suffix of~$\pi$ is tight, that is, for every~$b\in[n+1]$, it holds that 
\begin{equation}\label{eq:tight-suffixes}
    H_{\alpha^\pi}(S_b^\pi)=\max_{y\in\cP_{S_b^\pi}}\scp{\alpha^\pi,y}=w(S_b^\pi).
\end{equation}
To achieve this, we construct the prices backwards along~$\pi$, starting from~$\alpha=0$.
When we reach position~$b$ in $\pi$, the suffix~$S_{b+1}^\pi$ is tight and~$\alpha_{\pi(b)}=0$, so~$H_\alpha(S_b^\pi)=H_\alpha(S_{b+1}^\pi)=w(S_{b+1}^\pi)<w(S_b^\pi)$, since by downward-closedness, job~$\pi(b)$ can be given rate zero without changing~$\scp{\alpha,y}$.
We raise~$\alpha_{\pi(b)}$ until~$S_b^\pi$ is tight, which is possible because~$H_\alpha(S_b^\pi)$ is continuous in~$\alpha_{\pi(b)}$ and tends to infinity, as job~$\pi(b)$ can be processed at a positive rate.
The later suffixes $S_{b+1}^\pi,\ldots,S_n^\pi$ do not contain~$\pi(b)$, so they remain tight (see \Cref{lem:reverse-order-prices}).
On a single machine, we have~$H_\alpha(U)=\max_{j\in U}\alpha_j$ for nonempty~$U\subseteq J$, and the construction yields the prices~$\alpha^\pi_{\pi(b)}=w(S_b^\pi)$ of \cref{subsubsec:single-machine-warm-up}.

\begin{lemma}\label{lem:bound-fixed-order}
    For every~$\pi\in\mathcal S_n$ and every~$p\in\RR^n_{\geq0}$, we have~$\scp{\alpha^\pi,p}\leq\opt_\pi(p)$.
\end{lemma}
\begin{proof}[Proof sketch]
    Every schedule for~$p$ that completes the jobs in order~$\pi$ satisfies~$U(t)\in\{S_1^\pi,\ldots,S_{n+1}^\pi\}$, and hence~$H_{\alpha^\pi}(U(t))=w(U(t))$ by~\eqref{eq:tight-suffixes}, at every time~$t$.
    By~\eqref{eq:overview-budget}, its weighted completion time is at least~$\scp{\alpha^\pi,p}$.
\end{proof}
This lower bound holds for every rate polytope, and we need support submodularity only for the first inequality in~\eqref{ineq:fixed-order-bound}.
\Cref{lem:bound-fixed-order} also follows from LP duality, since~$\alpha^\pi$ extends to a feasible dual solution of a linear program that models the problem of scheduling the jobs in the order~$\pi$ (\Cref{rem:fixed-order-lp}).

\subsubsection{Greedy Schedules and Support Submodularity}
\label{subsubsec:greedy}

For the first inequality in~\eqref{ineq:fixed-order-bound}, we need a schedule for~$p$ whose budget decreases at least as fast as its weighted completion time accrues.
The \emph{$\alpha$-greedy schedule} uses at every time~$t$ the rate vector~$y^{U(t)}\in\argmax_{y'\in\cP_{U(t)}}\scp{\alpha,y'}$, where~$U(t)$ is its own set of unfinished jobs, so its budget decreases at the largest possible rate~$H_\alpha(U(t))$.
By~\eqref{eq:overview-budget}, its weighted completion time is at most~$\scp{\alpha,p}$, provided that every set~$U\subseteq J$ satisfies
\begin{equation}
    H_\alpha(U)\geq w(U).
    \label{eq:overview-all-set-domination}
\end{equation}
For~$\alpha=\alpha^\pi$, the suffix equalities~\eqref{eq:tight-suffixes} give~\eqref{eq:overview-all-set-domination} only for the suffixes of~$\pi$, but the $\alpha^\pi$-greedy schedule need not complete the jobs in the order~$\pi$.
Our key insight is that support submodularity extends~\eqref{eq:overview-all-set-domination} from the suffixes of~$\pi$ to all sets.
This is the only place where we use support submodularity.

\begin{lemma}\label{lem:overview-tight-chain}
    Consider a support-submodular PSP instance, and let~$\pi\in\mathcal S_n$.
    Then~$H_{\alpha^\pi}(U)\geq w(U)$ for all~$U\subseteq J$.
    Hence, for every~$p\in\RR^n_{\geq0}$, the $\alpha^\pi$-greedy schedule for~$p$ has weighted completion time at most~$\scp{\alpha^\pi,p}$, and~$\opt(p)\leq\scp{\alpha^\pi,p}$.
\end{lemma}
\begin{proof}[Proof sketch]
    Fix~$U\subseteq J$, and add the jobs in~$U$ one at a time to the empty set, in reverse order of their positions in~$\pi$.
    When job~$j=\pi(b)$ is added, the set of jobs added so far is~$U\cap S_{b+1}^\pi\subseteq S_{b+1}^\pi$.
    By submodularity of~$H_{\alpha^\pi}$ and~\eqref{eq:tight-suffixes}, it holds that
    \begin{equation*}
        H_{\alpha^\pi}\bigl((U\cap S_{b+1}^\pi)\cup\{j\}\bigr)-H_{\alpha^\pi}(U\cap S_{b+1}^\pi)
        \geq H_{\alpha^\pi}(S_b^\pi)-H_{\alpha^\pi}(S_{b+1}^\pi)
        =w_j.
    \end{equation*}
    Telescoping over the jobs in~$U$ gives~$H_{\alpha^\pi}(U)\geq w(U)$.
    Therefore, while jobs remain unfinished, the $\alpha^\pi$-greedy schedule decreases its budget at rate~$H_{\alpha^\pi}(U(t))\geq w(U(t))>0$.
    It thus completes all jobs, and by~\eqref{eq:overview-budget}, its weighted completion time is at most~$\scp{\alpha^\pi,p}$.
\end{proof}

On a single machine, the $\alpha^\pi$-greedy schedule always processes the unfinished job with the largest price, which is the first unfinished job in~$\pi$, so it is exactly the schedule of \cref{subsubsec:single-machine-warm-up} that processes the jobs in the order~$\pi$.

\subsubsection{Completing the Proofs}
\label{subsubsec:overview-completing}

\Cref{lem:bound-fixed-order,lem:overview-tight-chain} establish~\eqref{ineq:fixed-order-bound} for every order~$\pi$, so every support-submodular PSP instance satisfies, for every~$p\in\RR^n_{\geq0}$,
\begin{equation}
    \opt(p)=\min_{\pi\in\mathcal S_n}\scp{\alpha^\pi,p}.
    \label{eq:overview-opt-min-linear}
\end{equation}
As a minimum of finitely many linear functions,~$p\mapsto\opt(p)$ is concave, which proves the forward direction of \cref{thm:main-concave} (see \cref{thm:support-submodular-psp-concavity}) and, by \cref{subsubsec:reduction-concavity}, the $2$-competitiveness in \cref{thm:main-completion}.
Moreover, the $\alpha^\pi$-greedy schedule for an order~$\pi$ attaining the minimum is optimal.
If an optimal schedule for~$p$ completes the jobs in the order~$\pi$, then~$\opt_\pi(p)=\opt(p)$, so~\eqref{ineq:fixed-order-bound} gives~$\opt(p)=\scp{\alpha^\pi,p}$, and  with \cref{lem:overview-tight-chain}, this makes~$\alpha^\pi$ a supergradient of~$p\mapsto\opt(p)$ at~$p$.

For the backward direction of \Cref{thm:main-concave}, suppose that~$H_\alpha(S\cup\{i,j\})+H_\alpha(S)>H_\alpha(S\cup\{i\})+H_\alpha(S\cup\{j\})$ for some~$\alpha\in\RR^n_{\geq0}$, $S\subseteq J$, and distinct jobs~$i,j\notin S$.
Since every weight vector $w\in\RR_{>0}^n$ satisfies $w(S\cup\{i,j\})+w(S)=w(S\cup\{i\})+w(S\cup\{j\})$, we can choose weights such that~$H_\alpha(S\cup\{i,j\})>w(S\cup\{i,j\})$, whereas the sets of a chain~$S=S_1\supsetneq\cdots\supsetneq S_{m+1}=\emptyset$ are tight and~$H_\alpha(U)\leq w(U)$ for~$U\in\{S\cup\{i\},S\cup\{j\}\}$.
We then construct~$p=p^{(i)}+p^{(j)}$ such that all sets of unfinished jobs of the optimal schedules for~$p^{(i)}$ and~$p^{(j)}$ are among~$S\cup\{i\}$, $S\cup\{j\}$, and the sets of the chain, whereas some schedule for~$p$ first uses a maximizer of~$\scp{\alpha,y}$ over~$\cP_{S\cup\{i,j\}}$ and then maximizers over the sets of the chain.
By~\eqref{eq:overview-budget}, this gives~$\opt(p^{(i)})+\opt(p^{(j)})\geq\scp{\alpha,p}>\opt(p)$, so~$p\mapsto\opt(p)$ is not superadditive and thus not concave (see \Cref{subsec:concavity-implies-support-submodularity}). Hence, if~$p\mapsto\opt(p)$ is concave for all $w\in\RR_{>0}^n$, then the polytope is support submodular. However, concavity for one weight vector (e.g.\ unit weights) does not imply support submodularity of the polytope (\Cref{prop:triangle-concave}).

\subsection{Weighted Flow Time for Support-Submodular PSP and Beyond}
\label{subsec:overview-flow}

We sketch the proof of \Cref{thm:main-flow}, which is based on a potential-function analysis.
We compare PF with speed~$2+\eps$ against an arbitrary fixed feasible schedule~$y^*=(y^*(t))_{t\geq0}$ with unit speed, and we write~$y^{\textsc{pf}}(t)\in\cP$ for the rate vector of PF at time~$t$, so PF processes each job~$j$ at rate~$(2+\eps)\,y_j^{\textsc{pf}}(t)$.
For~$t\geq0$, let~$A(t)$ and~$A^*(t)$ denote the sets of active jobs at time~$t$ in the PF schedule and in~$y^*$, respectively, so that the weighted flow time of PF is~$\PF_{2+\eps}=\int_0^\infty w(A(t))\dd t$ and that of~$y^*$ is~$\int_0^\infty w(A^*(t))\dd t$.
We construct an absolutely continuous potential function~$\Phi\colon\RR_{\geq0}\to\RR_{\geq0}$ that starts and ends at zero and show that, at almost every time~$t\geq0$, it holds that
\begin{equation}
    w(A(t))+\frac{\dd}{\dd t}\Phi(t)\leq\gamma\cdot w(A^*(t)),
    \qquad\text{where}\quad \gamma\coloneqq\frac{2+\eps}{\eps}.
    \label{eq:overview-potential-inequality}
\end{equation}
Since~$\Phi$ is absolutely continuous, it is the integral of its derivative, which exists almost everywhere.
Hence, integrating this inequality over time bounds~$\PF_{2+\eps}$ by~$\gamma$ times the weighted flow time of~$y^*$, and taking the infimum over all feasible schedules~$y^*$ proves \Cref{thm:main-flow}.
In particular,~$\Phi$ does not jump when jobs are released or completed, so these events, which usually require a separate argument~\cite{ImMP11}, need no special treatment in our analysis.

The potential measures how far PF lags behind~$y^*$.
The \emph{lag} of job~$j$ at time~$t$ is defined by $z_j(t)\coloneqq\max\{0,\,p_j(t)-p_j^*(t)\}$, where~$p_j(t)$ and~$p_j^*(t)$ are the remaining processing requirements of~$j$ at time~$t$ in PF's schedule and in~$y^*$, respectively, with both set to~$0$ before~$r_j$.
The \emph{lag instance} at time~$t$ has processing requirements~$z(t)$ and all release dates~$0$.
In many potential-function analyses, the potential is a multiple of the algorithm's objective value on the lag instance~\cite{ImMP11}.
The PF analysis of~\cite{ImKM18} uses a potential built from the lags and the PF rates on sets of jobs ordered by release date, and controlling its change requires PF-monotonicity, which excludes unrelated machines~\cite{JLM2026}.
Instead, we use the \emph{optimal} weighted completion time of the lag instance and define~$\Phi(t)\coloneqq\gamma\cdot\OPT_0(z(t))$, where~$\OPT_0$ is the function~$\opt$ of \cref{subsec:overview-completion}, for which all jobs are released at time~$0$.
This lets the analysis use the price vectors of \cref{subsec:overview-completion} as supergradients of~$\OPT_0$.
For every~$p\in\RR^n_{\geq0}$, fix an optimal schedule for processing requirements~$p$ and release dates~$0$, let~$\pi_p$ be an order in which this schedule completes the jobs, and let~$\alpha(p)\coloneqq\alpha^{\pi_p}$, which is a supergradient of~$\OPT_0$ at~$p$ as discussed in \cref{subsubsec:overview-completing}.

The main ingredient is the following lemma, where~$L(t)\coloneqq\{j \mid z_j(t)>0\}$ is the set of lagging jobs.
\begin{lemma}
    The potential~$\Phi$ is absolutely continuous, and at almost every time~$t$ with~$w(A(t))>0$, 
    \begin{equation*}
        \frac{\dd}{\dd t}\Phi(t)\leq\gamma\cdot\Bigl(w(L(t))-(2+\eps)\frac{w(L(t))^2}{2w(A(t))}\Bigr).
    \end{equation*}
\end{lemma}
\begin{proof}[Proof sketch]
    Each lag~$z_j$ is Lipschitz continuous, since both schedules process~$j$ at bounded rates and~$z_j$ does not jump at~$r_j$.
    Moreover,~$\OPT_0$ is Lipschitz continuous as a minimum of finitely many linear functions by~\eqref{eq:overview-opt-min-linear}.
    Hence,~$\Phi$ is Lipschitz continuous and thus absolutely continuous.
    Fix a time~$t$ with~$w(A(t))>0$ at which~$\Phi$ and~$z$ are differentiable.
    Since~$\alpha(z(t))$ is a supergradient of~$\OPT_0$ at~$z(t)$, we have~$\frac{\dd}{\dd t}\OPT_0(z(t))=\scp{\alpha(z(t)),\frac{\dd}{\dd t}z(t)}$.
    A job~$j\notin L(t)$ satisfies~$z_j(t)=0$. Since~$z_j$ is nonnegative, the value~$z_j(t)=0$ is a minimum of~$z_j$, and since~$z_j$ is differentiable at~$t$, this gives~$\frac{\dd}{\dd t}z_j(t)=0$.
    A job~$j\in L(t)$ satisfies~$\frac{\dd}{\dd t}z_j(t)=y_j^*(t)-(2+\eps)y_j^{\textsc{pf}}(t)$.
    Hence,
    \begin{equation*}
        \frac{\dd}{\dd t}\OPT_0(z(t))
        =\scp{\alpha(z(t)),y^*(t)\odot\bm1_{L(t)}}-(2+\eps)\scp{\alpha(z(t)),y^{\textsc{pf}}(t)\odot\bm1_{L(t)}}.
    \end{equation*}
    Since~$y^*(t)\odot\bm1_{L(t)}\in\cP_{L(t)}$ by downward-closedness, the first term is at most~$H_{\alpha(z(t))}(L(t))$.
    Jobs without lag have processing requirement~$0$ in the lag instance, so they complete at time~$0$ and precede the lagging jobs in~$\pi_{z(t)}$.
    Thus,~$L(t)$ is a suffix of~$\pi_{z(t)}$, and~\eqref{eq:tight-suffixes} gives~$H_{\alpha(z(t))}(L(t))=w(L(t))$.
    For the second term, \cref{lem:overview-tight-chain} gives~$\scp{\alpha(z(t)),y^{\textsc{pf}}(t)\odot\bm1_{L(t)}}\geq\OPT_0(y^{\textsc{pf}}(t)\odot\bm1_{L(t)})$.
    The same first-order optimality argument, applied to the jobs in~$L(t)$, gives~$\OPT_0(y^{\textsc{pf}}(t)\odot\bm1_{L(t)})\geq\frac12\frac{w(L(t))^2}{w(A(t))}$.
\end{proof}

Since~$L(t)\subseteq A(t)$ and~$A(t)\setminus L(t)\subseteq A^*(t)$, adding $\frac{w(A(t))}{2\eps}\bigl((2+\eps)\frac{w(L(t))}{w(A(t))}-2\bigr)^2 \ge 0$ turns this bound into~$w(A(t))+\frac{\dd}{\dd t}\Phi(t)\leq\gamma\cdot \bigl(w(A(t)) - w(L(t))\bigr) = \gamma\cdot w(A(t)\setminus L(t))\leq\gamma\cdot w(A^*(t))$, which is~\eqref{eq:overview-potential-inequality}.

\paragraph{Beyond support submodularity.}
The proof uses the prices~$\alpha(z(t))$ only through three properties of the map~$\alpha \colon \RR^n_{\ge 0} \to \RR^n_{\ge 0}$, which hold for all~$p,q\in\RR^n_{\geq0}$:
\begin{enumerate}[label=(\roman*),nosep]
	\item $\bar{C}(p)\coloneqq\scp{\alpha(p),p}\ge\OPT_0(p)$,
	\item $\scp{\alpha(p),p}\leq\scp{\alpha(q),p}$, which makes~$\bar{C}$ concave with supergradient~$\alpha(p)$ at~$p$, and
	\item $H_{\alpha(p)}(\supp(p))\leq w(\supp(p))$, where~$\supp(p)\coloneqq\{j\in J\mid p_j>0\}$.
\end{enumerate}
Allowing a factor~$c$ in (iii) gives a \emph{$c$-approximate certificate} (\Cref{def:c-approximate-cert}), and with the potential $\gamma\cdot \bar{C}(z(t))$ for a suitable~$\gamma$, the same analysis shows that PF is $(2c+\eps)$-speed $O(c/\eps)$-competitive (\Cref{thm:c-approximate-flow-time}).
We construct such certificates for every downward-closed polytope between a binary packing polytope with at most~$c$ ones per column and its integer hull (\Cref{thm:c-sparse-c-approx}).
Their prices come from relaxing the polytope into one single machine per packing constraint, on all of which the Smith order of~$p$ is optimal simultaneously.
These prices satisfy~\eqref{eq:overview-all-set-domination}, so, as in \cref{subsubsec:greedy}, the $\alpha(p)$-greedy schedule for~$p$ shows that~$F(p)\geq\OPT_0(p)$.
Graph scheduling satisfies this with~$c=2$, so PF is $(4+\eps)$-speed $O(1/\eps)$-competitive for this problem for every~$\eps>0$ (\Cref{thm:main-flow-graph-scheduling}).

\subsection{A Randomized Lower Bound for General PSP}
\label{subsec:overview-randomized-LB}

The factor~$2$ for weighted completion time on support-submodular polytopes cannot extend to general PSP, even for randomized nonclairvoyant algorithms (\Cref{thm:randomized-LB}).

Our construction uses graph scheduling on a path with four vertices, replacing its three edges by groups~$A$, $B$, and~$C$ of~$N$, $3N$, and~$2N$ parallel unit-weight jobs, respectively.
We choose random processing requirements such that every nonclairvoyant algorithm completes in expectation at most one job per unit of used capacity of the two inner vertices, and at most~$2N$ of these completions are attributed to the vertex shared by~$A$ and~$B$.
Hence, at most~$\min\{2t,t+2N\}$ jobs complete by time~$t$ in expectation, and the expected total completion time is at least~$10N^2$.
Processing each group at rate~$\frac12$ in nondecreasing order of processing requirements achieves~$5N^2+O(N)$ in expectation.
After~$A$ has completed, however, $B$ and~$C$ are constrained only by their shared vertex, and with probability~$1-O(1/N)$ group~$B$ then processes shorter jobs than~$C$.
Processing~$B$ before~$C$ during a time interval of length~$\Theta(N)$ thus reduces the total completion time by~$\Omega(N^2)$, and Yao's principle yields the ratio~$10/(5-9\cdot10^{-5})>2+10^{-5}$.

\subsection{Dual Fitting for General PSP}
\label{subsec:overview-general-psp}

Finally, we give an overview of the proof of \Cref{thm:general-psp-completion}.
For general PSP, the optimal-value function need not admit any concave approximation within a constant factor (see \cref{prop:unbounded-concavity}).
Thus, the concavity route used above cannot yield a constant-factor guarantee for general PSP.

We instead analyze PF directly by fitting a solution to the dual of the standard \emph{mean-busy-time relaxation}, as in previous analyses of PF~\cite{JLM2026,ImKM18}. Throughout this subsection, fix a PSP instance with arbitrary release dates, and let~$\pf$ and~$\opt$ denote the weighted completion times of PF and of an optimal schedule.
If we write the rate polytope as $\cP=\{y\in\RR^n_{\geq0}\mid By\leq\bm{1}\}$ with~$B\in\RR^{D\times n}_{\geq0}$, the continuous dual of this relaxation is as follows:
\begin{alignat}{3}
    \DLP\qquad \text{maximize}\quad
        & \sum_{j\in J}\alpha_j-\sum_{d\in[D]}\int_0^\infty\beta_{dt}\dd t \notag\\*
    \text{subject to}\quad
        & \alpha_j-w_jt\leq p_j\sum_{d\in[D]}B_{dj}\beta_{dt}
        &&\qquad \forall j\in J,\ \forall t\geq r_j, \notag \\*
        & \alpha,\beta\geq0. && \notag
\end{alignat}
Here,~$\alpha_j$ is the value assigned to job~$j$, while~$\beta_{dt}$ is the price of resource~$d$ at time~$t$.
The corresponding primal relaxation charges each job its mean busy time and lower-bounds~$\opt$.
Hence, by weak duality, every feasible dual solution has value at most~$\opt$, and it suffices to construct a feasible pair~$(\dualSa,\dualSb)$ with value at least~$\pf/4$.
We do so by ensuring that the job values contribute at least~$\pf/2$, while the resource prices cost at most~$\pf/4$.

Let~$y_j^{\textsc{pf}}(t)$ and~$C_j$ be the rate and completion time of job~$j$ under PF, let $A(t)$ be the set of active jobs at time~$t$, and let~$\eta_d(t)$ be the KKT multiplier of resource~$d$ in PF's convex program at time~$t$.
At every time~$t$, the KKT conditions give the two standard identities
$\frac{w_j}{y_j^{\textsc{pf}}(t)}=\sum_{d\in[D]}B_{dj}\eta_d(t)$ for every job $j \in A(t)$ and $\sum_{d\in[D]}\eta_d(t)=w(A(t))$.
As in \cref{subsec:general-psp-dual-fitting}, we remove zero-size jobs and assume~$p_j>0$. We use the following dual assignment.
\begin{alignat*}{3}
    \dualSa_j
    &\coloneqq w_j\int_0^{C_j}\min\biggl\{1,\frac{p_j}{4s\cdot y_j^{\textsc{pf}}(s)}\biggr\}\dd s
    &\quad& \text{for every }j\in J, \\
    \dualSb_{dt}
    &\coloneqq \int_t^\infty\frac{\eta_d(s)}{4s}\dd s
    && \text{for every }d\in[D]\text{ and }t>0.
\end{alignat*}
As in \cref{subsec:general-dual-fitting}, the minimum equals~$1$ whenever~$s=0$ or~$y_j^{\textsc{pf}}(s)=0$, and at time~$t=0$, we choose prices~$\dualSb_{d0}$ large enough to satisfy the dual constraints at time zero, which does not affect the objective integral.

The weighted completion time~$w_jC_j$ can be viewed as job~$j$ collecting credit at rate~$w_j$ until it completes.
In~$\dualSa_j$, we retain this full credit at time $t < C_j$ when the normalized rate~$y_j^{\textsc{pf}}(t)/p_j$ is at most~$1/(4t)$.
If the job runs faster, we retain only the fraction~$p_j/(4t\cdot y_j^{\textsc{pf}}(t))$.
The resource variable~$\dualSb_{dt}$ collects the future discounted KKT prices needed to support these credits.

We first check dual feasibility.
For~$t\geq C_j$, the dual constraint holds because~$\dualSa_j\leq w_jC_j\leq w_jt$.
For~$r_j\leq t<C_j$ with~$t>0$, the first KKT identity gives
\begin{align*}
    \dualSa_j-w_jt
    &\leq w_j\int_t^{C_j}\min\biggl\{1,\frac{p_j}{4s\cdot y_j^{\textsc{pf}}(s)}\biggr\}\dd s
    \leq p_j\int_t^{C_j}\frac{1}{4s}\frac{w_j}{y_j^{\textsc{pf}}(s)}\dd s  \\
    &=p_j\sum_{d\in[D]}B_{dj}\int_t^{C_j}\frac{\eta_d(s)}{4s}\dd s
    \leq p_j\sum_{d\in[D]}B_{dj}\dualSb_{dt}.
\end{align*}

Next, we show that the dual objective value of $(\dualSa,\dualSb)$ is at least~$\pf/4$.
Exchanging the order of integration and using the second KKT identity~$\sum_{d\in[D]}\eta_d(t)=w(A(t))$, the total resource cost is
\begin{align}
    \sum_{d\in[D]}\int_0^\infty\dualSb_{dt}\dd t
    &= \sum_{d\in[D]}\int_0^\infty\int_t^\infty \frac{\eta_d(s)}{4s}\dd s\dd t \notag\\
    &= \frac14\int_0^\infty\sum_{d\in[D]}\eta_d(s)\dd s
    = \frac14\int_0^\infty w(A(s))\dd s
    \leq \frac14\pf. \label{eq:overview-general-beta-cost}
\end{align}
Thus, the resource variables consume at most one quarter of PF's completion-time cost.

It remains to bound the job values~$\dualSa_j$ from below.
Fix a job~$j$.
Recall that~$j$ collects credit at rate~$w_j$ until~$C_j$, in total~$w_jC_j$, and that~$\dualSa_j$ keeps only the fraction~$\min\{1,p_j/(4t\cdot y_j^{\textsc{pf}}(t))\}$ of the credit collected at time~$t$.
Hence,~$j$ loses credit at time~$t$ only while its normalized rate~$y_j^{\textsc{pf}}(t)/p_j$ exceeds~$1/(4t)$, and the faster~$j$ is processed, the more credit it loses.
Losing credit thus requires processing, and~$j$ receives only~$p_j$ units of processing until~$C_j$, that is,~$\int_0^{C_j}y_j^{\textsc{pf}}(t)/p_j\dd t=1$.
The crux of the analysis is this tradeoff between the processing that~$j$ accumulates and the credit that it loses.
\Cref{lem:volume-general} quantifies it with the AM--GM inequality and shows that~$j$ keeps at least half of its credit, that is,~$\dualSa_j\geq w_jC_j/2$.
Summing over all jobs gives~$\sum_{j\in J}\dualSa_j\geq\pf/2$, and together with~\eqref{eq:overview-general-beta-cost}, the fitted dual has value at least~$\pf/2-\pf/4=\pf/4$, which proves \cref{thm:general-psp-completion}.

\paragraph{Comparison with the previous factor~$27$.}
The earlier dual-fitting analyses of Im, Kulkarni, and Munagala~\cite{ImKM18} and Jäger, Lindermayr, and Megow~\cite{JLM2026} use a hard global threshold.
At each time~$t$, they choose a weighted~$\lambda$-quantile~$\zeta(t)$ of the normalized rates~$y_j^{\textsc{pf}}(t)/p_j$ of the unfinished jobs~$U(t)$.
Jobs at or below the threshold, of total weight at least~$\lambda w(U(t))$, contribute to the job variables, whereas jobs at or above it, of weight at least~$(1-\lambda)w(U(t))$, pay for the resource variables.
The resource variable at time~$t$ aggregates the prices~$\zeta(s)\eta_d(s)/\kappa$, for a parameter~$\kappa\geq1$, from all future times~$s\geq t$.
This yields job value at least~$\lambda\cdot\pf$ and resource cost at most~$\pf/((1-\lambda)\kappa)$.
Because this construction uses a $\kappa$-restricted capacity dual, converting its value back to the unit-capacity benchmark incurs an additional factor~$\kappa$.
Optimizing at~$\lambda=2/3$ and~$\kappa=9$ gives the factor~$27$.
By contrast, our construction determines each job's credit from its own normalized rate and uniformly discounts PF's KKT prices by~$1/(4t)$.
It therefore fits the unit-capacity dual directly ($\kappa=1$), avoiding both the global-quantile tradeoff and the capacity loss.

\section{Preliminaries}
\label{sec:preliminaries}

\subsection{Polytope Scheduling}
\label{subsec:polytope-scheduling}

In the Polytope Scheduling Problem (PSP), we are given a set~$J=[n]=\{1,\dots,n\}$ of~$n$ jobs.
Each job~$j \in J$ has a \emph{weight}~$w_j > 0$, a \emph{processing requirement}~$p_j \geq 0$, and a \emph{release date}~$r_j \geq 0$.
Moreover, we are given a \emph{rate polytope}
\begin{align*}
    \cP = \bigl\{\, y \in \RR^n_{\ge 0} \bigm| By \leq \bm{1} \,\bigr\},
\end{align*}
where the nonnegative matrix~$B \in \RR_{\geq 0}^{D \times n}$ has no zero column, which ensures that~$\cP$ is bounded.
We write~$B_j$ for the column of~$B$ that corresponds to job~$j$.
Since~$B$ is nonnegative,~$\cP$ is downward-closed, that is, if~$y \in \cP$ and~$0 \leq y' \leq y$ componentwise, then~$y' \in \cP$.
Moreover, since the right-hand side is positive, every job can be processed at a positive rate, that is, for every~$j \in J$ there exists some~$y \in \cP$ with~$y_j > 0$.
We write~$y_j^{\max}\coloneqq\max\{y_j\mid y\in\cP\}$ for the maximum rate of job~$j$, so that~$0<y_j^{\max}<\infty$.
Some scheduling models, such as resource scheduling with substitutes, lead to a compact, convex, and downward-closed set of feasible rate vectors that is not a polytope.
We handle such sets by polyhedral approximation in \cref{subsec:multidim-ssPSP}.

A \emph{schedule} is a (measurable) function~$y \colon \RR_{\geq 0} \to \RR^n_{\geq 0}$ that assigns to every time~$t$ a vector of \emph{rates}~$y(t) = (y_j(t))_{j \in J}$.
For a \emph{speed}~$s \geq 1$, a schedule is \emph{$s$-speed feasible} if it satisfies the following two conditions.
\begin{description}
	\item[(S1) Instantaneous polytope constraints:] at every time~$t \geq 0$, we have~$y(t) \in \cP$.
	\item[(S2) Job completion:] for every job~$j \in J$, there exists a time~$t \geq r_j$ such that~$s \int_{r_j}^{t} y_j(u) \dd u \geq p_j$.
\end{description}
Equivalently, an $s$-speed feasible schedule for processing requirements~$(p_j)_{j \in J}$ is a $1$-speed feasible schedule for processing requirements~$(p_j/s)_{j \in J}$.
We simply call $1$-speed feasible schedules \emph{feasible}.
Since every job can be processed at a positive rate, every instance admits a feasible schedule, for example, one that processes the jobs one after another.
The \emph{completion time}~$C_j$ of job~$j$ in an $s$-speed feasible schedule is the first time at which it has accumulated its full processing requirement, that is, $C_j \coloneqq \min\{ t\geq r_j \mid s \int_{r_j}^t y_j(u) \dd u \geq p_j\}$.
The \emph{flow time}~$F_j$ is the amount of time job~$j$ spends in the system, that is,~$F_j \coloneqq C_j - r_j$.
We consider the objectives of minimizing the (total) \emph{weighted completion time}~$\sum_{j\in J} w_j C_j$ and minimizing the (total) \emph{weighted flow time}~$\sum_{j\in J} w_j F_j$.
Setting the rate of a job to zero before its release date or after its completion time preserves feasibility for every speed, because~$\cP$ is downward-closed, and it does not change any completion time.
Hence, we assume without loss of generality that every schedule satisfies~$y_j(t) = 0$ for every job~$j \in J$ and every time~$t \notin [r_j, C_j)$.

For any time~$t \geq 0$, we write~$A(t) = \{j \in J \mid r_j \leq t < C_j\}$ for the set of \emph{active} jobs and $U(t) = \{j \in J \mid t < C_j\}$ for the set of \emph{unfinished} jobs.
Thus,~$A(t) \subseteq U(t)$, and~$U(t) \setminus A(t)$ consists of the jobs that have not yet been released by time~$t$.

We explicitly allow jobs with processing requirement~$p_j = 0$.
By definition, such a job completes at its release date, that is,~$C_j = r_j$ in every schedule.
Hence, it is never active, always receives rate zero, and contributes~$w_j r_j$ to the weighted completion time and nothing to the weighted flow time.
Allowing such jobs is convenient for our analysis.

For~$J'\subseteq J$, we write~$w(J') = \sum_{j \in J'} w_j$ for the total weight of a subset~$J'$.
Moreover, let~$\bm{1}_{J'}\in\{0,1\}^n$ be the indicator vector of~$J'$, and for~$y\in\RR^n$, let~$y\odot\bm{1}_{J'}$ denote the coordinate-wise product, that is, the vector that agrees with~$y$ on~$J'$ and is zero elsewhere.
We have the standard identities
\begin{align*}
    \sum_{j\in J} w_j C_j = \int_0^\infty w(U(t))\dd t \quad \text{ and } \quad \sum_{j\in J} w_j F_j = \int_0^\infty w(A(t))\dd t.
\end{align*}

\paragraph{Online algorithms for PSP.}
So far, we have described PSP as an offline problem, in which the entire instance is known in advance.
In the online setting, the instance is instead revealed over time.
An online algorithm learns about a job~$j$ only at its release date~$r_j$, at which point the weight~$w_j$ and the column~$B_j$ become known.
The algorithm constructs the schedule over time, that is, at every time~$t$ it has to fix the rate vector~$y(t)$ based only on the jobs released until time~$t$, without any knowledge of future jobs, and it cannot revise this decision later.
A \emph{clairvoyant} online algorithm additionally learns the processing requirement~$p_j$ at time~$r_j$, whereas a \emph{nonclairvoyant} online algorithm does not learn~$p_j$ until job~$j$ completes at time~$C_j$.
Even though an online algorithm does not know the full matrix~$B$ at time~$t$, it can choose a rate vector~$y(t) \in \cP$, because unreleased jobs have rate zero, and thus the constraints~$By(t) \leq \bm{1}$ only involve the columns of released jobs.
In our analysis, we therefore always work with the polytope~$\cP$ of the entire instance.

\begin{definition}[Competitive ratio and speed augmentation]
    Fix either objective and a class of instances.
    For an instance~$I$, let~$\alg(I)$ denote the objective value of the schedule produced by an online algorithm~$\alg$, and let~$\opt(I)$ denote the optimal objective value among all feasible schedules for~$I$.
    If~$I$ is clear from the context, we omit it.
    The algorithm is \emph{$c$-competitive} if~$\alg(I)\leq c\cdot\opt(I)$ for every instance~$I$ in the class.
    Its \emph{competitive ratio} is the infimum of all such~$c$.

    With \emph{speed augmentation~$s\geq1$}, the algorithm constructs an $s$-speed feasible schedule, whereas the offline optimum remains restricted to feasible schedules, that is, to speed~$1$.
    Let~$\alg_s(I)$ denote the objective value of the schedule produced by the algorithm with speed~$s$.
    We call~$\alg$ \emph{$s$-speed $c$-competitive} if~$\alg_s(I)\leq c\cdot\opt(I)$ for every instance~$I$ in the class.
    For randomized algorithms, both inequalities are required in expectation over the algorithm's internal randomness, for every fixed input~$I$ independent of that randomness.
\end{definition}

\subsection{Proportional Fairness}
\label{subsec:proportional-fairness}

The Proportional Fairness algorithm (PF) has been analyzed for PSP by Im, Kulkarni, and Munagala~\cite{ImKM18} and by Jäger, Lindermayr, and Megow~\cite{JLM2026}.
At every time~$t \geq 0$, PF determines the rates of the active jobs~$A(t)$ by solving the \emph{Eisenberg--Gale convex program}~\cite{eisenberg1959consensus}, which for a subset~$S \subseteq J$ of jobs can be written as follows.
\begin{alignat}{2}
    \text{maximize}\quad
        & \sum_{j \in S} w_j \log(y_j)
        && \tag{$\text{CP}$}\label{psp-pf-cp} \\
    \text{subject to}\quad
        & \sum_{j\in S} B_{dj}y_j \leq 1
        &\quad& \forall d \in [D] \notag \\
        & y_j \geq 0
        && \forall j\in S \notag
\end{alignat}
Maximizing this objective is equivalent to maximizing the weighted Nash social welfare~\cite{nash1950bargaining}.
Since~$\cP$ is bounded and the objective is strictly concave, \eqref{psp-pf-cp} has a unique optimal solution, which we denote by~$y(S) = (y_j(S))_{j \in S}$.
PF sets~$y_j^{\textsc{pf}}(t) = y_j(A(t))$ for every job~$j\in A(t)$ and~$y_j^{\textsc{pf}}(t) = 0$ for every other job.
With speed augmentation~$s \geq 1$, PF uses the same rule, so its rates still lie in~$\cP$, but jobs accumulate processing $s$ times faster.

PF is a nonclairvoyant online algorithm.
Indeed, to compute the rates at time~$t$, it only needs the set~$A(t)$ together with the weights~$w_j$ and the columns~$B_j$ of the jobs in~$A(t)$.
This information is available at time~$t$, because an online algorithm learns~$w_j$ and~$B_j$ at time~$r_j$ and observes when a job completes.
In particular, PF never uses the processing requirements.
Moreover, since~$A(t)$ only changes when a job is released or completes, PF only needs to recompute the rates at these events and uses them until the next event.

\paragraph{Polynomial-time implementation.}
Since each job is released once and completes once, PF solves at most~$2n$ instances of~\eqref{psp-pf-cp}.
Their optimal solutions can be irrational, but approximate solutions, which can be computed in polynomial time, suffice for all our guarantees up to a factor of~$1+\eps$ (see \cref{sec:pf-polynomial-time}).

\paragraph{KKT conditions.}
The optimal solution to~\eqref{psp-pf-cp} is characterized by the KKT conditions~\cite{BoydV04}.
The objective tends to~$-\infty$ as any~$y_j$ approaches zero, so every job in~$S$ receives a strictly positive rate at an optimal solution.
Thus, the nonnegativity constraints are inactive, and their Lagrange multipliers vanish.
Let~$\eta(S) = (\eta_1(S), \dots, \eta_D(S)) \geq 0$ be any vector of optimal Lagrange multipliers for the resource constraints~$By \leq \bm{1}$.
Stationarity gives, for every~$j \in S$,
\begin{align}
    \frac{w_j}{y_j(S)} &= \sum_{d=1}^D B_{dj}\,\eta_d(S),
    \label{eq:kkt-stationarity}
\end{align}
while complementary slackness gives, for every~$d \in [D]$,
\begin{align}
    \eta_d(S) \Bigl(1 - \sum_{j \in S} B_{dj}\,y_j(S)\Bigr) &= 0.
    \label{eq:kkt-cs}
\end{align}
Together with primal and dual feasibility, these conditions characterize optimality.

We use the following consequences of the KKT conditions in our analysis of PF.
\begin{lemma}[{\cite[Lemma~3.3 and Proposition~4.1]{ImKM18}}]
    \label{lem:pf-optimality}
    Let~$S\subseteq J$.
    \begin{enumerate}[label=\textup{(\roman*)}]
        \item Every vector~$\eta(S)$ of optimal Lagrange multipliers for~\eqref{psp-pf-cp} on~$S$ satisfies~$\sum_{d=1}^D \eta_d(S) = w(S)$.
        \item Every~$\hat y\in\cP$ satisfies~$\sum_{j\in S}w_j\hat y_j/y_j(S)\leq w(S)$.
    \end{enumerate}
    In particular,~$y_j(S)\geq w_jy_j^{\max}/w(S)$ for every~$j\in S$.
\end{lemma}
\begin{proof}
    For brevity, write~$y$ and~$\eta$ for~$y(S)$ and~$\eta(S)$.
    Using~\eqref{eq:kkt-cs} and then~\eqref{eq:kkt-stationarity} gives
    \[
        \sum_{d=1}^D \eta_d
        = \sum_{d=1}^D \eta_d \sum_{j \in S} B_{dj}\,y_j
        = \sum_{j \in S} y_j \sum_{d=1}^D B_{dj}\,\eta_d
        = \sum_{j \in S} y_j \cdot \frac{w_j}{y_j}
        = \sum_{j \in S} w_j
        = w(S),
    \]
    which proves~(i).
    For~(ii), let~$\hat y\in\cP$.
    Using~\eqref{eq:kkt-stationarity}, the nonnegativity of~$B$,~$\hat y$, and~$\eta$, the feasibility of~$\hat y$, and~(i), we obtain
    \[
        \sum_{j\in S}\frac{w_j}{y_j}\,\hat y_j
        = \sum_{j\in S}\hat y_j\sum_{d=1}^D B_{dj}\,\eta_d
        = \sum_{d=1}^D \eta_d\sum_{j\in S}B_{dj}\,\hat y_j
        \leq \sum_{d=1}^D \eta_d\sum_{j\in J}B_{dj}\,\hat y_j
        \leq \sum_{d=1}^D \eta_d
        = w(S).
    \]
    Finally, let~$j\in S$.
    The vector~$y_j^{\max}\bm{1}_{\{j\}}$ lies in~$\cP$
    by the definition of~$y_j^{\max}$ and since~$\cP$ is downward-closed,
    so~(ii) for this vector gives~$w_jy_j^{\max}/y_j\leq w(S)$.
\end{proof}
Part~(ii) is the first-order optimality condition of~\eqref{psp-pf-cp}.
It states that if every job~$j\in S$ is assigned the weight~$w_j/y_j(S)$ per unit of processing rate, then no feasible rate vector covers more total weight than the PF rates, which cover exactly~$w(S)$.

\paragraph{PF-monotone PSP.}
Im, Kulkarni, and Munagala~\cite{ImKM18} studied a class of polytopes in which the PF rates behave monotonically under changes of the job set.
For such polytopes, the PF rate of a job never decreases when another job leaves the system.

\begin{definition}[PF-monotone PSP]
    \label{def:monPSP}
    For fixed positive job weights, a PSP instance and its rate polytope are \emph{PF-monotone} if~$y_i(S\setminus\{j\}) \geq y_i(S)$ for every~$S\subseteq J$, every~$j\in S$, and every~$i \in S\setminus\{j\}$.
\end{definition}

\section{Support-Submodular PSP}
\label{sec:ssPSP}

We introduce the class of \emph{support-submodular polytope scheduling} instances.
This class captures several well-studied scheduling models, including polymatroid scheduling, unrelated machine scheduling, generalized flow scheduling, and scheduling with concave speedup curves, as we verify in \Cref{subsec:lattice-ssPSP,subsec:unrel-ssPSP,subsec:multidim-ssPSP}.
On the other hand, graph scheduling, multicommodity flow scheduling, and all-or-nothing scheduling are in general not support submodular (see \cref{subsec:non-support-submodular-problems,sec:non-support-submodular-PSP}).

\subsection{The Support Function and Submodularity}
\label{subsec:support-function}

Let~$\cP \subseteq \RR^n_{\geq 0}$ be a polytope.
The \emph{support function} of~$\cP$ is the function $h_{\cP} \colon \RR^n \to \RR$ defined by
\begin{align*}
    h_{\cP}(\alpha) \coloneqq \max_{y \in \cP} {\langle \alpha, y \rangle},
\end{align*}
where~$\langle \alpha, y \rangle = \sum_{j=1}^n \alpha_j y_j$.
In our scheduling context,~$\alpha_j$ can be interpreted as weight per
unit of processing rate assigned to job~$j$.
Thus,~$h_{\cP}(\alpha)$ measures the maximum total weight that a feasible
rate allocation can cover under these coefficients.

Our submodularity condition is not imposed on~$h_\cP$ as a function of~$\alpha$, but rather on the following related set function.
For a subset~$U \subseteq J$, the \emph{restricted polytope}~$\cP_U$ consists of the points of~$\cP$ that vanish outside~$U$, that is, $\cP_U \coloneqq \{ y \in \cP \mid y_j = 0 \ \forall j \notin U \}$.
Whenever the polytope is clear from the context, we write~$h_U$ for~$h_{\cP_U}$.
For a fixed direction~$\alpha \in \RR^n_{\geq 0}$, we define the set function
\begin{align*}
    H_\alpha \colon 2^J &\to \RR_{\geq0} \\
    U &\mapsto h_U(\alpha),
\end{align*}
which measures, for each job set~$U$, the maximum value achievable by the jobs in~$U$.
As in \cref{def:ssPSP-intro}, we also call~$H_\alpha$ the support function.

A set function~$f \colon 2^J \to \RR$ is \emph{submodular} if for all~$U, V \subseteq J$, we have $f(U) + f(V) \geq f(U \cup V) + f(U \cap V)$,
or equivalently, if for all~$U \subseteq V \subsetneq J$ and all~$j \in J \setminus V$, we have
$f(U \cup \{j\}) - f(U) \geq f(V \cup \{j\}) - f(V)$.
The latter formulation captures \emph{diminishing marginal returns}, meaning that the contribution of adding a new element does not increase as the existing set grows.

\begin{definition}[Support-submodular PSP, \Cref{def:ssPSP-intro}]
    \label{def:ssPSP}
    A PSP instance and its rate polytope are \emph{support submodular} if for every vector~$\alpha \in \RR^n_{\geq 0}$, the set function~$U \mapsto H_\alpha(U)$ is submodular.
\end{definition}

\begin{figure}[tb]
\centering
\begin{tikzpicture}[
    x={(-1.05cm,-0.8cm)}, y={(0cm,2.0cm)}, z={(2.1cm,-0.22cm)},
    ax/.style={-latex, thin, black!60},
    pedge/.style={thick, line join=round},
    face12/.style={fill=col6, fill opacity=0.35},
    face23/.style={fill=col3, fill opacity=0.35},
    face2/.style={col4, line width=2pt, line cap=round},
    bound/.style={fill=black!15, fill opacity=0.45},
    lab/.style={font=\footnotesize},
]
    \newcommand{\sspolyback}{%
        \draw[ax] (0,0,0) -- (1.4,0,0) node[lab, anchor=north east, inner sep=1pt] {$y_1$};
        \draw[ax] (0,0,0) -- (0,1.3,0) node[lab, anchor=south] {$y_2$};
        \draw[ax] (0,0,0) -- (0,0,1.35) node[lab, anchor=west] {$y_3$};
        \fill[face12] (0,0,0) -- (1,0,0) -- (0,1,0) -- cycle;
        \fill[face23] (0,0,0) -- (0,1,0) -- (0,0,1) -- cycle;
        \draw[face2] (0,0,0) -- (0,1,0);
        \fill[bound] (1,0,0) -- (0,1,0) -- (0,0,1) -- cycle;
    }
    \newcommand{\sspolylabels}{%
        \node[lab, text=col6, anchor=east] at (0.75,0.4,0) {$\cP_{\{1,2\}}$};
        \node[lab, text=col3!85!black] at (0.02,0.3,0.48) {$\cP_{\{2,3\}}$};
        \node[lab, text=col4!85!black, anchor=east] at (0.12,0.82,0) {$\cP_{\{2\}}$};
    }

    \sspolyback
    \draw[pedge] (1,0,0) -- (0,1,0) -- (0,0,1) -- cycle;
    \sspolylabels
    \node[lab] at (0.7cm,-1.75cm) {$y_1+y_2+y_3\leq 1$};

    \begin{scope}[xshift=7cm]
        \sspolyback
        \draw[thin, dashed] (1,0,0) -- (0,0,1);
        \fill[col5, fill opacity=0.25] (1,0,0) -- (1,0,1) -- (0,0,1) -- cycle;
        \draw[pedge] (1,0,0) -- (0,1,0) -- (0,0,1);
        \draw[pedge, col5] (1,0,0) -- (1,0,1) -- (0,0,1) (0,1,0) -- (1,0,1);
        \sspolylabels
        \fill[col5] (1,0,1) circle (2.2pt)
            node[lab, anchor=north west, inner sep=2pt] {$(1,0,1)$};
    \end{scope}
    \node[lab] at (7.7cm,-1.75cm) {$y_1+y_2\leq 1,\ y_2+y_3\leq 1$};
\end{tikzpicture}
\caption{Two polytopes in~$\RR^3_{\geq0}$ with the same restrictions~$\cP_{\{1,2\}}$, $\cP_{\{2,3\}}$, and~$\cP_{\{2\}}$.
    For~$\alpha=\bm 1$, these restrictions give $H_\alpha(\{1,2\})=H_\alpha(\{2,3\})=H_\alpha(\{2\})=1$, so the submodularity inequality $H_\alpha(\{1,2\})+H_\alpha(\{2,3\})\geq H_\alpha(\{1,2,3\})+H_\alpha(\{2\})$ holds if and only if $y_1+y_2+y_3\leq 1$ for all~$y\in\cP$, where the gray triangle is the set of points in~$\RR^3_{\geq0}$ with~$y_1+y_2+y_3=1$.
    The left polytope is a polymatroid and hence support submodular (\Cref{prop:polymatroid-support-submodular}).
    Since jobs~$1$ and~$3$ share no constraint, the right polytope additionally contains the orange tetrahedron with apex~$(1,0,1)$, where~$y_1+y_2+y_3=2$, so it is not support submodular.}
\label{fig:support-submodularity}
\end{figure}
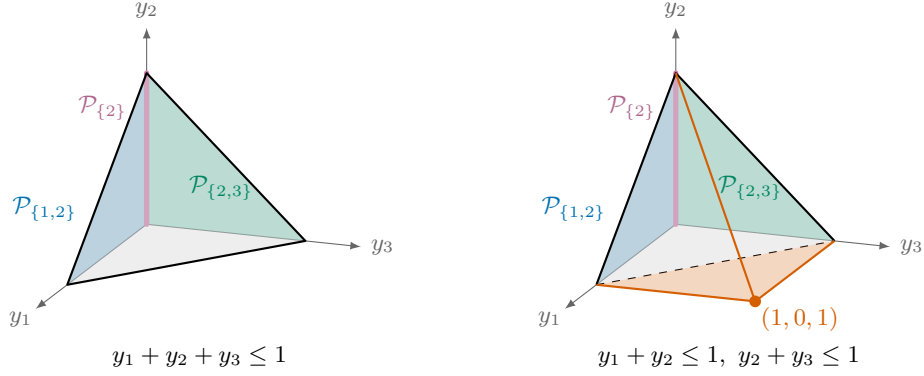

Support submodularity is a condition on the polytope alone, independent of the job weights and processing requirements.
This is in contrast to PF-monotonicity (cf.~\Cref{def:monPSP} and~\cite{ImKM18}), which is defined in terms of an optimal solution to the Eisenberg--Gale convex program and hence depends on the weights of the instance.
\Cref{fig:support-submodularity} illustrates this definition with two polytopes in~$\RR^3_{\geq0}$.

\subsection{Polymatroids and Generalized Flow Scheduling}
\label{subsec:lattice-ssPSP}

If~$\cP$ is downward closed, then~$H_\alpha(U)=h_\cP(\alpha\odot\bm{1}_U)$ for every~$U\subseteq J$, because setting the coordinates of a point of~$\cP$ outside~$U$ to zero keeps it in~$\cP$ and does not change its~$\alpha\odot\bm{1}_U$-value.
Kashiwabara and Takabatake~\cite{KT03} studied the stronger condition that~$h_\cP$ is submodular on the whole lattice~$\RR^n_{\geq0}$, that is,
\begin{equation}
    \label{eq:lattice-submodular}
    h_\cP(\alpha)+h_\cP(\alpha')
    \geq h_\cP(\alpha\vee\alpha')+h_\cP(\alpha\wedge\alpha')
    \qquad \forall\alpha,\alpha'\in\RR^n_{\geq0}\,,
\end{equation}
where~$\vee$ and~$\wedge$ denote the componentwise maximum and minimum.

\begin{lemma}
    \label{lem:lattice-submodular}
    If the rate polytope~$\cP$ of a PSP instance satisfies~\eqref{eq:lattice-submodular}, then~$\cP$ is support submodular and PF-monotone for all positive job weights.
\end{lemma}

For PF-monotonicity, the weighted reciprocal PF rates~$w_j/y_j(S)$ uniquely minimize a convex function whose only nonseparable term is~$h_\cP$, and~\eqref{eq:lattice-submodular} forces this minimizer to increase coordinate-wise when jobs are added.

\begin{proof}
    Let~$\alpha\in\RR^n_{\geq0}$ and~$U,W\subseteq J$.
    Since~$\alpha\geq0$, we have~$\alpha\odot\bm{1}_U\vee\alpha\odot\bm{1}_W=\alpha\odot\bm{1}_{U\cup W}$ and~$\alpha\odot\bm{1}_U\wedge\alpha\odot\bm{1}_W=\alpha\odot\bm{1}_{U\cap W}$.
    As~$H_\alpha(V)=h_\cP(\alpha\odot\bm{1}_V)$ for every~$V\subseteq J$,~\eqref{eq:lattice-submodular} yields~$H_\alpha(U)+H_\alpha(W)\geq H_\alpha(U\cup W)+H_\alpha(U\cap W)$, so~$\cP$ is support submodular.

    For PF-monotonicity, let~$w\in\RR^n_{>0}$, and for a nonempty job set~$S\subseteq J$, let~$y(S)$ be the PF rate vector (see \Cref{subsec:proportional-fairness}), extended by zero to a point of~$\cP$.
    Its coordinates on~$S$ are positive, and the first-order optimality condition (\cref{lem:pf-optimality}(ii)) states that~$\sum_{j\in S}w_jy_j/y_j(S)\leq w(S)$ for every~$y\in\cP$.
    Define~$q^S\in\RR^n_{\geq0}$ by $q^S_j\coloneqq w_j/y_j(S)$ for~$j\in S$ and~$q^S_j\coloneqq0$ for~$j\notin S$.
    Let~$Q_S$ be the set of vectors~$q\in\RR^n_{\geq0}$ with~$q_j>0$ exactly for~$j\in S$, and define~$\Psi_S\colon Q_S\to\RR$ by~$\Psi_S(q)\coloneqq h_\cP(q)-\sum_{j\in S}w_j\log q_j$.

    We first show that~$q^S$ is the unique minimizer of~$\Psi_S$.
    Let~$q\in Q_S$.
    Since~$y(S)\in\cP$, we have $h_\cP(q)\geq\sum_{j\in S}q_jy_j(S)$, and hence~$\Psi_S(q)\geq\sum_{j\in S}\bigl(q_jy_j(S)-w_j\log q_j\bigr)$.
    Each summand is a strictly convex function of~$q_j>0$ that is uniquely minimized at~$q_j=w_j/y_j(S)=q^S_j$.
    For~$q=q^S$, the inequality holds with equality, because the first-order condition gives~$h_\cP(q^S)\leq w(S)=\sum_{j\in S}q^S_jy_j(S)$.
    Thus~$\Psi_S(q)\geq\Psi_S(q^S)$, with equality only if~$q=q^S$, so~$q^S$ is the unique minimizer of~$\Psi_S$.

    Now let~$\emptyset\neq S\subseteq T\subseteq J$.
    We show~$q^S\leq q^T$ by Topkis' monotone comparative statics argument~\cite{Topkis78}.
    First,~$q^S\wedge q^T\in Q_S$ and~$q^S\vee q^T\in Q_T$.
    The logarithmic terms of~$\Psi_S(q^S\wedge q^T)+\Psi_T(q^S\vee q^T)$ and of~$\Psi_S(q^S)+\Psi_T(q^T)$ coincide, because~$\min\{q^S_j,q^T_j\}\cdot\max\{q^S_j,q^T_j\}=q^S_jq^T_j$ for~$j\in S$ and~$(q^S\vee q^T)_j=q^T_j$ for~$j\in T\setminus S$.
    Hence,~\eqref{eq:lattice-submodular} applied to~$q^S$ and~$q^T$ gives
    \begin{equation*}
        \Psi_S(q^S\wedge q^T)+\Psi_T(q^S\vee q^T)\leq \Psi_S(q^S)+\Psi_T(q^T)\,.
    \end{equation*}
    Since~$q^S$ and~$q^T$ minimize~$\Psi_S$ and~$\Psi_T$, respectively, each term on the left is at least the corresponding term on the right, so~$\Psi_S(q^S\wedge q^T)=\Psi_S(q^S)$.
    By uniqueness,~$q^S\wedge q^T=q^S$, that is,~$q^S_i\leq q^T_i$ and hence~$y_i(S)\geq y_i(T)$ for every~$i\in S$.
    For~$T\subseteq J$, $j\in T$, and~$S=T\setminus\{j\}\neq\emptyset$, this is the inequality of \cref{def:monPSP}.
\end{proof}

\paragraph{Polymatroids.}
\label{subsec:polymatroid-ssPSP}
A \emph{polymatroid}~\cite[Section~14.2]{KorteVygen} is a polytope of the form
\begin{align*}
    \cP_f = \Bigl\{ y \in \RR_{\geq 0}^n \Bigm| \sum_{j \in U} y_j \leq f(U) \quad \forall U \subseteq J \Bigr\},
\end{align*}
where~$f \colon 2^J \to \RR_{\geq 0}$ is a monotone submodular set function satisfying~$f(\emptyset)=0$.
A set function is monotone if~$f(U) \leq f(V)$ whenever~$U \subseteq V$.
PSP with a polymatroid
captures, e.g., scheduling on related machines~\cite{FeldmanMNP2008,ImKM18}, video-on-demand (multicast)~\cite{BikhchandaniVSV11,ImKM18}, single-sink flow scheduling~\cite{Megiddo74,ImKM18}, and scheduling with restricted assignment~\cite{JLM2026}.
We refer to~\cite{ImKM18} for a more detailed description of these problems.

\begin{proposition}
    \label{prop:polymatroid-support-submodular}
    The support function of every polymatroid satisfies~\eqref{eq:lattice-submodular}.
    In particular, every polymatroid is support submodular.
\end{proposition}
\begin{proof}
    Let~$\alpha\in\RR^n_{\geq0}$, number the jobs so that~$\alpha_1\geq\cdots\geq\alpha_n$, and set~$\alpha_{n+1}\coloneqq0$.
    By Edmonds' greedy theorem for polymatroids~\cite[Theorem~14.11]{KorteVygen}, the vector~$y$ with~$y_k=f([k])-f([k-1])$ maximizes~$\langle\alpha,y\rangle$ over the polymatroid~$\cP_f$.
    Summation by parts gives
    \begin{align*}
        h_{\cP_f}(\alpha)
        = \sum_{k=1}^n \alpha_k \bigl(f([k])-f([k-1])\bigr)
        = \sum_{k=1}^n (\alpha_k-\alpha_{k+1})\, f([k])
        = \int_0^\infty f\bigl(\{\alpha\geq\theta\}\bigr)\dd\theta\,,
    \end{align*}
    where the last equality uses the level set~$\{\alpha\geq\theta\}\coloneqq\{j\in J\mid\alpha_j\geq\theta\}$, which equals~$[k]$ for $\theta\in(\alpha_{k+1},\alpha_k]$ and is empty for~$\theta>\alpha_1$, and~$f(\emptyset)=0$.
    For~$\alpha,\alpha'\in\RR^n_{\geq0}$ and~$\theta>0$, we have $\{\alpha\vee\alpha'\geq\theta\}=\{\alpha\geq\theta\}\cup\{\alpha'\geq\theta\}$ and~$\{\alpha\wedge\alpha'\geq\theta\}=\{\alpha\geq\theta\}\cap\{\alpha'\geq\theta\}$.
    Applying the submodularity of~$f$ to these sets and integrating over~$\theta$ yields~\eqref{eq:lattice-submodular}.
    Support submodularity follows from \cref{lem:lattice-submodular}, which also recovers the PF-monotonicity of polymatroids~\cite{JainV10,ImKM18}.
\end{proof}

\paragraph{Generalized flow scheduling.}
\label{subsec:genflow-ssPSP}
In \emph{generalized flow scheduling}~\cite{ChenIP25}, the jobs send a single commodity through a network whose arcs may lose or gain flow, which models resources that serve different jobs at different efficiencies.
We are given a directed multigraph~$G=(V,E)$ whose nodes include the jobs, that is,~$J\subseteq V$.
Every arc~$e\in E$ has a capacity~$c_e>0$ and a gain factor~$\gamma_e>0$, so that~$\gamma_e x_e$ units of flow leave~$e$ at its head if~$x_e$ units enter it at its tail.
For a flow~$x\in\RR^E$ and a node~$v\in V$, let~$\operatorname{div}x(v) \coloneqq \sum_{e\in\delta^+(v)} x_e - \sum_{e\in\delta^-(v)} \gamma_e x_e$ be the net outflow, or divergence, of~$x$ at~$v$, where~$\delta^+(v)$ and~$\delta^-(v)$ denote the sets of arcs leaving and entering~$v$.
Every node~$v\in V$ has an excess capacity~$b_v\in\RR$, with~$b_j\geq0$ for every job~$j\in J$, and the nodes of a given set~$V_=\subseteq V\setminus J$ have a fixed excess.
A flow~$x$ is \emph{feasible} if~$0\leq x\leq c$, $\operatorname{div}x(v)\leq b_v$ for every~$v\in V$, and~$\operatorname{div}x(v)=b_v$ for every~$v\in V_=$.
In particular, a node~$v\notin V_=$ with~$b_v\geq0$ can absorb any amount of flow.
Every job is processed at a rate of at most its net outflow, so the rate polytope is
\begin{align*}
    \cP = \Bigl\{ y\in\RR^n_{\geq0} \Bigm| \exists \text{ feasible flow } x\colon\ y_j\leq\operatorname{div}x(j)\ \forall j\in J \Bigr\}\,.
\end{align*}
It is downward closed and bounded, since~$y_j\leq\operatorname{div}x(j)\leq b_j$, and as in PSP we assume that every job has a positive rate in some point of~$\cP$.
For~$V_==\emptyset$, this is the model of Chen, Im, and Petety~\cite{ChenIP25}, up to taking the downward closure.

With unit gains,~$V_==V\setminus(J\cup\{t\})$ for a sink~$t$, $b_v=0$ for~$v\notin J$, and~$b_j=\sum_{e\in\delta^+(j)}c_e$ for~$j\in J$, we obtain single-sink flow scheduling, whose rate polytope is a polymatroid~\cite{Megiddo74,ImKM18}.
Gains additionally allow job-dependent speeds.
For example, two jobs~$1,2$ with~$b_1=s_1$ and~$b_2=s_2$, a node~$t\notin V_=$ with~$b_t=0$, an arc~$(1,2)$ with capacity~$s_1$ and gain~$s_2/s_1$, and an arc~$(2,t)$ with capacity~$s_2$ and unit gain give the rate polytope~$\{y\in\RR^2_{\geq0}\mid y_1/s_1+y_2/s_2\leq1\}$ of a single machine with speeds~$s_1,s_2>0$, which is not a polymatroid if~$s_1\neq s_2$.

Chen, Im, and Petety~\cite[Theorem~2.16]{ChenIP25} showed, building on Fleischer~\cite{Fleischer10}, that generalized flow scheduling satisfies the linear substitutes condition, by proving an inequality of the form~\eqref{eq:lattice-submodular} for their dual profit.
We prove lattice submodularity directly by LP duality to apply our framework; \Cref{sec:comparison-chen-im-petety} relates it to their substitutes condition.

\begin{proposition}
    \label{prop:genflow-support-submodular}
    The rate polytope of generalized flow scheduling satisfies~\eqref{eq:lattice-submodular}.
    Hence, generalized flow scheduling is support submodular and PF-monotone for all positive job weights.
\end{proposition}
\begin{proof}
    We first express~$h_\cP(\alpha)$ for~$\alpha\in\RR^n_{\geq0}$ by LP duality.
    If~$y\in\cP$ with feasible flow~$x$, then the vector~$(\operatorname{div}x(j))_{j\in J}$ also lies in~$\cP$ and dominates~$y$.
    Since~$\alpha\geq0$, it follows that~$h_\cP(\alpha)$ is the optimal value of the linear program that maximizes~$\langle\alpha,y\rangle$ over~$x\in\RR^E$ and~$y\in\RR^n$ with~$0\leq x\leq c$ and~$0\leq y_j\leq b_j$ for all~$j\in J$, subject to~$\operatorname{div}x(j)-y_j=0$ for~$j\in J$, $b_v-\operatorname{div}x(v)=0$ for~$v\in V_=$, and~$b_v-\operatorname{div}x(v)\geq0$ for~$v\in N\coloneqq V\setminus(J\cup V_=)$.
    This program is feasible because~$\cP\neq\emptyset$, and bounded because of the bounds on~$x$ and~$y$.
    We dualize the node constraints with multipliers~$\lambda\in\Lambda\coloneqq\{\lambda\in\RR^V\mid\lambda_v\leq0\ \forall v\in N\}$ and keep the bounds.
    Since~$\sum_{v\in V}\lambda_v\operatorname{div}x(v)=\sum_{e=(v,w)\in E}(\lambda_v-\gamma_e\lambda_w)x_e$, the Lagrangian~$\langle\alpha,y\rangle+\sum_{j\in J}\lambda_j(\operatorname{div}x(j)-y_j)-\sum_{v\in V\setminus J}\lambda_v(b_v-\operatorname{div}x(v))$ equals
    \begin{align*}
        \sum_{e=(v,w)\in E}(\lambda_v-\gamma_e\lambda_w)x_e
        -\sum_{v\in V\setminus J}b_v\lambda_v
        +\sum_{j\in J}(\alpha_j-\lambda_j)y_j\,,
    \end{align*}
    and its maximum over~$0\leq x\leq c$ and~$0\leq y_j\leq b_j$ is
    \begin{align*}
        \varphi(\alpha,\lambda)\coloneqq
        \sum_{e=(v,w)\in E}c_e(\lambda_v-\gamma_e\lambda_w)_+
        -\sum_{v\in V\setminus J}b_v\lambda_v
        +\sum_{j\in J}b_j(\alpha_j-\lambda_j)_+\,,
    \end{align*}
    so strong LP duality gives~$h_\cP(\alpha)=\min_{\lambda\in\Lambda}\varphi(\alpha,\lambda)$, where the minimum is attained.

    Next, we show that~$\varphi$ is submodular jointly in~$(\alpha,\lambda)$, that is,
    \begin{align}
        \label{eq:genflow-lattice}
        \varphi(\alpha\vee\alpha',\lambda\vee\lambda')
        +\varphi(\alpha\wedge\alpha',\lambda\wedge\lambda')
        \leq\varphi(\alpha,\lambda)+\varphi(\alpha',\lambda')
        \qquad\forall\alpha,\alpha'\in\RR^n,\ \lambda,\lambda'\in\RR^V\,.
    \end{align}
    The middle term of~$\varphi$ is linear, and a term that depends on a single coordinate, such as the term of a loop, satisfies~\eqref{eq:genflow-lattice} with equality.
    Every other term is a nonnegative multiple of~$(\xi_1-\nu\xi_2)_+$ for two coordinates~$\xi_1,\xi_2$ of~$(\alpha,\lambda)$ and a constant~$\nu>0$, namely~$c_e(\lambda_v-\gamma_e\lambda_w)_+$ for an arc~$e=(v,w)$ with~$v\neq w$ and~$b_j(\alpha_j-\lambda_j)_+$ for a job~$j$.
    It therefore suffices to show that
    \begin{align*}
        \bigl(\max\{\xi_1,\xi_1'\}-\nu\max\{\xi_2,\xi_2'\}\bigr)_+
        +\bigl(\min\{\xi_1,\xi_1'\}-\nu\min\{\xi_2,\xi_2'\}\bigr)_+
        \leq(\xi_1-\nu\xi_2)_++(\xi_1'-\nu\xi_2')_+
    \end{align*}
    for all~$\xi_1,\xi_1',\xi_2,\xi_2'\in\RR$.
    If the pairs~$(\xi_1,\xi_2)$ and~$(\xi_1',\xi_2')$ are comparable, this holds with equality.
    Otherwise, by symmetry,~$\xi_1>\xi_1'$ and~$\xi_2<\xi_2'$.
    Since~$\nu>0$, the arguments~$\xi_1-\nu\xi_2'$ and~$\xi_1'-\nu\xi_2$ on the left lie between the arguments~$a\coloneqq\xi_1-\nu\xi_2$ and~$a'\coloneqq\xi_1'-\nu\xi_2'$ on the right and have the same sum.
    Hence they equal~$\theta a+(1-\theta)a'$ and~$(1-\theta)a+\theta a'$ for some~$\theta\in[0,1]$, and convexity of the positive part gives the claim.

    We now prove~\eqref{eq:lattice-submodular}.
    Let~$\alpha,\alpha'\in\RR^n_{\geq0}$, and let~$\lambda,\lambda'\in\Lambda$ satisfy~$h_\cP(\alpha)=\varphi(\alpha,\lambda)$ and~$h_\cP(\alpha')=\varphi(\alpha',\lambda')$.
    The set~$\Lambda$ is closed under componentwise maxima and minima, so~\eqref{eq:genflow-lattice} gives
    \begin{align*}
        h_\cP(\alpha)+h_\cP(\alpha')
        \geq\varphi(\alpha\vee\alpha',\lambda\vee\lambda')
        +\varphi(\alpha\wedge\alpha',\lambda\wedge\lambda')
        \geq h_\cP(\alpha\vee\alpha')+h_\cP(\alpha\wedge\alpha')\,.
    \end{align*}
    The remaining claims follow from \cref{lem:lattice-submodular}.
\end{proof}

The proof uses a single multiplier~$\lambda_v$ per node, because all jobs send the same commodity.
If every job routes its own commodity between its own source and sink, as in multicommodity flow scheduling, then support submodularity fails in general (\Cref{prop:mcf-not-gamma-concave}).

\subsection{Unrelated Machine Scheduling}
\label{subsec:unrel-ssPSP}

In the \emph{unrelated machine scheduling} problem, we have a set~$M$ of machines, and each job~$j \in J$ can be processed on machine~$i \in M$ at speed~$s_{ij} \geq 0$.
At each time, a fractional assignment~$x_{ij} \geq 0$ specifies the fraction of machine~$i$ allocated to job~$j$, subject to each machine allocating at most one unit of capacity and each job receiving at most one unit of capacity, that is,
\begin{align*}
    \sum_{j \in J} x_{ij} \leq 1 \quad \forall i \in M
    \quad \text{ and } \quad \sum_{i \in M} x_{ij} \leq 1 \quad \forall j \in J.
\end{align*}
The resulting rate of job~$j$ is $y_j = \sum_{i \in M} s_{ij} x_{ij}$.
Hence, for a subset of jobs~$U \subseteq J$, the restricted rate polytope is
\begin{alignat*}{2}
    \cP_U = \Biggl\{ y\in\RR^n_{\geq0} \; \Bigm| \;
    &y_j = \sum_{i \in M} s_{ij} x_{ij}\ \forall j \in J,\ \sum_{j \in U} x_{ij} \leq 1\ \forall i \in M,\ \\
    &\sum_{i \in M} x_{ij} \leq 1\ \forall j \in U,\ \\
    &\sum_{i \in M} x_{ij} = 0 \ \forall j \in J \setminus U,\
    x_{ij} \ge 0 \ \forall i \in M,\, j \in J \Biggr\}.
\end{alignat*}

Two prominent special cases of unrelated machine scheduling are \emph{related machine scheduling}, in which each machine~$i$ has a single speed~$s_i=s_{ij}$ for all jobs~$j$, and \emph{restricted assignment scheduling}, where~$s_{ij}\in \{0,1\}$.
While both are polymatroidal~\cite{ImKM18,JLM2026}, we emphasize that unrelated machine scheduling is not a special case of polymatroid scheduling, because unrelated machines are not PF-monotone in general~\cite[Proposition~4.13]{JLM2026}, whereas polymatroids are PF-monotone~\cite{JainV10}.
For the same reason, \cref{lem:lattice-submodular} shows that the support function of an unrelated machine polytope is in general not submodular on the whole lattice~$\RR^n_{\geq0}$.
Nevertheless, we now show that unrelated machine scheduling is support submodular.

For any~$\alpha \in \RR^n_{\geq 0}$, we have
\begin{align*}
    h_U(\alpha)
    = \max_{y \in \cP_U} {\langle \alpha, y \rangle}
    = \max_{x \in \RR_{\geq 0}^{M \times U}} \Bigl\{ \sum_{j \in U} \sum_{i \in M} \alpha_j s_{ij} x_{ij} \Bigm| \sum_{j \in U} x_{ij} \leq 1\ \forall i \in M,\quad \sum_{i \in M} x_{ij} \leq 1\ \forall j \in U \Bigr\}.
\end{align*}
This is precisely the LP relaxation of the maximum-weight matching problem on the complete bipartite graph~$M \times U$, where edge~$(i,j)$ has weight~$\alpha_j s_{ij}$.
The fractional bipartite matching polytope is integral~\cite[Theorem~11.4]{KorteVygen}, so~$h_U(\alpha)$ equals the value of a maximum-weight matching on~$M \times U$.
We now show that~$H_\alpha\colon U \mapsto h_U(\alpha)$ is submodular for every~$\alpha \geq 0$, that is, the maximum-weight matching value has diminishing returns.
This is a classical fact: Shapley~\cite{Shapley62} showed that agents on the same side of an assignment market are substitutes, which is precisely this diminishing-returns property; see also \textcite[Proposition~6.2]{JalalyT21}.
In the language of combinatorial auctions, $H_\alpha$ is an OXS valuation, that is, an OR of XOR-of-singletons valuations, one per machine. Such valuations even satisfy the gross substitutes condition~\cite{LehmannLN_ACM01}, which implies submodularity~\cite{GulStacchetti99_Gross_Substitutes}.
We include a self-contained proof for completeness.
\begin{proposition}
    \label{prop:unrel-machines-support-submodular}
    Let~$w_{ij} \geq 0$ be edge weights on the complete bipartite graph~$M \times J$.
    Then the set function assigning to each~$U \subseteq J$ the maximum weight of a matching in~$M \times U$ is submodular.
    Consequently, unrelated machine scheduling with preemption and migration is support submodular.
\end{proposition}
\begin{proof}
    Let~$g(U)$ denote the maximum weight of a matching in~$M\times U$.
    Submodularity is equivalent to the local diminishing-returns inequalities, so it suffices to prove for all~$U\subseteq J$ and all distinct $j,k\in J\setminus U$ that
    \begin{align*}
        g(U\cup \{j\}) + g(U\cup \{k\})
        \geq g(U) + g(U\cup \{j,k\}).
    \end{align*}
    Let~$P$ be a maximum-weight matching in~$M\times U$ and let~$Q$ be a maximum-weight matching in~$M\times (U\cup \{j,k\})$.
    Then~$g(U) = w(P)$ and~$g(U\cup \{j,k\}) = w(Q)$.

    Consider the symmetric difference~$P \symmdiff Q$.
    This graph decomposes into vertex-disjoint alternating paths and even cycles.
    Since~$j$ and~$k$ are unmatched in~$P$, each is either isolated in~$P \symmdiff Q$ or an endpoint of an alternating path.
    Moreover, their possibly trivial path components are distinct. Indeed, if they shared a path, it would have even length because both vertices lie on the job side of the bipartition, forcing one of them to be matched in~$P$, a contradiction.

    We now construct two new matchings~$P'$ and~$Q'$ by redistributing edges along each component of~$P \symmdiff Q$.
    If the path containing~$j$ is nontrivial, assign the edge incident to~$j$ to~$P'$ and alternate edges along the path between~$P'$ and~$Q'$.
    Similarly, if the path containing~$k$ is nontrivial, assign the edge incident to~$k$ to~$Q'$ and alternate thereafter.
    On all remaining components, keep edges from~$P$ in~$P'$ and edges from~$Q$ in~$Q'$ (edges in $P\cap Q$ are kept in both~$P'$ and~$Q'$).

    By construction,~$P'$ is a matching in~$M\times (U\cup \{j\})$ and~$Q'$ is a matching in~$M\times (U\cup \{k\})$.
    Every edge of~$P \symmdiff Q$ belongs to exactly one of~$P'$ and~$Q'$, while every edge of~$P \cap Q$ belongs to both, exactly as for~$P$ and~$Q$.
    Thus~$w(P') + w(Q') = w(P) + w(Q)$.
    Therefore,
    \begin{align*}
        g(U) + g(U\cup \{j,k\})
        = w(P)+w(Q)
        = w(P')+w(Q')
        \leq g(U\cup \{j\}) + g(U\cup \{k\}).
    \end{align*}
    Hence~$g$ is submodular. For~$w_{ij} = \alpha_j s_{ij}$, we have~$g = H_\alpha$ by the discussion above, so unrelated machine scheduling is support submodular.
\end{proof}

\subsection{Multidimensional Resource Scheduling with Substitutes}
\label{subsec:multidim-ssPSP}

In \emph{multidimensional scheduling}~\cite{GhodsiZHKSS10}, there are~$D$ infinitely divisible resources, where resource~$d$ has total supply~$R_d > 0$.
At each time, the resources are allocated among the active jobs. If job~$j$ receives resource vector~$x_j =
(x_{j1},\dots,x_{jD}) \geq 0$, feasibility requires~$\sum_{j \in J} x_{jd} \leq R_d$ for all~$d \in [D]$.
By rescaling, we may assume~$R_d = 1$ for all~$d$.
The processing rate of job~$j$ is determined by a nondecreasing concave utility function~$u_j \colon \RR_{\geq 0}^D \to \RR_{\geq 0}$ with~$u_j(0) = 0$, giving the set of feasible rate vectors
\begin{align*}
    \cP = \Bigl\{ y \in \RR_{\geq 0}^n \Bigm| y_j \leq u_j(x_j) \ \forall j\in J,\quad \sum_{j \in J} x_{jd} \leq 1 \ \forall d \in [D], \quad x_j \geq 0 \ \forall j \in J \Bigr\}.
\end{align*}
This set is convex and downward closed, but in general not a polytope.
For example, for~$n=2$, $D=1$, and~$u_1(x)=u_2(x)=\sqrt{x}$, it is a quarter disk.
Below, we approximate it by polytopes that preserve support submodularity.
\Cref{def:ssPSP} applies to convex downward-closed sets such as~$\cP$, with maxima replaced by suprema.

Im, Kulkarni, and Munagala~\cite{ImKM18} study a structured subclass called \emph{resource scheduling with substitutes} (RS-S), in which the utility functions take the form
\begin{align*}
    u_j(x_j) = \Bigl( \sum_{d \in [D]} \bigl(f_{jd}(x_{jd})\bigr)^{\rho_j} \Bigr)^{1/\rho_j'},
\end{align*}
where the~$f_{jd}$ are increasing, smooth, concave functions with~$f_{jd}(0) = 0$, and the parameters satisfy~$\rho_j \in (0,1]$ and~$\rho_j' \geq \rho_j$.
The parameter~$\rho_j$ controls the degree of substitutability between resources, where smaller values correspond to resources that are harder to substitute for one another.
Every RS-S instance is PF-monotone~\cite{ImKM18}.

\emph{RS-S with separable utilities} is the special case where~$\rho_j = \rho_j' = 1$ for all~$j \in J$.
Here the utilities reduce to the separable form~$u_j(x_j) = \sum_{d \in [D]} f_{jd}(x_{jd})$.
This includes the single-resource case~$D = 1$, known as \emph{scheduling with concave speedup curves}, where each job's rate is an increasing concave function of the one resource it receives.
We show that the support function of RS-S with separable utilities satisfies~\eqref{eq:lattice-submodular}.
For piecewise-linear~$f_{jd}$, RS-S with separable utilities is a special case of generalized flow scheduling, in which a unit of rate sent along a linear piece of slope~$\sigma$ consumes~$1/\sigma$ units of the resource.
General utilities follow by approximation.

\begin{lemma}
    \label{lem:RS-S-generalized-flow}
    For all~$j\in J$ and~$d\in[D]$, let~$\tilde f_{jd}\colon[0,1]\to\RR_{\geq0}$ be increasing, concave, and piecewise linear with~$\tilde f_{jd}(0)=0$.
    Then
    \begin{align*}
        \tilde\cP\coloneqq\Bigl\{ y \in \RR_{\geq 0}^n \Bigm| y_j \leq \sum_{d\in[D]}\tilde f_{jd}(x_{jd}) \ \forall j\in J,\quad \sum_{j \in J} x_{jd} \leq 1 \ \forall d \in [D], \quad x \geq 0 \Bigr\}
    \end{align*}
    is the rate polytope of a generalized flow scheduling instance.
    In particular,~$h_{\tilde\cP}$ satisfies~\eqref{eq:lattice-submodular}.
\end{lemma}
\begin{proof}
    Since~$\tilde f_{jd}$ is increasing, concave, and piecewise linear on~$[0,1]$, it consists of linear pieces $k\in[K_{jd}]$ with lengths~$\ell_{jdk}>0$ satisfying~$\sum_{k}\ell_{jdk}=1$ and slopes~$\sigma_{jd1}\geq\sigma_{jd2}\geq\cdots\geq\sigma_{jdK_{jd}}>0$.
    We first show that
    \begin{equation}
        \label{eq:piecewise-linear-fill}
        \tilde f_{jd}(s)=\max\Bigl\{\sum_{k}\sigma_{jdk}u_k \Bigm| \sum_k u_k=s,\ 0\leq u_k\leq\ell_{jdk}\ \forall k\in[K_{jd}]\Bigr\}
        \qquad\forall s\in[0,1]\,.
    \end{equation}
    Filling the pieces in order defines a feasible vector~$\bar u$ with~$\sum_k\sigma_{jdk}\bar u_k=\tilde f_{jd}(s)$, because~$\tilde f_{jd}(0)=0$.
    Let~$m$ be the smallest index with~$\sum_{k\leq m}\ell_{jdk}\geq s$, so that~$\bar u_k=\ell_{jdk}$ for~$k<m$ and~$\bar u_k=0$ for~$k>m$.
    Every feasible~$u$ then satisfies~$u_k\leq\bar u_k$ for~$k<m$ and~$u_k\geq\bar u_k$ for~$k>m$, so the monotone slopes give~$\sum_k\sigma_{jdk}(u_k-\bar u_k)\leq\sigma_{jdm}\sum_k(u_k-\bar u_k)=0$.

    We now construct the instance.
    Its directed multigraph has the node set~$J\cup R\cup\{t\}$, where $R\coloneqq\{v_d\mid d\in[D]\}$ contains one resource node~$v_d$ for every~$d\in[D]$, and the nodes~$v_1,\dots,v_D,t$ are pairwise distinct and not in~$J$.
    For every~$j\in J$, $d\in[D]$, and~$k\in[K_{jd}]$, it has an arc from~$j$ to~$v_d$ with capacity~$\sigma_{jdk}\ell_{jdk}$ and gain~$1/\sigma_{jdk}$, and for every~$d\in[D]$, it has an arc from~$v_d$ to~$t$ with capacity~$1$ and unit gain.
    We set~$V_=\coloneqq R$, $b_v\coloneqq0$ for~$v\in R\cup\{t\}$, and~$b_j\coloneqq\sum_{d\in[D]}\tilde f_{jd}(1)$ for~$j\in J$.
    For a flow~$z$, write~$z_{jdk}$ for its value on the arc of piece~$k$ from~$j$ to~$v_d$, and~$z_d$ for its value on the arc from~$v_d$ to~$t$.

    Let~$z$ be a feasible flow, and set~$x_{jd}\coloneqq\sum_k z_{jdk}/\sigma_{jdk}$.
    Flow conservation at~$v_d$ gives~$\sum_{j\in J}x_{jd}=z_d\leq1$, so in particular~$x_{jd}\leq1$.
    The capacities give~$0\leq z_{jdk}/\sigma_{jdk}\leq\ell_{jdk}$, so~\eqref{eq:piecewise-linear-fill} with~$u_k=z_{jdk}/\sigma_{jdk}$ yields~$\sum_k z_{jdk}\leq\tilde f_{jd}(x_{jd})$.
    Since no arc enters a job node, $\operatorname{div}z(j)=\sum_{d}\sum_k z_{jdk}\leq\sum_d\tilde f_{jd}(x_{jd})$.
    Hence, every rate vector of the instance lies in~$\tilde\cP$.

    Conversely, let~$y\in\tilde\cP$ with allocation~$x$.
    Then~$x_{jd}\leq1$, and~\eqref{eq:piecewise-linear-fill} provides~$u_{jdk}\in[0,\ell_{jdk}]$ with~$\sum_k u_{jdk}=x_{jd}$ and~$\sum_k\sigma_{jdk}u_{jdk}=\tilde f_{jd}(x_{jd})$.
    The flow with~$z_{jdk}\coloneqq\sigma_{jdk}u_{jdk}$ and~$z_d\coloneqq\sum_{j\in J}x_{jd}$ respects all capacities, conserves flow at every~$v_d\in R$, and satisfies~$\operatorname{div}z(t)=-\sum_d z_d\leq0$ and~$\operatorname{div}z(j)=\sum_d\tilde f_{jd}(x_{jd})$.
    Since~$\tilde f_{jd}$ is increasing, $y_j\leq\operatorname{div}z(j)\leq b_j$ for every job~$j$, so~$z$ is feasible and~$y$ is a rate vector of the instance.

    Finally, every job has a positive rate in some point of~$\tilde\cP$, because~$\tilde f_{jd}(1)>0$.
    Thus \cref{prop:genflow-support-submodular} applies to~$\tilde\cP$.
\end{proof}

For general utilities,~$\cP$ need not be a polytope.
One can approximate~$\cP$ by a packing polytope, e.g., the downward closure of the convex hull of finitely many points of~$\cP$~\cite{ImKM18,Bronstein08}, but such an approximation need not preserve support submodularity.
Instead, we approximate the functions~$f_{jd}$ by piecewise-linear functions, which preserves separability.

\begin{proposition}
    \label{prop:RS-S-polytope-approximation}
    Let~$\cP$ be the feasible rate body of an RS-S instance with separable utilities.
    For every~$\eps\in(0,1)$, there exists a packing polytope~$\cP_\eps$ with~$(1-\eps)\cP\subseteq\cP_\eps\subseteq\cP$ whose support function satisfies~\eqref{eq:lattice-submodular}.
    In particular,~$\cP_\eps$ is support submodular.
\end{proposition}
\begin{proof}
    Let~$\delta \coloneqq \min_{j\in J} u_j(\frac1n\bm{1})$, which is positive since the~$f_{jd}$ are increasing.
    Fix~$\eps\in(0,1)$ and let~$\tilde{f}_{jd}$ be the piecewise-linear interpolant of~$f_{jd}$ on a grid of~$[0,1]$ that contains~$1/n$ and is fine enough that~$\|\tilde{f}_{jd}-f_{jd}\|_\infty \leq \eps\delta/D$ for all~$j,d$. Such a grid exists since each~$f_{jd}$ is continuous on~$[0,1]$.
    Each~$\tilde{f}_{jd}$ is increasing and concave with~$\tilde{f}_{jd}(0)=0$, because~$f_{jd}$ has these properties.
    Thus we obtain a polytope~$\cP_\eps \coloneqq \{y\in \RR_{\geq0}^n \mid y_j \leq \sum_{d\in [D]} \tilde{f}_{jd}(x_{jd}) \ \forall j\in J,\ \sum_{j\in J} x_{jd}\leq 1 \ \forall d\in [D],\ x_j\geq 0\ \forall j\in J\}$ satisfying~$\cP_\eps\subseteq\cP$, as~$\tilde{f}_{jd}\leq f_{jd}$.
    By \cref{lem:RS-S-generalized-flow}, $\cP_\eps$ is the rate polytope of a generalized flow scheduling instance, so its support function satisfies~\eqref{eq:lattice-submodular}, and~$\cP_\eps$ is support submodular by \cref{lem:lattice-submodular}.
    Since~$\tilde{f}_{jd}(1/n)=f_{jd}(1/n)$, the allocation~$x_{jd}=1/n$ shows~$\delta\bm{1}\in\cP_\eps$.
    Moreover, for~$y\in\cP$, the approximation~$\|\tilde{f}_{jd}-f_{jd}\|_\infty \leq \eps\delta/D$ shows~$(y-\eps\delta\bm{1})_+\in\cP_\eps$.
    As~$(1-\eps)(y-\eps\delta\bm{1})_+ + \eps\delta\bm{1} \geq (1-\eps)y$, convexity and downward-closedness of~$\cP_\eps$ give~$(1-\eps)\cP\subseteq\cP_\eps$.
\end{proof}

\begin{proposition}
    \label{prop:RS-S-support-submodular}
    The support function of every RS-S instance with separable utilities, that is, $\rho_j = \rho_j'=1$ for all~$j\in J$, satisfies~\eqref{eq:lattice-submodular}.
    In particular, every such instance is support submodular.
\end{proposition}
\begin{proof}
    The body~$\cP$ is bounded, since~$y_j\leq\sum_{d\in[D]}f_{jd}(1)$ for all~$y\in\cP$, so~$h_\cP$ is finite.
    For~$\eps\in(0,1)$, let~$\cP_\eps$ be as in \cref{prop:RS-S-polytope-approximation}.
    Then~$(1-\eps)h_\cP(\alpha)\leq h_{\cP_\eps}(\alpha)\leq h_\cP(\alpha)$ for every~$\alpha\in\RR^n_{\geq0}$, so~$h_{\cP_\eps}$ converges pointwise to~$h_\cP$ as~$\eps\to0$, and~\eqref{eq:lattice-submodular} carries over from~$\cP_\eps$ to~$\cP$.
    Since~$\cP$ is downward closed, the first part of the proof of \cref{lem:lattice-submodular} applies with maxima replaced by suprema and shows that~$\cP$ is support submodular.
\end{proof}

In \cref{subsec:rho-leq-1}, we show that RS-S with nonseparable utilities need not be support submodular, even if only a single job~$j$ has~$\rho_j \in (0,1)$.
Together with the PF-monotonicity of RS-S~\cite{ImKM18}, this distinguishes the classes support-submodular PSP and PF-monotone PSP.

Since~$\cP$ need not be a polytope, we cannot directly apply \cref{thm:main-completion,thm:main-flow} to~$\cP$, but we can apply them to the polytopes~$\cP_\eps$.

\begin{proposition}
    \label{prop:RS-S-PF-guarantees}
    For RS-S with separable utilities, Proportional Fairness is~$2$-competitive for minimizing weighted completion time when all jobs are released at time~$0$.
    Moreover, for every~$\kappa>2$, Proportional Fairness with speed augmentation~$\kappa$ is~$\frac{\kappa}{\kappa-2}$-competitive for minimizing weighted flow time with arbitrary release dates.
\end{proposition}
\begin{proof}
    Fix~$\eps\in(0,1)$ and let~$\cP_\eps$ be given by \cref{prop:RS-S-polytope-approximation}.
    Suppose first that all jobs are released at time~$0$.
    Every schedule feasible for~$\cP_\eps$ is feasible for~$\cP$, and slowing down a feasible schedule for~$\cP$ by a factor~$1-\eps$ gives a feasible schedule for~$\cP_\eps$ whose weighted completion time is larger by the factor~$1/(1-\eps)$, so~$\OPT(\cP_\eps)\leq \OPT(\cP)/(1-\eps)$.
    By \Cref{thm:main-completion}, PF on~$\cP_\eps$ is therefore~$\frac{2}{1-\eps}$-competitive with respect to~$\cP$.
    To pass to the limit~$\eps\to0$, we first bound the PF rates away from zero uniformly in~$\eps$. Let~$\delta>0$ be as in the proof of \cref{prop:RS-S-polytope-approximation}, so that~$\delta\bm{1}\in\cP_\eps\subseteq\cP$ for every~$\eps$. If~$y$ is the PF rate vector for a nonempty active set~$A$ on any of these bodies, then the first-order optimality condition (\cref{lem:pf-optimality}(ii), which extends to convex bodies~\cite[Proposition~4.1]{ImKM18}) applied to~$\delta\bm{1}_{\{j\}}$ gives~$w_j\delta/y_j\leq w(A)$, so~$y_j\geq\delta w_j/w(J)$ for every~$j\in A$.

    Fix such a set~$A$, let~$y_\eps$ and~$y^*$ be its PF rate vectors on~$\cP_\eps$ and~$\cP$, respectively, and write $f_A(y)\coloneqq\sum_{j\in A}w_j\log y_j$. Since~$(1-\eps)\cP\subseteq\cP_\eps$, optimality gives
    \begin{equation*}
        f_A(y_\eps)\geq f_A((1-\eps)y^*)=f_A(y^*)+w(A)\log(1-\eps)\,.
    \end{equation*}
    By compactness of~$\cP$ and the uniform positive lower bounds, every sequence~$\eps\to0$ has a subsequence along which~$y_\eps$ converges to a feasible vector~$\bar y$ with positive coordinates on~$A$. Taking limits gives~$f_A(\bar y)\geq f_A(y^*)$, so uniqueness of the PF maximizer implies~$\bar y=y^*$. Thus~$y_\eps\to y^*$ for every active set~$A$.

    The PF rates change only at arrival and completion events. If several jobs complete simultaneously in the PF schedule on~$\cP$, then on~$\cP_\eps$ they may complete one after another, but the last of them completes within time~$o(1)$ of the first as~$\eps\to0$, because its remaining processing requirement tends to zero while its rate stays bounded away from zero.
    By induction over the event times of PF on~$\cP$, the remaining processing requirements and thus all completion times of PF on~$\cP_\eps$ converge to those of PF on~$\cP$.
    Thus the objective value of PF on~$\cP_\eps$ converges to that of PF on~$\cP$ as~$\eps\to0$, and PF is~$2$-competitive on~$\cP$.

    For weighted flow time with arbitrary release dates, the same approximation transfers \cref{cor:flow-time-support-submodular-psp} to~$\cP$.
    Since~$(1-\eps)\cP\subseteq\cP_\eps$, the scaled polytope~$\cP_\eps/(1-\eps)$ contains~$\cP$, and it is support submodular because scaling a polytope by a positive factor scales every support function by this factor.
    Moreover, scaling the polytope scales the PF rates by the same factor, so PF on~$\cP_\eps/(1-\eps)$ with speed~$\kappa$ produces the same schedule as PF on~$\cP_\eps$ with speed~$\kappa/(1-\eps)$.
    Hence, for every~$\kappa>2$, \Cref{cor:flow-time-support-submodular-psp} shows that PF on~$\cP_\eps$ with speed~$\kappa/(1-\eps)$ is~$\frac{\kappa}{\kappa-2}$-competitive for minimizing weighted flow time with respect to~$\cP$.
    As~$\eps\to0$, the same continuity argument shows that PF on~$\cP$ with speed~$\kappa$ is~$\frac{\kappa}{\kappa-2}$-competitive.
\end{proof}

\subsection{Non-Support-Submodular Scheduling Problems}
\label{subsec:non-support-submodular-problems}

Not every polytope is support submodular.
In \cref{sec:non-support-submodular-PSP} we prove that graph scheduling, multicommodity flow scheduling, all-or-nothing scheduling, and RS-S with nonseparable utilities are not support submodular in general.
In this subsection, we define graph scheduling, for which we nevertheless obtain flow-time guarantees in \cref{subsec:sparse-binary-polytopes}.

In \emph{graph scheduling}~\cite{DinitzM20,LindermayrLM26,RohwedderS25,RohwedderS26}, we are given a loopless multigraph~$G=(V,E)$.
Each vertex~$v\in V$ is a processor that can work on at most one job at a time, and each edge~$e=uv\in E$ is a job that must be processed on both of its endpoints simultaneously.
A job that requires only a single processor~$v$ is modeled by an edge from~$v$ to a new vertex that is incident to no other edge.
Hence, the jobs processed at any point in time form a matching of~$G$, and since the scheduler may time-share between matchings, the rate polytope is the matching polytope.
Writing~$\delta(v)$ for the set of edges incident to~$v\in V$ and~$E(S)$ for the set of edges with both endpoints in~$S\subseteq V$, Edmonds' matching polytope theorem~\cite[Theorem~11.16]{KorteVygen} gives
\begin{align*}
    \cP_G = \left\{ y\in \RR_{\geq0}^E \;\middle|\;
    \begin{aligned}
        \sum_{e\in\delta(v)}y_e &\leq 1 &&\forall v\in V,\\
        \sum_{e\in E(S)}y_e &\leq \frac{|S|-1}{2} &&\forall S\subseteq V\text{ with }|S|\text{ odd}
    \end{aligned}
    \right\}.
\end{align*}
For~$U\subseteq E$ and~$\alpha\in\RR^E_{\geq0}$, the support function~$H_\alpha(U)$ is the maximum~$\alpha$-weight of a matching of~$G$ contained in~$U$.

In \cref{subsec:graph-scheduling} we show that~$\cP_G$ is support submodular if and only if any two edges in the same connected component of~$G$ share an endpoint, that is, if and only if every connected component of~$G$ is a star (all its edges share a common vertex) or a triangle (possibly with parallel edges).
Thus, support submodularity holds exactly when the rate polytope equals~$\{y\in\RR^E_{\geq0}\mid \sum_{e\in E_i}y_e\leq 1\ \forall i\}$, where~$E_1,\dots,E_m$ are the edge sets of the connected components, that is, when the instance is a collection of independent single machines.

In these non-support-submodular problems, the recurring obstruction is complementarity. A job that needs two resources simultaneously, such as an edge in graph scheduling, which occupies both of its endpoints, can block two jobs that do not compete with each other. Displacing the blocking job pays off only if both blocked jobs are present, so the marginal value of one blocked job can rise once the other one is added (cf.~\Cref{fig:support-submodularity}). In contrast, in unrelated machine scheduling, a job can draw its rate from any of the machines, so the machines act as substitutes.

\section{Weighted Completion Time for Support-Submodular PSP}
\label{sec:completion_time_results}

In this section, we prove \Cref{thm:main-concave,thm:main-completion}.
The main insight is that support submodularity of the rate polytope is equivalent to concavity of the map~$p\mapsto\OPT(p)$ for all weight vectors, and that this concavity alone suffices to show that Proportional Fairness is~$2$-competitive for minimizing weighted completion time.
Additionally, for arbitrary release dates, we show that PF is~$3$-competitive.

Except in the proof of \cref{thm:main-completion}, all jobs in this section are released at time~$0$, and~$\OPT(p)$ denotes the optimal weighted completion time for processing requirements~$p$, that is, the infimum of the weighted completion times of all feasible schedules for~$p$.

\subsection{Fixed-Order Bounds}
\label{subsec:fixed-order-bounds}

Let~$\mathcal S_n$ denote the set of all permutations of~$J=[n]$, which we also call \emph{orders}.
For~$\pi\in\mathcal S_n$ and~$b\in[n]$, we define the suffix~$S_b^\pi\coloneqq\{\pi(b),\ldots,\pi(n)\}$, and we set~$S_{n+1}^\pi\coloneqq\emptyset$.
A schedule \emph{completes the jobs in the order~$\pi$} if~$C_{\pi(1)}\leq C_{\pi(2)}\leq\cdots\leq C_{\pi(n)}$.
For~$p\in\RR^n_{\geq0}$, let~$\OPT_\pi(p)$ denote the infimum of the weighted completion times of the feasible schedules for~$p$ that complete the jobs in the order~$\pi$, where~$\inf\emptyset\coloneqq\infty$.
Every feasible schedule completes the jobs in the order obtained by sorting them by completion time, breaking ties arbitrarily, and hence
\begin{equation}
    \OPT(p)=\min_{\pi\in\mathcal S_n}\OPT_\pi(p).
    \label{eq:opt-fixed-order}
\end{equation}

We compare weighted completion times with linear functions of~$p$ via the following relations.
Consider a feasible schedule for~$p$ with rate vectors~$y(t)$ and sets~$U(t)$ of unfinished jobs.
Since the rate of every job is zero after its completion, we have~$y(t)\in\cP_{U(t)}$ for every time~$t$, and every job~$j$ receives exactly~$p_j$ units of processing, that is,~$\int_0^\infty y(t)\dd t=p$.
Hence, for every~$\alpha\in\RR^n_{\geq0}$, we have
\begin{equation}
    \sum_{j\in J}w_jC_j=\int_0^\infty w(U(t))\dd t
    \qquad\text{and}\qquad
    \langle\alpha,p\rangle=\int_0^\infty\langle\alpha,y(t)\rangle\dd t\leq\int_0^\infty H_\alpha(U(t))\dd t.
    \label{eq:budget-identity}
\end{equation}
Thus, the weighted completion time is at least~$\langle\alpha,p\rangle$ if~$H_\alpha(U(t))\leq w(U(t))$ at every time~$t$, and at most~$\langle\alpha,p\rangle$ if~$\langle\alpha,y(t)\rangle\geq w(U(t))$ at every time~$t$.

For~$\alpha\in\RR^n_{\geq0}$, we call a set~$S\subseteq J$ \emph{tight for~$\alpha$} if~$H_\alpha(S)=w(S)$.
The following lemma constructs prices for which every suffix of a given order is tight.
The construction proceeds backwards along the order.
Once the prices of all later jobs are fixed, we increase the price of the current job until its suffix is tight.

\begin{lemma}[Prices with tight suffixes]
    \label{lem:reverse-order-prices}
    For every order~$\pi\in\mathcal S_n$, there is a unique vector~$\alpha^\pi\in\RR^n_{\geq0}$ such that every suffix of~$\pi$ is tight for~$\alpha^\pi$, that is,~$H_{\alpha^\pi}(S_b^\pi)=w(S_b^\pi)$ for all~$b\in[n+1]$.
\end{lemma}
\begin{proof}
    Fix~$\pi$ and start with~$\alpha=0$.
    We fix the coordinates~$\alpha_{\pi(n)},\ldots,\alpha_{\pi(1)}$ in this order, so coordinates that are not yet fixed are zero.
    Since every~$y\in\cP_S$ satisfies~$y_i=0$ for all~$i\notin S$, the value~$H_\alpha(S)$ depends only on the coordinates~$\alpha_i$ with~$i\in S$.
    We maintain the invariant that, before the step for~$b\in[n]$, the suffixes~$S_{b+1}^\pi,\ldots,S_{n+1}^\pi$ are tight.
    For~$b=n$, this holds since~$H_\alpha(\emptyset)=0=w(\emptyset)$.

    Consider the step for~$b$, and let~$j\coloneqq\pi(b)$.
    Since~$\alpha_j=0$, setting the rate of~$j$ to zero does not change~$\langle\alpha,y\rangle$ and preserves feasibility by downward-closedness of $\cP$.
    Thus, before~$\alpha_j$ is fixed, we have
    \begin{equation*}
        H_\alpha(S_b^\pi)=H_\alpha(S_{b+1}^\pi)=w(S_{b+1}^\pi)<w(S_b^\pi).
    \end{equation*}
    With the other coordinates fixed,~$H_\alpha(S_b^\pi)$ is a continuous and convex function of~$\alpha_j$, since it is the maximum of~$\langle\alpha,y\rangle$ over the finitely many vertices~$y$ of~$\cP_{S_b^\pi}$.
    Moreover,~$y_j^{\max}\bm{1}_{\{j\}}\in\cP_{S_b^\pi}$ by downward-closedness, so~$H_\alpha(S_b^\pi)\geq\alpha_jy_j^{\max}$ tends to infinity as~$\alpha_j$ increases, because~$y_j^{\max}>0$.
    By the intermediate value theorem, we can therefore choose~$\alpha_j\geq0$ such that~$S_b^\pi$ is tight.
    This value is unique, because a convex function on~$[0,\infty)$ that is smaller than~$w(S_b^\pi)$ at~$0$ attains the value~$w(S_b^\pi)$ at most once.
    No later suffix contains~$j$, so the suffixes~$S_{b+1}^\pi,\ldots,S_{n+1}^\pi$ remain tight.
    After the step for~$b=1$, every suffix of~$\pi$ is tight, and we let~$\alpha^\pi$ be the resulting vector.

    For uniqueness, let~$\alpha'\in\RR^n_{\geq0}$ be a vector for which every suffix of~$\pi$ is tight.
    We show by downward induction on~$b\in[n+1]$ that~$\alpha'$ agrees with~$\alpha^\pi$ on~$S_b^\pi$, which for~$b=1$ gives~$\alpha'=\alpha^\pi$.
    For~$b=n+1$, there is nothing to show, since~$S_{n+1}^\pi=\emptyset$.
    If~$\alpha'$ agrees with~$\alpha^\pi$ on~$S_{b+1}^\pi$, then the value~$\alpha_j=\alpha'_{\pi(b)}$ makes~$S_b^\pi$ tight in the step for~$b$ above, because~$H_{\alpha'}(S_b^\pi)$ depends only on the coordinates of~$\alpha'$ in~$S_b^\pi$.
    Hence,~$\alpha'_{\pi(b)}=\alpha^\pi_{\pi(b)}$ by the uniqueness of this value.
\end{proof}

The vector~$\alpha^\pi$ depends only on~$\pi$, the weights, and the rate polytope, but not on the processing requirements.

The set of unfinished jobs of a schedule that completes the jobs in the order~$\pi$ is always a suffix of~$\pi$, which gives the following lower bound for every rate polytope.

\begin{lemma}[Fixed-order lower bound, \Cref{lem:bound-fixed-order}]
    \label{lem:5-bound-fixed-order}
    Let~$\pi\in\mathcal S_n$ and~$\alpha\in\RR^n_{\geq0}$ such that~$H_\alpha(S_b^\pi)\leq w(S_b^\pi)$ for all~$b\in[n]$.
    Then, for every~$p\in\RR^n_{\geq0}$, every feasible schedule for~$p$ that completes the jobs in the order~$\pi$ has weighted completion time at least~$\langle\alpha,p\rangle$.
    In particular,~$\langle\alpha^\pi,p\rangle\leq\OPT_\pi(p)$.
\end{lemma}
\begin{proof}
    Fix such a schedule with rate vectors~$y(t)$ and sets~$U(t)$ of unfinished jobs, and let~$t\geq0$.
    Let~$b$ be the smallest index in~$[n]$ with~$C_{\pi(b)}>t$, or~$b\coloneqq n+1$ if~$C_{\pi(n)}\leq t$.
    Since~$C_{\pi(1)}\leq\cdots\leq C_{\pi(n)}$, we have~$U(t)=\{j\in J\mid C_j>t\}=S_b^\pi$.
    Hence,~$H_\alpha(U(t))\leq w(U(t))$, by assumption if~$b\leq n$ and because~$H_\alpha(\emptyset)=0=w(\emptyset)$ if~$b=n+1$.
    Now~\eqref{eq:budget-identity} gives
    \begin{equation*}
        \langle\alpha,p\rangle
        \leq\int_0^\infty H_\alpha(U(t))\dd t
        \leq\int_0^\infty w(U(t))\dd t
        =\sum_{j\in J}w_jC_j.
    \end{equation*}
    Since every suffix of~$\pi$ is tight for~$\alpha^\pi$ by \cref{lem:reverse-order-prices}, taking the infimum over all feasible schedules for~$p$ that complete the jobs in the order~$\pi$ proves~$\langle\alpha^\pi,p\rangle\leq\OPT_\pi(p)$.
\end{proof}

\begin{remark}
    \label{rem:fixed-order-lp}
    \Cref{lem:5-bound-fixed-order} also follows from weak LP duality.
    Let~$\cP=\{y\in\RR^n_{\geq0}\mid By\leq\bm{1}\}$ with~$B\in\RR^{D\times n}_{\geq0}$ as in \cref{subsec:polytope-scheduling}.
    Every schedule that completes the jobs in the order~$\pi$ induces a feasible solution of the following linear program~$\LP_\pi(p)$ with the same objective value, where $\tau_b\coloneqq C_{\pi(b)}-C_{\pi(b-1)}$, with~$C_{\pi(0)}\coloneqq0$, is the length of the time interval during which exactly the jobs in~$S_b^\pi$ are unfinished, and~$x_{jb}$ is the amount of job~$j$ processed in this interval.
    Hence,~$\LP_\pi(p)$ is at most the weighted completion time of every such schedule, and~$\alpha$ extends to a feasible solution of the dual~$\DLP_\pi(p)$.
    \smallskip

    \begin{minipage}[t]{0.49\textwidth}
    \centering $\LP_\pi(p)$
    \vspace*{-0.3cm}
    \begin{alignat*}{2}
        \min\quad
        &\sum_{b=1}^n w(S_b^\pi) \tau_b \\
        \text{s.t.}\quad
        &\sum_{b=1}^n x_{jb} \ge p_j
            &&\quad \forall j\in J \\
        &\sum_{j\in S_b^\pi} B_{dj}x_{jb}\leq \tau_b
            &&\quad \forall d\in[D],\ b\in[n] \\
        &x_{jb}=0
            &&\quad \forall b\in[n],\ j\notin S_b^\pi \\
        &x, \tau\geq0
    \end{alignat*}
    \end{minipage}\hfill
    \vrule width 0.4pt\hfill
    \begin{minipage}[t]{0.49\textwidth}
    \centering $\DLP_\pi(p)$
    \vspace*{-0.3cm}
    \begin{alignat*}{2}
        \max\quad
        &\sum_{j\in J} p_j\alpha_j \\
        \text{s.t.}\quad
        &\sum_{d=1}^D\beta_{db}\leq w(S_b^\pi)
            &&\quad \forall b\in[n] \\
        &\alpha_j\leq\sum_{d=1}^D B_{dj}\beta_{db}
            &&\quad \forall b\in[n],\ j\in S_b^\pi \\
        &\alpha, \beta\geq0
    \end{alignat*}
    \end{minipage}

    The first dual constraint bounds the total resource price in interval~$b$ by~$w(S_b^\pi)$, the rate at which the weighted completion time accrues in that interval.
    To see that~$\alpha$ can be extended to a dual feasible solution~$(\alpha,\beta)$, consider for every~$b\in[n]$ the linear program~$\max\{\langle\alpha,y\rangle\mid y\in\cP_{S_b^\pi}\}$, whose optimal value is~$H_\alpha(S_b^\pi)$.
    Set~$\beta_b=(\beta_{db})_{d\in[D]}$ to an optimal dual solution of this linear program.
    Dual feasibility implies that~$\sum_{d=1}^D B_{dj}\beta_{db}\geq\alpha_j$ for all~$j\in S_b^\pi$.
    Moreover, strong duality implies that~$\sum_{d=1}^D\beta_{db}=H_\alpha(S_b^\pi)\leq w(S_b^\pi)$, where the inequality is the assumption of \cref{lem:5-bound-fixed-order}.
    This shows that~$(\alpha,\beta)$ is feasible for~$\DLP_\pi(p)$, so weak duality gives~$\langle\alpha,p\rangle\leq\LP_\pi(p)$, which is at most the weighted completion time of every feasible schedule for~$p$ that completes the jobs in the order~$\pi$.
\end{remark}

\subsection{Greedy Schedules and Concavity}
\label{subsec:greedy-concavity}

\Cref{lem:5-bound-fixed-order} bounds~$\langle\alpha^\pi,p\rangle$ from above by~$\OPT_\pi(p)$ for every order~$\pi$.
We now bound~$\langle\alpha^\pi,p\rangle$ from below by~$\OPT(p)$ for support-submodular polytopes.
By~\eqref{eq:budget-identity}, a feasible schedule for~$p$ has weighted completion time at most~$\langle\alpha^\pi,p\rangle$ if~$\langle\alpha^\pi,y(t)\rangle\geq w(U(t))$ at every time~$t$.
Since~$y(t)\in\cP_{U(t)}$, this requires~$H_{\alpha^\pi}(U(t))\geq w(U(t))$, which holds with equality whenever~$U(t)$ is a suffix of~$\pi$, by \cref{lem:reverse-order-prices}.
However, the schedule that we construct below need not complete the jobs in the order~$\pi$, so its sets of unfinished jobs need not be suffixes of~$\pi$.
The following lemma extends the inequality~$H_{\alpha^\pi}(U)\geq w(U)$ from the suffixes of~$\pi$ to all sets~$U\subseteq J$, and it is the only place where we use support submodularity.

\begin{lemma}[Suffix-to-all-sets propagation]
    \label{lem:suffix-tightness-implication}
    Consider a support-submodular PSP instance, and let~$\pi\in \mathcal{S}_n$ and~$\alpha\in\RR^n_{\geq0}$ such that every suffix of~$\pi$ is tight for~$\alpha$.
    Then~$H_\alpha(U) \geq w(U)$ for all~$U\subseteq J$.
\end{lemma}
\begin{proof}
    Fix~$U\subseteq J$, and let~$b\in[n]$ with~$\pi(b)\in U$.
    Since~$U\cap S_b^\pi=(U\cap S_{b+1}^\pi)\cup\{\pi(b)\}$ and $U\cap S_{b+1}^\pi\subseteq S_{b+1}^\pi$, submodularity of~$H_\alpha$ and tightness of the suffixes give
    \begin{equation*}
        H_\alpha(U\cap S_b^\pi)-H_\alpha(U\cap S_{b+1}^\pi)
        \geq H_\alpha(S_b^\pi)-H_\alpha(S_{b+1}^\pi)
        = w_{\pi(b)}.
    \end{equation*}
    For~$b$ with~$\pi(b)\notin U$, the left-hand side is zero, since then~$U\cap S_b^\pi=U\cap S_{b+1}^\pi$.
    Summing over all~$b\in[n]$, the left-hand side telescopes to~$H_\alpha(U)-H_\alpha(\emptyset)=H_\alpha(U)$, so~$H_\alpha(U)\geq w(U)$.
\end{proof}

The schedule for the upper bound uses at every time~$t$ a rate vector that maximizes~$\langle\alpha^\pi,y\rangle$ over $y\in\cP_{U(t)}$.
Formally, for~$\alpha\in\RR^n_{\geq0}$, fix for every~$U\subseteq J$ a rate vector~$y^U\in\cP_U$ with~$\langle\alpha,y^U\rangle=H_\alpha(U)$.
For~$p\in\RR^n_{\geq0}$, the \emph{$\alpha$-greedy schedule} uses at every time~$t$ the rate vector~$y^{U(t)}$, where~$U(t)$ is its own set of unfinished jobs at time~$t$.
That is, whenever its set of unfinished jobs is~$U$, it uses the rate vector~$y^U$ until the next job in~$U$ completes.
We denote its weighted completion time by~$\alphagr(p)$.

\begin{lemma}[$\alpha$-greedy schedule]
    \label{lem:alpha-greedy-schedule}
    Let~$\alpha\in \RR_{\geq0}^n$ such that~$H_\alpha(U) \geq w(U)$ for all~$U\subseteq J$.
    Then, for every~$p\in \RR_{\geq0}^n$, the $\alpha$-greedy schedule for~$p$ is feasible, and~$\OPT(p)\leq\alphagr(p) \leq \langle \alpha, p\rangle$.
\end{lemma}
\begin{proof}
    While the set of unfinished jobs is some~$U\neq\emptyset$, the schedule uses the rate vector~$y^U$, which satisfies~$\langle\alpha,y^U\rangle=H_\alpha(U)\geq w(U)>0$.
    Since~$y^U_i=0$ for all~$i\notin U$, some job~$j\in U$ has~$y^U_j>0$, so the next job in~$U$ completes after finite time.
    Thus, all jobs complete after at most~$n$ such intervals, and the schedule is feasible, which gives~$\OPT(p)\leq\alphagr(p)$.
    Moreover,~$\langle\alpha,y^{U(t)}\rangle=H_\alpha(U(t))\geq w(U(t))$ at every time~$t$, so~\eqref{eq:budget-identity} gives
    \begin{equation*}
        \alphagr(p)
        =\int_0^\infty w(U(t))\dd t
        \leq\int_0^\infty\langle\alpha,y^{U(t)}\rangle\dd t
        =\langle\alpha,p\rangle. \qedhere
    \end{equation*}
\end{proof}

For~$\alpha=\alpha^\pi$, \Cref{lem:reverse-order-prices,lem:suffix-tightness-implication,lem:alpha-greedy-schedule} together prove \cref{lem:overview-tight-chain}.
Combined with the fixed-order lower bound, they give the forward direction of \cref{thm:main-concave}.

\begin{theorem}
    \label{thm:support-submodular-psp-concavity}
    Consider a support-submodular PSP instance, and let~$p\in\RR^n_{\geq0}$.
    Then
    \begin{equation}
        \OPT(p)=\min_{\pi\in\mathcal S_n}\langle\alpha^\pi,p\rangle.
        \label{eq:opt-min-linear}
    \end{equation}
    For every order~$\pi$ attaining this minimum, the $\alpha^\pi$-greedy schedule for~$p$ is optimal, so an optimal schedule for~$p$ exists.
    Conversely, every order in which some optimal schedule for~$p$ completes the jobs attains this minimum.
    In particular, the optimal-value function~$p\mapsto\OPT(p)$ is concave on~$\RR^n_{\geq0}$.
\end{theorem}
\begin{proof}
    Let~$\pi\in\mathcal S_n$.
    Every suffix of~$\pi$ is tight for~$\alpha^\pi$ by \cref{lem:reverse-order-prices}, so \cref{lem:suffix-tightness-implication} gives~$H_{\alpha^\pi}(U)\geq w(U)$ for all~$U\subseteq J$.
    Hence, \cref{lem:alpha-greedy-schedule,lem:5-bound-fixed-order} give
    \begin{equation}
        \OPT(p)\leq\mathrm{GR}_{\alpha^\pi}(p)\leq\langle\alpha^\pi,p\rangle\leq\OPT_\pi(p).
        \label{eq:fixed-order-sandwich}
    \end{equation}
    Taking the minimum over~$\pi\in\mathcal S_n$ and using~\eqref{eq:opt-fixed-order}, we obtain
    \begin{equation*}
        \OPT(p)
        \leq\min_{\pi\in\mathcal S_n}\langle\alpha^\pi,p\rangle
        \leq\min_{\pi\in\mathcal S_n}\OPT_\pi(p)
        =\OPT(p),
    \end{equation*}
    so equality holds throughout, which proves~\eqref{eq:opt-min-linear}.
    If~$\pi$ attains the minimum in~\eqref{eq:opt-min-linear}, then the first two inequalities in~\eqref{eq:fixed-order-sandwich} are equalities, so the $\alpha^\pi$-greedy schedule is optimal.
    If some optimal schedule completes the jobs in the order~$\pi$, then~$\OPT_\pi(p)=\OPT(p)$, so all inequalities in~\eqref{eq:fixed-order-sandwich} are equalities, and~$\pi$ attains the minimum.
    Finally, the vectors~$\alpha^\pi$ do not depend on~$p$, so~$p\mapsto\OPT(p)$ is the minimum of finitely many linear functions and hence concave.
\end{proof}

\subsection{Proof of \texorpdfstring{\Cref{thm:main-completion}}{Theorem~1.3}}
\label{subsec:main-proof-completion}

We now show that concavity of the optimal-value function implies the guarantees of \cref{thm:main-completion}.
Jäger, Lindermayr, and Megow~\cite{JLM2026} proved this implication in the language of superadditivity via dual fitting. The proof of \Cref{lem:subinstance-of-structured-instance} that we give below is simpler and self-contained.
We will reuse this for the flow-time objective in \cref{sec:flow-time}.
\begin{lemma}
    \label{lem:subinstance-of-structured-instance}
    For every nonempty~$A \subseteq J$, let~$y \in \cP$ be a maximizer of~$\sum_{j \in A} w_j \log y_j$.
    Then, for every~$L\subseteq A$, we have
    \begin{align*}
        \OPT(y \odot \bm 1_L) \geq \frac{w(L)^2}{2 w(A)}.
    \end{align*}
    In particular, since~$\OPT$ is coordinate-wise nondecreasing, we have $\OPT(y) \geq \OPT(y \odot \bm 1_A) \geq \frac12 w(A)$.
\end{lemma}
\begin{proof}
    Let~$\hat{y}(s)\in \cP$ for~$s\geq0$ be the rates of an arbitrary schedule for the instance with processing requirements~$\hat{p} \coloneqq y \odot \bm{1}_{L}$.
    The jobs~$j \in J\setminus L$ have zero processing requirement and complete immediately, so we may assume that~$\hat{y}_j(s) = 0$ for all~$j\in J\setminus L$ and~$s > 0$.
    Let $\hat{p}_j(s) \coloneqq \hat{p}_j - \int_0^s \hat{y}_j(s') \dd s'$ be the remaining processing requirement of $j$ at time~$s$ in this schedule, and let $R(s) \coloneqq \sum_{j \in L} w_j \cdot \frac{\hat p_j(s)}{\hat p_j}$ be the remaining fractional weight.
    Then, at any time~$s$, the remaining weight is bounded from below by $R(s)$, so the weighted completion time is bounded from below by the weighted mean busy time $\int_0^\infty R(s) \dd s$.

    We have~$R(0)=w(L)$ and, for almost every~$s\geq0$,
    \begin{align*}
        \frac{\dd}{\dd s} R(s)
        = - \sum_{j \in L} \frac{w_j}{\hat{p}_j} \hat{y}_j(s)
        = -  \sum_{j \in A} \frac{w_j}{y_j} \hat{y}_j(s).
    \end{align*}
    Since~$\hat{y}(s) \in \cP$ and~$y_j=y_j(A)$ for every~$j\in A$, the first-order optimality condition (\cref{lem:pf-optimality}(ii)) implies that for every~$s \ge 0$
    \begin{align*}
        \sum_{j \in A} \frac{w_j}{y_j} \hat{y}_j(s)
        \leq \sum_{j \in A} w_j
        = w(A) .
    \end{align*}

    Consequently, $R(s) \geq w(L) - w(A) s$ for $s \geq 0$.
    Moreover, $R(s) \geq 0$, and thus
    \begin{align*}
        \int_0^\infty R(s) \dd s
        \geq \int_{0}^\infty \max\bigl\{0,\, w(L) - w(A) s\bigr\}\dd s
        = \int_{0}^{w(L) / w(A)} \bigl(w(L) - w(A) s\bigr) \dd s
        = \frac{w(L)^2}{2 w(A)} .
    \end{align*}

    Hence, for every schedule $\hat{y}$ for processing requirements $\hat{p} = y \odot \bm{1}_L$, the weighted completion time is at least $\frac{w(L)^2}{2w(A)}$.
    In particular,~$\OPT(y\odot\bm{1}_L)\geq\frac{w(L)^2}{2w(A)}$.
\end{proof}

\maintheoremcompletiontime*
\begin{proof}
    In this proof, jobs may have arbitrary release dates~$r_j \ge 0$.
    Recall that for any time~$t$, the set~$U(t)\coloneqq\{j\in J\mid t<C_j\}$ denotes the unfinished jobs and~$A(t)\coloneqq\{j\in J\mid r_j\leq t<C_j\}$ denotes the active jobs in the PF schedule.
    Using \cref{lem:subinstance-of-structured-instance} for $A = L = A(t)$ at every time~$t$ with~$A(t)\neq\emptyset$ gives
    \begin{align*}
        \PF(p) &= \int_0^\infty w(U(t)) \dd t
        = \int_0^T w(A(t)) \dd t + \sum_{j\in J} w_j r_j
        \leq 2 \int_0^T \OPT_0(y^{\textsc{pf}}(t)) \dd t + \sum_{j\in J} w_j r_j ,
    \end{align*}
    where $T \coloneqq \max_{j \in J} C_j$ and $\OPT_0$ denotes the optimal weighted completion time of the instance in which all jobs are released at time~$0$, i.e., the optimal-value function~$\OPT$ of the rest of this section.
    If~$T=0$, the claim is immediate. Hence assume that~$T>0$.
    By \cref{thm:support-submodular-psp-concavity}, support submodularity implies that $p\mapsto \OPT_0(p)$ is concave.
    Therefore, using Jensen's inequality and positive homogeneity, we obtain
    \[\int_0^T \opt_0(y^{\textsc{pf}}(t)) \dd t \le T \cdot \opt_0\Bigl(\frac 1 T \int_0^T y^{\textsc{pf}}(t) \dd t \Bigr) = \opt_0\Bigl(\int_0^T y^{\textsc{pf}}(t) \dd t\Bigr) = \opt_0(p) \le \opt(p),\]
    where the last inequality holds because releasing all jobs at time~$0$ cannot increase the optimal weighted completion time.
    Moreover, we have~$\sum_{j\in J} w_j r_j \leq \OPT(p)$, because in every feasible schedule, each job completes no earlier than its release date.
    This implies that~$\PF(p) \leq 3\cdot \OPT(p)$.
    Furthermore, if all jobs are released at time~$0$, then the guarantee improves to~$\PF(p)\leq 2\cdot \OPT(p)$.
\end{proof}

\subsection{Concavity for All Weights Implies Support Submodularity}
\label{subsec:concavity-implies-support-submodularity}

We now prove the converse of \cref{thm:support-submodular-psp-concavity}.
Recall that a job with processing requirement zero completes at time~$0$ and therefore never belongs to the set of unfinished jobs.

\begin{theorem}
    \label{thm:concavity-implies-support-submodularity}
    Let~$\cP \subseteq \RR^n_{\geq0}$ be a downward-closed rate polytope such that, for every weight vector~$w \in \RR^n_{>0}$, the optimal-value function~$p\mapsto\OPT(p)$ of the PSP instance with rate polytope~$\cP$ and weights~$w$ is concave on~$\RR^n_{\geq0}$.
    Then~$\cP$ is support submodular.
\end{theorem}

The proof applies the comparison~\eqref{eq:budget-identity} to a violated submodularity inequality.
If~$H_\alpha$ is not submodular, we construct weights and a decomposition~$p=p^{(i)}+p^{(j)}$ such that, in every optimal schedule for either piece, every set~$U(t)$ of unfinished jobs satisfies~$H_\alpha(U(t))\leq w(U(t))$.
As in the proof of \cref{lem:5-bound-fixed-order}, this bounds the sum of their weighted completion times from below by~$\langle\alpha,p\rangle$.
In contrast, some schedule for~$p$ satisfies~$\langle\alpha,y(t)\rangle\geq w(U(t))$ at every time~$t$, with strict inequality on a time interval of positive length, so its weighted completion time is strictly below~$\langle\alpha,p\rangle$, contradicting concavity.

Since~$\cP$ need not be support submodular here, \Cref{thm:support-submodular-psp-concavity} does not guarantee that optimal schedules exist.
The following lemma shows that they exist for every rate polytope.

\begin{lemma}[Existence of optimal schedules]
    \label{lem:optimal-schedule-exists}
    For every~$p\in\RR^n_{\geq0}$, some feasible schedule for~$p$ has weighted completion time~$\OPT(p)$.
\end{lemma}
\begin{proof}
    By~\eqref{eq:opt-fixed-order}, some order~$\pi\in\mathcal S_n$ satisfies~$\OPT_\pi(p)=\OPT(p)<\infty$.
    Consider a feasible schedule for~$p$ that completes the jobs in the order~$\pi$, set~$C_{\pi(0)}\coloneqq0$, and for~$b\in[n]$, let~$\tau_b\coloneqq C_{\pi(b)}-C_{\pi(b-1)}$ and let~$x_{jb}$ be the amount of processing that job~$j$ receives during~$[C_{\pi(b-1)},C_{\pi(b)})$.
    Then~$(\tau,x)\geq0$ satisfies the linear constraints
    \begin{equation}
        \sum_{b=1}^n x_{jb}=p_j
        \quad\text{for all } j\in J,
        \qquad
        Bx_{\cdot b}\leq\tau_b\bm{1}
        \quad\text{and}\quad
        x_{jb}=0
        \quad\text{for all } b\in[n] \text{ and } j\notin S_b^\pi,
        \label{eq:fixed-order-constraints}
    \end{equation}
    and the weighted completion time of the schedule is~$\sum_{a=1}^n w_{\pi(a)}(\tau_1+\cdots+\tau_a)=\sum_{b=1}^n w(S_b^\pi)\tau_b$.
    Hence, the linear program that minimizes~$\sum_{b=1}^n w(S_b^\pi)\tau_b$ subject to~\eqref{eq:fixed-order-constraints} and~$(\tau,x)\geq0$ is feasible and has optimal value at most~$\OPT_\pi(p)=\OPT(p)$.
    Its objective is nonnegative, so it has an optimal solution~$(\tau,x)$.
    If~$\tau_b=0$, then~$Bx_{\cdot b}\leq0$ implies~$x_{\cdot b}=0$, because~$B\geq0$ has no zero column.
    Consider the schedule that, for~$b=1,\ldots,n$ with~$\tau_b>0$, uses the rate vector~$x_{\cdot b}/\tau_b\in\cP_{S_b^\pi}$ for~$\tau_b$ time units, and that sets the rate of every job to zero after its completion.
    Since~$x_{\pi(a)b}=0$ for~$b>a$ by~\eqref{eq:fixed-order-constraints}, the intervals for~$b\leq a$ provide job~$\pi(a)$ with~$\sum_{b=1}^n x_{\pi(a)b}=p_{\pi(a)}$ units of processing, so it completes by time~$\tau_1+\cdots+\tau_a$.
    Thus, this schedule is feasible and has weighted completion time at most~$\sum_{b=1}^n w(S_b^\pi)\tau_b\leq\OPT(p)$.
\end{proof}

The following lemma, which uses the maximum rates~$y_j^{\max}$ from \cref{subsec:polytope-scheduling}, lets us force the completion orders of the two pieces.

\begin{lemma}[Forcing a completion order]
    \label{lem:concavity-forced-order}
    Let~$w\in\RR^n_{>0}$ and~$p\in\RR^n_{\geq0}$, and let~$u,v\in J$ be two jobs with~$p_v>0$ and
    \begin{align*}
        \frac{p_u}{p_v} < \frac{y_u^{\max}}{y_v^{\max}} \cdot \frac{w_u}{w(J)}.
    \end{align*}
    Then every optimal schedule for~$p$ satisfies~$C_u<C_v$.
\end{lemma}
\begin{proof}
    Suppose a schedule has~$C_u\geq C_v$, and set~$d\coloneqq p_u/y_u^{\max}$.
    We process~$u$ alone at its maximum rate for~$d$ time units, then replay the original schedule with a delay of~$d$ and rate zero for~$u$.
    Downward-closedness of $\cP$ ensures feasibility.
    Job~$u$ now completes at time~$d$, and every other completion time increases by at most~$d$, except that zero-size jobs still complete at time~$0$.
    Since~$C_u\geq C_v\geq p_v/y_v^{\max}$, the change in weighted completion time is at most
    \begin{align*}
        w(J \setminus \{u\}) d - w_u (C_u - d)
        =w(J)d-w_uC_u
        \leq w(J)\frac{p_u}{y_u^{\max}}-w_u\frac{p_v}{y_v^{\max}}
        <0.
    \end{align*}
    Thus the original schedule is not optimal.
\end{proof}

\begin{proof}[Proof of \cref{thm:concavity-implies-support-submodularity}]
    Suppose that~$\cP$ is not support submodular.
    We construct positive weights and processing vectors~$p=p^{(i)}+p^{(j)}$ such that~$\OPT(p^{(i)})+\OPT(p^{(j)})>\OPT(p)$.
    This contradicts concavity, since~$\OPT$ is positively homogeneous by time scaling, and every concave positively homogeneous function is superadditive.

    Let~$\alpha\in\RR^n_{\geq0}$ be such that~$H_\alpha$ is not submodular, which exists by assumption.
    By the local diminishing-returns characterization of submodularity, there are a set~$S$ and distinct jobs~$i,j\notin S$ with
    \begin{align}
        \delta\coloneqq H_\alpha(S\cup\{i,j\})+H_\alpha(S)-H_\alpha(S\cup\{i\})-H_\alpha(S\cup\{j\})>0.
        \label{eq:concavity-support-violation}
    \end{align}
    We first observe that~$S\neq\emptyset$, which we need when forcing completion orders below. To see this, note that
    if $y \in \cP_{\{i,j\}}$ attains~$H_\alpha(\{i,j\})$, then downward-closedness of~$\cP$ gives
    $H_\alpha(\{i,j\}) = \langle \alpha, y \rangle = \alpha_i y_i + \alpha_j y_j \leq H_\alpha(\{i\})+H_\alpha(\{j\})$, a contradiction to \eqref{eq:concavity-support-violation} for $S = \emptyset$.

    As a maximum of finitely many linear functions of~$\alpha$, $H_\alpha(U)$ depends continuously on~$\alpha$ for each $U \subseteq J$.
    For distinct vertices~$y,y'$ of~$\cP$, the equation~$\langle\alpha,y-y'\rangle=0$ defines a hyperplane, and the union of these finitely many hyperplanes has empty interior.
    Hence, we may perturb~$\alpha$ into~$\RR^n_{>0}$ such that it avoids all these hyperplanes and still satisfies the strict inequality~\eqref{eq:concavity-support-violation}.
    From now on,~$\alpha$ denotes the perturbed vector and~$\delta>0$ the corresponding value in~\eqref{eq:concavity-support-violation}.
    Since every~$\cP_U$ is a face of~$\cP$, its vertices are vertices of~$\cP$, so avoiding these hyperplanes ensures that $y\mapsto \langle\alpha,y\rangle$ has a unique maximizer over every~$\cP_U$.

    We set~$S_0\coloneqq S\cup\{i,j\}$, $S_1\coloneqq S$, and~$m\coloneqq|S|$.
    Let~$y^{(0)}$ be the unique maximizer of~$y\mapsto\langle\alpha,y\rangle$ over~$\cP_{S_0}$.
    For~$k=1,\ldots,m$, we choose the maximizer~$y^{(k)}$ of~$y\mapsto \langle\alpha,y\rangle$ over~$\cP_{S_k}$, select~$\pi(k)\in S_k$ with~$y^{(k)}_{\pi(k)}>0$, and set~$S_{k+1}\coloneqq S_k\setminus\{\pi(k)\}$.
    Such a positive coordinate exists because~$\alpha>0$ and every job in the nonempty set~$S_k$ can receive a positive rate.
    Moreover,~$H_\alpha(S_{k+1})<H_\alpha(S_k)$, because otherwise a maximizer over~$\cP_{S_{k+1}}$ would also maximize over~$\cP_{S_k}$, contradicting uniqueness since its coordinate~$\pi(k)$ is zero.
    By construction we have~$S_k=\{\pi(k),\ldots,\pi(m)\}$ for all $k \in [m]$, $S_{m+1} = \emptyset$, and
    \begin{align}
        H_\alpha(S_1)>H_\alpha(S_2)>\cdots>H_\alpha(S_{m+1})=0.
        \label{eq:concavity-chain-decrease}
    \end{align}

    We define the weights by
    \begin{align*}
        w_{\pi(k)}&\coloneqq H_\alpha(S_k)-H_\alpha(S_{k+1})
        &&\text{for } k\in[m],\\*
        w_\ell&\coloneqq H_\alpha(S\cup\{\ell\})-H_\alpha(S)+\delta/4
        &&\text{for } \ell\in\{i,j\},
    \end{align*}
    and~$w_\ell\coloneqq1$ for~$\ell\notin S_0$.
    These weights are positive by~\eqref{eq:concavity-chain-decrease} and monotonicity of~$H_\alpha$.
    Telescoping and~\eqref{eq:concavity-support-violation} give
    \begin{align}
        w(S_k)&=H_\alpha(S_k) &&\text{for } k\in[m+1],
        \label{eq:concavity-chain-tight}\\*
        w(S\cup\{\ell\})&=H_\alpha(S\cup\{\ell\})+\delta/4 &&\text{for } \ell\in\{i,j\},
        \label{eq:concavity-small-sets}\\*
        w(S_0)&=H_\alpha(S_0)-\delta/2.
        \label{eq:concavity-large-set}
    \end{align}

    For a parameter~$\lambda>1$ that we choose later, define~$p\coloneqq\sum_{k=0}^m\lambda^k y^{(k)}$.
    Consider a schedule that uses rate~$y^{(k)}$ for~$\lambda^k$ time units, in the order~$k=0,\ldots,m$.
    All rates lie in~$\cP$, and the schedule gives each job exactly its processing requirement, so it is feasible.
    Note that for job~$\pi(k)$, we have~$y^{(l)}_{\pi(k)} = 0$ for~$l > k$ by construction, hence $p_{\pi(k)} = \sum_{l=0}^k\lambda^l y^{(l)}_{\pi(k)}$. Similarly, $p_i = y^{(0)}_i$ and $p_j = y^{(0)}_j$.
    This means that, for every~$k\in\{0,\ldots,m\}$, only jobs in~$S_k$ are unfinished while the schedule uses~$y^{(k)}$, hence $\OPT(p)
    \leq\sum_{k=0}^m\lambda^k w(S_k)$.
    Using~\eqref{eq:concavity-chain-tight} and~\eqref{eq:concavity-large-set} together with~$\langle\alpha,y^{(k)}\rangle=H_\alpha(S_k)$, we further obtain
    \begin{align}
        \OPT(p)
        \leq\sum_{k=0}^m\lambda^k w(S_k)
        =\sum_{k=0}^m\lambda^k H_\alpha(S_k)-\delta/2
        =\langle\alpha,p\rangle-\delta/2.
        \label{eq:concavity-upper-bound}
    \end{align}

    We split~$p$ by setting $p^{(\ell)}\coloneqq\frac12 p\odot\bm{1}_S+p\odot\bm{1}_{\{\ell\}}$ for $\ell \in \{i,j\}$, so that $p^{(i)}+p^{(j)}=p$.
    For~$k\in[m]$, we also have~$y^{(k)}_{\pi(k)}>0$ by construction.
    Thus, as~$\lambda\to\infty$, we have
    \begin{align*}
        p_{\pi(k)}=\lambda^k y^{(k)}_{\pi(k)}+O(\lambda^{k-1})
        \qquad \text{and} \qquad p_\ell=y^{(0)}_\ell \quad \text{for } \ell \in \{i,j\}.
    \end{align*}
    Consequently, for each~$\ell\in\{i,j\}$, we have
    \begin{align*}
        \frac{p^{(\ell)}_\ell}{p^{(\ell)}_{\pi(1)}}\to0
        \qquad \text{and} \qquad
        \frac{p^{(\ell)}_{\pi(k)}}{p^{(\ell)}_{\pi(k+1)}}\to0
        \quad\text{for } k\in[m-1].
    \end{align*}
    We therefore choose~$\lambda$ large enough that \cref{lem:concavity-forced-order} applies to every consecutive pair in~$\ell,\pi(1),\ldots,\pi(m)$, simultaneously for~$\ell=i$ and~$\ell=j$.
    This is possible because the right-hand side in the condition of
    \Cref{lem:concavity-forced-order} does not depend on $\lambda$ and is by construction positive.
    Thus, every optimal schedule for~$p^{(\ell)}$ completes these jobs in this order.

    Fix~$\ell\in\{i,j\}$ and an optimal schedule for~$p^{(\ell)}$, which exists by \cref{lem:optimal-schedule-exists}, with rates~$y^*(t)$ and unfinished sets~$U(t)$.
    Jobs outside~$S\cup\{\ell\}$ have processing requirement zero and thus complete at time $0$.
    As shown above,
    the fixed optimal schedule completes the jobs in $S\cup\{\ell\}$
    in order $\ell,\pi(1),\ldots,\pi(m)$. Therefore, $U(t)\in\{S\cup\{\ell\},S_1,\ldots,S_{m+1}\}$ for every time $t$.
    By~\eqref{eq:concavity-chain-tight} and~\eqref{eq:concavity-small-sets}, we have~$H_\alpha(U(t))\leq w(U(t))$ for every time $t$.
    Integrating over time gives
    \begin{align*}
        \OPT(p^{(\ell)})
        =\int_0^\infty w(U(t))\dd t
        \geq\int_0^\infty H_\alpha(U(t))\dd t
        \geq\int_0^\infty\langle\alpha,y^*(t)\rangle\dd t
        =\langle\alpha,p^{(\ell)}\rangle.
    \end{align*}
    We combine this bound with~\eqref{eq:concavity-upper-bound} to obtain
    \begin{equation*}
        \OPT(p^{(i)})+\OPT(p^{(j)})
        \geq\langle\alpha,p^{(i)}+p^{(j)}\rangle
        =\langle\alpha,p\rangle
        \ge \OPT(p) + \frac\delta2 > \OPT(p).
    \end{equation*}
    This completes the proof of the theorem.
\end{proof}

Together with \cref{thm:support-submodular-psp-concavity}, this proves \cref{thm:main-concave}.

Finally, we show that concavity of the optimal-value function for a single weight vector does not imply support submodularity.
In particular, we construct a PSP instance on the fractional matching polytope of a triangle that is not support submodular, but whose optimal-value function is concave for unit weights.
\begin{proposition}
    \label{prop:triangle-concave}
    Let~$G$ be a triangle with unit edge weights.
    Its fractional matching polytope $\cQ_G = \{y\in\RR^3_{\geq0}\mid y_1+y_2\leq1,\ y_2+y_3\leq1,\ y_1+y_3\leq1\}$ is not support submodular, while the optimal-value function~$p\mapsto \OPT(p)$ for~$\cQ_G$ is concave.
\end{proposition}
\begin{proof}
    For~$\alpha=\bm1$, the support function of~$\cQ_G$ satisfies~$H_\alpha(\{1,2\})=H_\alpha(\{1,3\})=H_\alpha(\{1\})=1$ and~$H_\alpha(\{1,2,3\})=3/2$, so~$\cQ_G$ is not support submodular.

    For concavity, recall that the proof of \cref{thm:support-submodular-psp-concavity} uses support submodularity only to obtain the inequalities~$H_{\alpha^\pi}(U)\geq w(U)$ for all~$U\subseteq J$ via \cref{lem:suffix-tightness-implication}.
    For~$\cQ_G$, we verify the corresponding inequalities directly.
    The vertices of~$\cQ_G$ are~$0$, the unit vectors, and~$(\frac12,\frac12,\frac12)$.
    For~$U\subseteq J$, the restriction~$(\cQ_G)_U$ is a face of~$\cQ_G$, so its vertices are the vertices of~$\cQ_G$ that lie in~$(\cQ_G)_U$, and~$(\frac12,\frac12,\frac12)$ is one of them only for~$U=J$.
    Fix~$\sigma\in\mathcal S_3$ and let~$\alpha\in\RR^3_{\geq0}$ be given by~$\alpha_{\sigma(1)}=3$, $\alpha_{\sigma(2)}=2$, and $\alpha_{\sigma(3)}=1$.
    Since~$\frac12(3+2+1)=3$, we obtain~$H_{\alpha}(U)=\max_{j\in U}\alpha_j$ for every nonempty~$U\subseteq J$.
    As the entries of~$\alpha$ are the distinct values~$1$, $2$, and~$3$, this gives~$H_{\alpha}(U)\geq|U|=w(U)$ for every nonempty $U\subseteq J$, with equality for the suffixes~$\{\sigma(3)\}$, $\{\sigma(2),\sigma(3)\}$, and~$J$ of~$\sigma$, and~$H_{\alpha}(\emptyset)=0=w(\emptyset)$.
    In particular,~$\alpha=\alpha^\sigma$ by \cref{lem:reverse-order-prices}, and \cref{lem:alpha-greedy-schedule,lem:5-bound-fixed-order} give~$\OPT(p)\leq\langle\alpha^\sigma,p\rangle\leq\OPT_\sigma(p)$ for every~$p\in\RR^3_{\geq0}$.
    Taking the minimum over~$\sigma\in\mathcal S_3$ and using~\eqref{eq:opt-fixed-order}, we obtain~$\OPT(p)=\min_{\sigma\in\mathcal S_3}\langle\alpha^\sigma,p\rangle$, which is concave as a minimum of linear functions.
\end{proof}

\section{Weighted Flow Time for Support-Submodular PSP and Beyond}
\label{sec:flow-time}

By \cref{thm:support-submodular-psp-concavity}, the optimal weighted completion time of a support-submodular PSP instance with release dates~$0$ is the minimum of the linear functions~$p\mapsto\langle\alpha^\pi,p\rangle$.
For~$p\in\RR^n_{\geq0}$, let the support be $\supp(p)\coloneqq \{j \in J \mid p_j > 0\}$.
If an optimal schedule for~$p$ completes the jobs in the order~$\pi$, then~$\pi$ attains this minimum, and~$\supp(p)$ is empty or a suffix of~$\pi$ (see the proof of \Cref{cor:flow-time-support-submodular-psp}), so~$H_{\alpha^\pi}(\supp(p)) = w(\supp(p))$ by \cref{lem:reverse-order-prices}.
We use these properties to show that PF is $(2+\eps)$-speed $O(1/\eps)$-competitive for minimizing weighted flow time on support-submodular polytopes.
Relaxing the equality~$H_{\alpha^\pi}(\supp(p))=w(\supp(p))$ to the inequality~$H_{\alpha(p)}(\supp(p))\leq c\cdot w(\supp(p))$ for suitable prices~$\alpha(p)$ yields a more general result, whose speed and competitive ratio depend on~$c$ and which covers polytopes beyond the support-submodular class.
We show that every downward-closed polytope between a binary packing polytope with column sparsity~$c$ and its integer hull admits a~$c$-approximate certificate.
As a special case, we obtain the first $O(1)$-speed~$O(1)$-competitive nonclairvoyant algorithm for weighted flow time in graph scheduling, which is not support submodular in general (see \Cref{prop:graph-scheduling-stars}).
Throughout this section, $\OPT_0(p)$ denotes the optimal weighted completion time of the instance with processing requirements~$p$ and all release dates equal to~$0$.

\begin{definition}[$c$-approximate certificate]
    \label{def:c-approximate-cert}
    Let~$\cP$ be a polytope and~$w\in\RR^n_{>0}$.
    A function $\alpha \colon \RR^n_{\ge 0} \to \RR^n_{\ge 0}$ is a \emph{$c$-approximate certificate} for~$\cP$ and~$w$ if for all $p \in \RR^n_{\ge 0}$ we have
    \begin{enumerate}[label=(\roman*)]
        \item $\langle \alpha(p), p \rangle \geq \opt_0(p)$,
        \item $\langle \alpha(p), p \rangle \leq \langle \alpha(q), p \rangle$ for all $q \in \RR^n_{\ge 0}$, and
        \item $H_{\alpha(p)}(\supp(p)) \leq c \cdot w(\supp(p))$.
    \end{enumerate}
    We say that~$\cP$ \emph{admits} a $c$-approximate certificate for~$w$ if such a function exists.
\end{definition}
For a $c$-approximate certificate~$\alpha$, define the function~$\bar{C} \colon \RR^n_{\geq0}\to \RR_{\geq0},\ p\mapsto \langle \alpha(p),p\rangle$.
Then (i) states that~$\bar{C}$ is bounded from below by~$\OPT_0$, (ii) implies that~$\bar{C}$ is concave (see \Cref{lem:F-Lipschitz}), and (iii) bounds the support function at the prices~$\alpha(p)$ in terms of the weights.

\begin{lemma}
    \label{lem:F-Lipschitz}
    Let $\alpha \colon \RR^n_{\ge 0} \to \RR^n_{\ge 0}$ be a $c$-approximate certificate and $\bar{C} \colon \RR^n_{\ge 0} \to \RR_{\ge 0},\ p \mapsto \langle \alpha(p), p\rangle$.
    Then $\bar{C}$ is concave, positively homogeneous, and Lipschitz continuous.
\end{lemma}
\begin{proof}
    By \cref{def:c-approximate-cert}(ii), $\bar{C}(p) = \min_{q \in \RR^n_{\ge 0}} {\langle \alpha(q), p \rangle}$, i.e., $\bar{C}$ is a pointwise minimum of linear functions and therefore concave and positively homogeneous.

    Recall that~$0<y_j^{\max} < \infty$ for every job~$j$.
    Define $M \coloneqq c\cdot \frac{w(J)}{\min_j y_j^{\max}}$.
    For every~$q\in\RR^n_{>0}$, we have~$\supp(q)=J$, so monotonicity of the support function and \cref{def:c-approximate-cert}(iii) give for every~$j\in J$ that $\alpha_j(q)\cdot y_j^{\max} = H_{\alpha(q)}(\{j\}) \leq H_{\alpha(q)}(J) \leq c\cdot w(J)$, hence~$\alpha(q) \in [0,M]^n$.
    By \cref{def:c-approximate-cert}(ii), $\bar{C}(p) \leq \langle \alpha(q), p\rangle = \bar{C}(q) + \langle \alpha(q), p - q\rangle \le \bar{C}(q) + M\|p-q\|_1$, and exchanging the roles of $p$ and $q$ gives $|\bar{C}(p) - \bar{C}(q)| \le M\|p-q\|_1$ for all $p, q \in \RR^n_{>0}$.
    For any~$p\in\RR^n_{\geq0}$ and~$\delta>0$, nonnegativity of the certificate vectors and \cref{def:c-approximate-cert}(ii) imply
    \[
        \bar{C}(p)\leq \bar{C}(p+\delta\bm{1})
        \leq \langle\alpha(p),p+\delta\bm{1}\rangle
        =\bar{C}(p)+\delta\langle\alpha(p),\bm{1}\rangle.
    \]
    Thus~$\bar{C}(p+\delta\bm{1})\to \bar{C}(p)$ as~$\delta\to0$.
    Applying the interior Lipschitz bound to~$p+\delta\bm{1}$ and~$q+\delta\bm{1}$ and taking~$\delta\to0$ proves the same bound for all~$p,q\in\RR^n_{\geq0}$.
\end{proof}

The goal of this section is to show the following theorem.
\begin{theorem}
    \label{thm:c-approximate-flow-time}
    For any~$\kappa > 2c$, Proportional Fairness with speed augmentation~$\kappa$ is~$\frac{\kappa}{\kappa-2c}$-competitive for minimizing weighted flow time on every polytope that admits a~$c$-approximate certificate for the job weights.
    In particular, PF is~$\left(2c + \eps\right)$-speed $O(c/\eps)$-competitive for every~$\eps>0$.
\end{theorem}

\subsection{Potential Function}
\label{subsec:potential-approximate-flow-time}

Let~$\alpha$ be a $c$-approximate certificate for the rate polytope~$\cP$ and the job weights~$w$, let~$\bar{C}(q)\coloneqq \langle\alpha(q),q\rangle$ for~$q\in\RR^n_{\geq0}$, and consider an instance with processing requirements~$p$ and arbitrary release dates.
We fix an arbitrary feasible schedule~$y^*$ for this instance, which has unit speed, and let~$\kappa$ be the speed of PF.
We compare PF with~$y^*$, and since~$y^*$ is arbitrary, this comparison yields a bound in terms of the optimal weighted flow time~$\OPT(p)$ of this instance. Let~$\PF_\kappa(p)$ denote the weighted flow time of PF with speed~$\kappa$.
For~$t\geq0$, let~$y^*(t)\in\cP$ and~$y^{\textsc{pf}}(t)\in\cP$ denote the rate vectors of~$y^*$ and of the PF schedule at time~$t$, and let~$p_j^*(t)$ and~$p_j(t)$ denote the corresponding remaining processing requirements of job~$j$, where we set~$p_j^*(t)=p_j(t)=0$ for~$t<r_j$.
We define the \emph{lag}~$z_j(t)$ as the work that PF is behind~$y^*$ on job~$j$ at time~$t$,
\begin{align*}
    z_j(t) \coloneqq \max\{0, p_j(t)-p_j^*(t)\}.
\end{align*}
We write~$L(t) \coloneqq \supp(z(t))$ for the set of jobs with positive lag at time~$t$.
Furthermore, let~$A^*(t)$ and~$A(t)$ denote the sets of active jobs at time~$t$ in~$y^*$ and in PF's schedule, respectively.
Since~$z_j(t)>0$ implies~$t\geq r_j$ and~$p_j(t)>0$, we have~$L(t)\subseteq A(t)$. Moreover, every~$j\in A(t)\setminus L(t)$ satisfies $p_j^*(t)\geq p_j(t)>0$ and is therefore active in~$y^*$ as well, so
\begin{align}
    A(t)\setminus L(t) \subseteq A^*(t). \label{eq:flow-late-jobs}
\end{align}

We prove \cref{thm:c-approximate-flow-time} by a potential-function argument.
Intuitively, the potential measures the cost of catching up, as it evaluates~$\bar{C}$ on the \emph{lag instance} at time~$t$, which has processing requirements~$z(t)$ and release dates~$0$.
For a constant~$\gamma$ fixed in the proof of the theorem, we define the potential as
\begin{align*}
    \Phi(t) \coloneqq \gamma\cdot \bar{C}(z(t)).
\end{align*}
The approximate certificate provides prices at every processing vector, and we use the prices~$\alpha(z(t))$ of the lag instance~$z(t)$, for which~$\bar{C}(z(t)) = \langle\alpha(z(t)),z(t)\rangle$.
Our goal is to show that~$w(A(t)) + \frac{\dd}{\dd t}\Phi(t) \leq \frac{\kappa}{\kappa-2c}\cdot w(A^*(t))$ at almost every time~$t\geq0$, because integrating this inequality over time yields the performance guarantee.
We obtain this in three steps. We first express the derivative of~$\Phi$ in terms of prices. We then bound the contribution of~$y^*$ from above using the $c$-approximation of the support function (\Cref{def:c-approximate-cert}(iii)). Finally, we bound the contribution of PF from below using the first-order optimality of the PF allocation as in \cref{lem:subinstance-of-structured-instance}.
\begin{lemma}
    \label{lem:F-derivative}
    The function~$t\mapsto \bar{C}(z(t))$ is absolutely continuous, and at almost every time~$t\geq0$, it is differentiable with
    \begin{align*}
        \frac{\dd}{\dd t} \bar{C}(z(t))
        \leq \sum_{j\in L(t)} \alpha_j(z(t)) \bigl(y^*_j(t) - \kappa \cdot y_j^{\textnormal{\textsc{pf}}}(t)\bigr).
    \end{align*}
\end{lemma}
\begin{proof}
    Let $C_j, C_j^*$ be the completion times in the PF schedule and in~$y^*$, respectively, and let $\mathcal{E} \coloneqq \{r_j \mid j \in J\} \cup \{C_j \mid j \in J\} \cup \{C_j^* \mid j \in J\}$ be the set of event times.
    Since all rate vectors lie in~$\cP$, every job~$j$ satisfies~$y_j^{\textsc{pf}}(t)\leq y_j^{\max}$ and~$y_j^*(t)\leq y_j^{\max}$ at every time~$t$, so~$p_j(\cdot)$ and~$p_j^*(\cdot)$ are Lipschitz continuous on~$[r_j,\infty)$.
    At the release date~$r_j$, both jump from~$0$ to~$p_j$, so~$p_j(\cdot)-p_j^*(\cdot)$ vanishes on~$[0,r_j]$ and is Lipschitz continuous on~$\RR_{\geq0}$, and hence so is~$z_j$.
    Since~$\bar{C}$ is Lipschitz continuous by \cref{lem:F-Lipschitz}, so is~$t\mapsto \bar{C}(z(t))$.
    Hence, it is absolutely continuous and therefore differentiable almost everywhere~\cite{Folland99}. 
    Moreover, for~$t\geq r_j$, we have~$p_j^*(t)=p_j-\int_{r_j}^t y_j^*(u)\dd u$, and thus~$\frac{\dd}{\dd t}p_j^*(t)=-y_j^*(t)$ at almost every time~$t>r_j$~\cite{Folland99}, and both sides vanish for~$t<r_j$.

    Fix a time~$t\notin\mathcal E$ at which~$z$, the composition~$t\mapsto \bar{C}(z(t))$, and~$p_j^*(\cdot)$ for every~$j\in J$ are differentiable with~$\frac{\dd}{\dd t}p_j^*(t)=-y_j^*(t)$, and let~$h>0$.
    Applying \cref{def:c-approximate-cert}(ii) at~$z(t+h)$ gives
    \begin{align*}
        \bar{C}(z(t+h))-\bar{C}(z(t))
        \;\leq\; \langle\alpha(z(t)),z(t+h)\rangle-\langle\alpha(z(t)),z(t)\rangle
        \;=\; \bigl\langle\alpha(z(t)),\,z(t+h)-z(t)\bigr\rangle.
    \end{align*}
    Dividing by~$h$ and letting~$h\to0$ yields~$\frac{\dd}{\dd t} \bar{C}(z(t)) \leq \langle\alpha(z(t)),\frac{\dd}{\dd t}z(t)\rangle$.

    It remains to evaluate the right-hand side.
    If~$j\in L(t)$, then~$p_j-p_j^*$ is continuous and positive at~$t$, hence~$z_j=p_j-p_j^*$ in a neighborhood of~$t$ and~$\frac{\dd}{\dd t}z_j(t)=-\kappa\cdot y_j^{\textsc{pf}}(t)+y_j^*(t)$.
    If~$j\notin L(t)$, then~$z_j(t)=0$, and since~$z_j$ is nonnegative and differentiable at~$t$, this is a minimum, so~$\frac{\dd}{\dd t}z_j(t)=0$.
    Summing over~$j\in J$ gives the claim.
\end{proof}

We now further bound the derivative of~$t\mapsto \bar{C}(z(t))$.
\begin{lemma}
    \label{lem:F-derivative-2}
    At almost every time~$t$ with~$w(A(t)) > 0$, we have
    \begin{align*}
        \frac{\dd}{\dd t} \bar{C}(z(t)) \leq c \cdot w(L(t)) - \kappa \cdot \frac{w(L(t))^2}{2\cdot w(A(t))}.
    \end{align*}
\end{lemma}
\begin{proof}
    Fix a time~$t$ with~$w(A(t))>0$ at which the bound of \cref{lem:F-derivative} holds, that is,
    \begin{align*}
        \frac{\dd}{\dd t} \bar{C}(z(t)) \leq \langle \alpha(z(t)), y^*(t)\odot\bm{1}_{L(t)}\rangle - \kappa \cdot \langle \alpha(z(t)), y^{\textsc{pf}}(t)\odot \bm{1}_{L(t)}\rangle.
    \end{align*}
    Since~$y^*(t)\odot \bm{1}_{L(t)}\in \cP_{L(t)}$, \Cref{def:c-approximate-cert}(iii) for~$p=z(t)$, whose support is~$L(t)$, gives
    \begin{align*}
        \langle \alpha(z(t)), y^*(t)\odot\bm{1}_{L(t)}\rangle \leq H_{\alpha(z(t))}(L(t)) \leq c\cdot w(L(t)).
    \end{align*}
    Furthermore, combining \cref{def:c-approximate-cert}(ii) and~(i) with \cref{lem:subinstance-of-structured-instance} for the PF rate vector~$y^{\textsc{pf}}(t)$ of~$A(t)$ and~$L(t)\subseteq A(t)$ gives
    \begin{align*}
        \langle \alpha(z(t)), y^{\textsc{pf}}(t)\odot \bm{1}_{L(t)}\rangle \geq \bar{C}(y^{\textsc{pf}}(t)\odot \bm{1}_{L(t)}) \geq \OPT_0(y^{\textsc{pf}}(t)\odot \bm{1}_{L(t)}) \geq \frac{w(L(t))^2}{2\cdot w(A(t))}.
    \end{align*}
    Combining these inequalities proves the claim.
\end{proof}

\begin{proof}[Proof of \cref{thm:c-approximate-flow-time}]
    We choose~$\gamma \coloneqq \frac{\kappa}{c(\kappa - 2c)}$, which is positive since~$\kappa > 2c$.
    Fix a time~$t$ with~$w(A(t))>0$ at which the bound of \cref{lem:F-derivative-2} holds, and let~$x \coloneqq w(L(t))/w(A(t))$. Then
    \begin{align}
        w(A(t)) + \frac{\dd}{\dd t} \Phi(t)
        &\leq w(A(t)) + c\gamma\cdot w(L(t)) - \gamma\kappa\cdot \frac{w(L(t))^2}{2\cdot w(A(t))} \notag \\
        &= w(A(t)) \cdot \Bigl(1 + c\gamma x - \frac12 \gamma\kappa x^2 \Bigr).
        \label{eq:flow-amortized-ineq}
    \end{align}
    By the choice of~$\gamma$, we have
    \begin{align*}
        c\gamma(1-x) - \Bigl(1+c\gamma x - \frac12 \gamma\kappa  x^2\Bigr)
        = \frac12 \gamma \kappa x^2 - 2c\gamma x + (c\gamma-1)
        = \frac12 \gamma \kappa \left(x-\frac{2c}{\kappa}\right)^2 \geq 0.
    \end{align*}
    Hence, \eqref{eq:flow-amortized-ineq}, $L(t)\subseteq A(t)$, and~\eqref{eq:flow-late-jobs} give
    \begin{align*}
        w(A(t)) + \frac{\dd}{\dd t} \Phi(t)
        \leq c\gamma \cdot w(A(t)\setminus L(t))
        \leq \frac{\kappa}{\kappa-2c} \cdot w(A^*(t)).
    \end{align*}
    If~$w(A(t))=0$, then~$L(t)\subseteq A(t)=\emptyset$, so \cref{lem:F-derivative} gives~$\frac{\dd}{\dd t} \Phi(t) \leq 0$ at almost every such~$t$, and the same bound holds.
    Finally,~$\Phi$ is absolutely continuous by \cref{lem:F-derivative}, $\Phi(0) = 0$, and~$\Phi(t)=0$ once both schedules have completed all jobs.
    Hence, integrating over time shows that~$\PF_\kappa(p)$ is at most~$\frac{\kappa}{\kappa-2c}$ times the weighted flow time~$\int_0^\infty w(A^*(t))\dd t$ of~$y^*$.
    Since~$y^*$ is an arbitrary feasible schedule, we obtain~$\PF_\kappa(p) \leq \frac{\kappa}{\kappa-2c} \cdot \OPT(p)$.
    Thus, PF with speed~$\kappa > 2c$ is~$\frac{\kappa}{\kappa-2c}$-competitive as claimed.
\end{proof}

For support-submodular PSP, assigning to every~$p$ the vector~$\alpha^\pi$ of an order~$\pi$ in which an optimal schedule for~$p$ completes the jobs gives a $1$-approximate certificate, so \cref{thm:c-approximate-flow-time} yields the following theorem, which implies \cref{thm:main-flow}.
\begin{theorem}
    \label{cor:flow-time-support-submodular-psp}
    For any~$\kappa>2$, Proportional Fairness with speed augmentation~$\kappa$ is~$\frac{\kappa}{\kappa-2}$-competitive for minimizing weighted flow time in support-submodular PSP.
    In particular, PF is~$(2+\eps)$-speed $O(1/\eps)$-competitive for every~$\eps>0$.
\end{theorem}
\begin{proof}
    For every~$p\in\RR^n_{\geq0}$, fix an optimal schedule for processing requirements~$p$ and release dates~$0$, which exists by \cref{thm:support-submodular-psp-concavity}.
    Let~$\pi_p$ be an order in which this schedule completes the jobs, and set~$\alpha(p)\coloneqq\alpha^{\pi_p}$.
    By \cref{thm:support-submodular-psp-concavity}, the order~$\pi_p$ attains the minimum in~\eqref{eq:opt-min-linear}, so~$\langle \alpha(p),p\rangle = \OPT_0(p) \leq \langle \alpha(q),p\rangle$ for all~$p,q\in\RR_{\geq0}^n$.
    This gives \cref{def:c-approximate-cert}(i) and~(ii).

    It remains to show \cref{def:c-approximate-cert}(iii) with~$c=1$.
    In every schedule for~$p$, a job~$k$ with~$p_k=0$ completes at time~$0$, whereas a job~$j$ with~$p_j>0$ completes no earlier than~$p_j/y_j^{\max}>0$.
    Hence, every job outside~$\supp(p)$ precedes every job in~$\supp(p)$ in~$\pi_p$, so~$\supp(p)$ is empty or a suffix of~$\pi_p$, and \cref{lem:reverse-order-prices} gives~$H_{\alpha(p)}(\supp(p)) = w(\supp(p))$.
    Therefore,~$p\mapsto\alpha(p)$ is a~$1$-approximate certificate, and the claim follows from \cref{thm:c-approximate-flow-time}.\qedhere
\end{proof}

\subsection{Application: Sparse Binary Packing Constraints}
\label{subsec:sparse-binary-polytopes}

We now identify a class of polytopes admitting approximate certificates.
This class is characterized by the property that each job occupies a bounded number of unit-capacity resources.
The corresponding packing constraints form a binary matrix with sparse columns, and we also allow any strengthening that preserves all integral allocations.

\begin{definition}[$c$-sparse binary polytope]
    \label{def:c-sparse-binary}
    A downward-closed polytope~$\cQ = \{y\in\RR_{\geq0}^n \mid By\leq \bm{1}\}$ is \emph{$c$-sparse binary} if~$B\in \{0,1\}^{D\times n}$ and every column of~$B$ contains at least one and at most~$c$ ones.
\end{definition}
Let~$\cQ_{\mathrm{int}}\coloneqq \mathrm{conv}(\cQ\cap\{0,1\}^n)$ denote the integer hull of~$\cQ$.
Let~$R_j$ denote the set of resources used by job~$j$, that is,~$R_j\coloneqq \{d\in [D] \mid B_{dj}=1\}$.
Similarly, let~$J_d$ be the set of jobs competing for resource~$d$, that is,~$J_d \coloneqq \{j\in J \mid B_{dj}=1\}$.
Thus, if~$\cQ$ is $c$-sparse binary, then~$1\leq |R_j|\leq c$ for every job~$j\in J$.

For~$1$-sparse binary polytopes, every job uses a single resource, hence the rate polytope describes a collection of independent single machines.
For~$c=2$, the motivating example is graph scheduling (\Cref{subsec:non-support-submodular-problems}).
For a loopless multigraph~$G=(V,E)$, the incidence matrix has exactly two ones per column, so the fractional matching polytope~$\cQ_G\coloneqq\{y\in\RR^E_{\geq0}\mid \sum_{e\in\delta(v)} y_e\leq1\ \forall v\in V\}$ is~$2$-sparse binary.
Its integral points are exactly the incidence vectors of matchings of~$G$, so~$(\cQ_G)_{\mathrm{int}}=\cP_G$ is the rate polytope of graph scheduling.
Hence, the following theorem with~$c=2$,~$\cQ=\cQ_G$, and~$\cP=\cP_G$ applies to graph scheduling.
\begin{theorem}
    \label{thm:c-sparse-c-approx}
    Let~$\cQ$ be a~$c$-sparse binary polytope.
    Every downward-closed polytope~$\cP$ with $\cQ_{\mathrm{int}}\subseteq\cP\subseteq\cQ$ admits a~$c$-approximate certificate for all weights.
\end{theorem}

Throughout the proof, fix~$\cQ$ and~$\cP$ as in the theorem and weights~$w\in\RR^n_{>0}$, and let~$H_\alpha$ denote the support function of~$\cP$.
We construct the certificate by relaxing the problem into~$D$ independent single-machine problems, one per resource.
Each such subproblem is solved by Smith's rule~\cite{smith1956various}.
Crucially, since every job occupies each of its resources at the same rate, a single global Smith order solves all of these subproblems simultaneously.
The sparsity of the constraint matrix enters only in the approximation of~$H_{\alpha(p)}(\supp(p))$, where it controls how often a job is charged across the relaxed subproblems.

For an order~$\pi$ of the jobs, we write~$j\preceq_\pi k$ if~$j$ is not later than~$k$ in~$\pi$, and we define the \emph{local suffix weight} of~$j\in J_d$ at resource~$d\in [D]$ by
\begin{align*}
    W_d^\pi(j) \coloneqq w(\{k\in J_d\mid j\preceq_\pi k\}).
\end{align*}
We define the price vector by
\begin{align*}
    \hat\alpha_j^\pi \coloneqq \sum_{d\in R_j} W_d^\pi(j) - (|R_j|-1)w_j.
\end{align*}
Since~$W_d^\pi(j)\geq w_j$ for every~$d\in R_j$, we have~$\hat\alpha_j^\pi\geq w_j>0$.
Now, let~$\pi_p$ be the Smith order of~$p$, that is,~$\pi_p$ orders~$J$ by nonincreasing ratio~$w_j/p_j$, with ties broken arbitrarily and~$w_j/0 = \infty$, so that the jobs outside~$\supp(p)$ come first.
We now define the certificate $\alpha(p) \coloneqq \hat\alpha^{\pi_p}$.

We first show that~$\alpha(p)$ satisfies \cref{def:c-approximate-cert}(ii).
\begin{lemma}
    \label{lem:inner-product-bound}
    For all~$p,q\in \RR_{\geq0}^n$, we have~$\langle \alpha(p),p\rangle \leq \langle\alpha(q),p\rangle$.
\end{lemma}
\begin{proof}
    For~$d\in[D]$ and an arbitrary order~$\pi$ of~$J$, we have
    \begin{align*}
        \sum_{j\in J_d} W_d^\pi(j) p_j
        = \sum_{j\in J_d} p_j \sum_{\substack{k\in J_d\\ j\preceq_\pi k}} w_k
        = \sum_{k\in J_d} w_k \sum_{\substack{j\in J_d\\ j\preceq_\pi k}} p_j
        \geq \sum_{k\in J_d} w_k C_k^{d,\pi}\,,
    \end{align*}
    where~$C_k^{d,\pi}$ is the completion time of job~$k$ when a single machine processes the jobs of~$J_d$ one at a time at unit rate in the order~$\pi$. The inequality accounts for zero-size jobs, which complete at time~$0$ even if their prefix sum is positive. By Smith's rule~\cite{smith1956various}, the actual weighted completion time is minimized by the restriction of~$\pi_p$ to~$J_d$. Since all zero-size jobs come first in~$\pi_p$, the inequality is an equality for this order, so~$\pi_p$ also minimizes the prefix-sum expression on the left.

    We have for every order~$\pi$ of~$J$ that
    \begin{align*}
        \langle\hat\alpha^\pi, p\rangle
        = \sum_{j\in J} \Bigl(\sum_{d\in R_j} W_d^\pi(j) - (|R_j|-1)w_j\Bigr) p_j
        = \sum_{d\in[D]} \sum_{j\in J_d} W_d^\pi(j)p_j - \sum_{j\in J}(|R_j|-1) w_jp_j.
    \end{align*}
    For every~$d$, the restriction of the Smith order~$\pi_p$ to~$J_d$ orders~$J_d$ by nonincreasing~$w_j/p_j$ and thus minimizes the~$d$-th summand. Since~$\sum_{j\in J} (|R_j|-1)w_jp_j$ does not depend on the order,~$\pi_p$ minimizes the whole expression.
    Thus~$\langle\hat\alpha^{\pi_p},p\rangle\leq\langle\hat\alpha^\pi,p\rangle$ for every order~$\pi$ of~$J$, in particular for~$\pi=\pi_q$.
\end{proof}

Next, we show that~$\alpha(p)$ satisfies \cref{def:c-approximate-cert}(iii).
\begin{lemma}
    \label{lem:c-sparse-support-UB}
    For every~$p\in \RR_{\geq0}^n$, we have~$H_{\alpha(p)}(\supp(p)) \leq c\cdot w(\supp(p))$.
\end{lemma}
\begin{proof}
    Define~$x_d \coloneqq w(J_d\cap \supp(p))$ for~$d\in [D]$.
    Every~$k\in J_d\setminus \supp(p)$ satisfies~$w_k/p_k = \infty$, hence~$k$ comes before any~$j\in J_d\cap \supp(p)$ in the Smith order~$\pi_p$.
    Consequently, for~$j\in J_d\cap \supp(p)$ the set~$\{k\in J_d\mid j\preceq_{\pi_p}k\}$ is contained in~$J_d\cap \supp(p)$, and therefore~$W_d^{\pi_p}(j)\leq x_d$.
    Thus, we obtain for every~$j\in \supp(p)$ that
    \begin{align*}
        \alpha_j(p) = \sum_{d\in R_j} W_d^{\pi_p}(j) - (|R_j|-1) w_j \leq \sum_{d\in R_j} x_d.
    \end{align*}
    Thus,~$x\in \RR_{\geq0}^D$ is feasible for the fractional covering LP
    \begin{align*}
        \min\Bigl\{ \sum_{d\in [D]} x_d \Bigm| x\geq 0,\ \sum_{d\in R_j} x_d\geq \alpha_j(p) \ \forall j\in \supp(p) \Bigr\}.
    \end{align*}
    The LP dual maximizes~$\langle\alpha(p),y\rangle$ over~$\cQ_{\supp(p)}$ and thus has value at least~$H_{\alpha(p)}(\supp(p))$, since $\cP\subseteq\cQ$.
    Weak duality and~$|R_j|\leq c$ therefore give
    \[
        H_{\alpha(p)}(\supp(p)) \leq \sum_{d\in [D]} x_d = \sum_{d\in [D]} w(J_d\cap \supp(p)) = \sum_{j\in \supp(p)} |R_j| w_j \leq c\cdot w(\supp(p)). \qedhere
    \]
\end{proof}

To show that~$\alpha$ is a~$c$-approximate certificate, it remains to show \cref{def:c-approximate-cert}(i).
We prove a lower bound on the support function for every order~$\pi$ of the jobs using an integral allocation, and then invoke the~$\alpha$-greedy schedule construction to bound~$\OPT_0(p)$ by~$\langle \alpha(p),p\rangle$.
Note that the following lemma holds for every binary matrix~$B$ without zero columns and does not require sparsity.
\begin{lemma}
    \label{lem:c-sparse-support-LB}
    For a binary constraint matrix~$B$ with no zero column, any order~$\pi$ of~$J$, and every~$U\subseteq J$, there exists~$y\in\cQ_U\cap\{0,1\}^n$ with~$\langle\hat\alpha^\pi,y\rangle\geq w(U)$.
    In particular,~$H_{\hat\alpha^\pi}(U) \geq w(U)$.
\end{lemma}
\begin{proof}
    We prove this by induction on~$|U|$.
    For~$|U|=0$, take~$y=0$.
    Now let~$|U|\geq 1$ and let~$j^*\in U$ be the~$\pi$-minimal job in~$U$.
    Let~$N \coloneqq \bigcup_{d\in R_{j^*}} J_d$ be the set of jobs sharing a resource with~$j^*$ and set~$U'\coloneqq U\setminus N$.

    We first bound~$w(U)$.
    Every~$k\in U$ satisfies~$j^*\preceq_\pi k$ by~$\pi$-minimality of~$j^*$, hence for every~$d\in R_{j^*}$ we have~$U\cap J_d \subseteq \{k\in J_d\mid j^*\preceq_\pi k\}$, and therefore,~$w(U\cap J_d) \leq W_d^\pi(j^*)$.
    Moreover, for each~$k\in U\cap N$ there is at least one~$d\in R_{j^*}$ such that~$k\in U\cap J_d$, and~$j^* \in U\cap J_d$ for all~$d\in R_{j^*}$, hence
    \begin{align*}
        \sum_{d\in R_{j^*}} w(U\cap J_d)
        \geq w(U\cap N) + (|R_{j^*}|-1) w_{j^*}.
    \end{align*}
    It follows that
    \begin{align*}
        w(U)
        = w(U') + w(U\cap N)
        \leq w(U') + \sum_{d\in R_{j^*}} W_d^\pi(j^*) - (|R_{j^*}|-1) w_{j^*}
        = w(U') + \hat\alpha_{j^*}^\pi.
    \end{align*}

    As~$|U'|<|U|$, the induction hypothesis gives~$y'\in\cQ_{U'}\cap\{0,1\}^n$ with~$\langle\hat\alpha^\pi,y'\rangle\geq w(U')$.
    No job of~$U'$ uses a resource of~$R_{j^*}$, so for~$d\in R_{j^*}$ the load of~$y'+\bm{1}_{\{j^*\}}$ on resource~$d$ equals~$1$, while for~$d\notin R_{j^*}$ it is unchanged.
    Hence~$y\coloneqq y'+\bm{1}_{\{j^*\}}\in\cQ_U\cap\{0,1\}^n$, and
    \[
        \langle\hat\alpha^\pi,y\rangle
        = \langle\hat\alpha^\pi,y'\rangle + \hat\alpha^\pi_{j^*}
        \geq w(U') + \hat\alpha^\pi_{j^*}
        \geq w(U).
    \]
    Since~$y\in\cQ_{\mathrm{int}}\subseteq\cP$ and~$\supp(y)\subseteq U$, this also gives~$H_{\hat\alpha^\pi}(U)\geq w(U)$.
\end{proof}

\begin{proof}[Proof of \cref{thm:c-sparse-c-approx}]
    The function~$\alpha$ satisfies \cref{def:c-approximate-cert}(ii) by \cref{lem:inner-product-bound}.
    Moreover, $\alpha$ satisfies~(iii) by \cref{lem:c-sparse-support-UB}.
    For every~$p\in \RR_{\geq0}^n$, \Cref{lem:c-sparse-support-LB} shows that~$\alpha(p)=\hat\alpha^{\pi_p}$ satisfies the hypothesis of \cref{lem:alpha-greedy-schedule}, so the feasible $\alpha(p)$-greedy schedule for~$p$ has weighted completion time at most~$\langle \alpha(p),p\rangle$.
    Hence~$\OPT_0(p) \leq \langle \alpha(p),p\rangle$, which is \cref{def:c-approximate-cert}(i).
    Hence~$\alpha$ is a~$c$-approximate certificate for~$\cP$.
\end{proof}

Combining \cref{thm:c-sparse-c-approx} with \cref{thm:c-approximate-flow-time}, we obtain the following result, which implies \cref{thm:main-flow-graph-scheduling}.
\begin{corollary}
    \label{cor:flow-time-c-sparse-binary}
    Let~$\cQ$ be a~$c$-sparse binary polytope and~$\cP$ a downward-closed polytope such that~$\cQ_{\mathrm{int}}\subseteq\cP\subseteq\cQ$.
    For any~$\kappa>2c$, Proportional Fairness with speed augmentation~$\kappa$ is~$\frac{\kappa}{\kappa-2c}$-competitive for minimizing weighted flow time on~$\cP$.
    In particular, for graph scheduling, PF is~$(4+\eps)$-speed~$O(1/\eps)$-competitive for every~$\eps>0$.
\end{corollary}

\section{Weighted Completion Time for General PSP}
\label{sec:general-psp}

We now turn to weighted completion time on general rate polytopes.
In \cref{subsec:randomized-LB}, we show that, in contrast to support-submodular PSP, no randomized nonclairvoyant algorithm is $2$-competitive, and in \cref{subsec:general-psp-dual-fitting}, we prove that PF is $4$-competitive.

\subsection{A Randomized Lower Bound for General PSP}
\label{subsec:randomized-LB}

On a single machine, no randomized nonclairvoyant algorithm is better than $2$-competitive for total completion time~\cite{MotwaniPT94}, and by \cref{thm:main-completion}, PF achieves this ratio on every support-submodular polytope.
We now show that general PSP is strictly harder.
Our instance uses only two resources and three groups~$A$, $B$, and~$C$ of $N$, $3N$, and $2N$ unit-weight jobs: jobs in~$A$ use the first resource, jobs in~$C$ use the second, and jobs in~$B$ use both simultaneously.
We draw the processing requirements from distributions that ensure that every nonclairvoyant algorithm completes at most one job per unit of consumed resource capacity in expectation, which forces an expected cost of at least~$10N^2$.
In contrast, processing each group at rate~$\frac12$ in shortest-first order costs~$5N^2+O(N)$ in expectation, and an exchange argument within a single time window improves this by~$\Omega(N^2)$.
By Yao's principle~\cite{Yao77}, this yields a competitive ratio strictly above~$2$.

\mainrandomizedLB*

\begin{proof}
    We construct a stochastic instance against which no preemptive (deterministic) nonclairvoyant scheduling policy is better than $(2+10^{-5})$-competitive, i.e., $\EX\bigl[\sum_{j \in J} C_j^\alg\bigr] \ge (2+10^{-5}) \EX[\OPT]$ for every scheduling policy~$\alg$, and then apply Yao's principle~\cite{Yao77} to conclude the randomized lower bound.

    Let $N \in \mathbb N$ and consider a multigraph with four vertices $\{1,2,3,4\}$ and edge set $J = A \mathbin{\dot\cup} B \mathbin{\dot\cup} C$ consisting of groups of sizes $|A| = N$, $|B| = 3N$, and $|C| = 2N$, where all edges from $A$ are incident to $1$ and $2$, all edges from $B$ are incident to $2$ and $3$, and all edges from $C$ are incident to $3$ and $4$.
    Thus, the rate polytope is
    \[
        \cP_N=\Biggl\{y\in\RR_{\geq0}^{n}\Bigm|\sum_{j\in A\cup B}y_j\leq1,\ \sum_{j\in B\cup C}y_j\leq1\Biggr\}.
    \]
    All jobs have unit weight and release date~$0$.
    Set~$\xi \coloneqq \frac12\ln3$ and sample the processing requirements~$p$ independently with
    \begin{align*}
        \pr[p_j>x]=e^{-\Lambda_j(x)},\qquad \text{where} \qquad
        \Lambda_j(x)=
        \begin{cases}
            x, & j\in A\cup C,\\
            x+\min\{x,\xi\}, & j\in B.
        \end{cases}
    \end{align*}
    Thus, for distinct~$i,j \in A \cup C$, we have~$\EX[p_j]=1$ and~$\EX[\min\{p_i,p_j\}]=\frac12$, and for distinct $i,j \in B$, integrating~$\pr[p_j>a]=e^{-\Lambda_j(a)}$ and~$\pr[\min\{p_i,p_j\}>a]=e^{-2\Lambda_j(a)}$ with~$e^{-2\xi}=\frac13$ gives
    \begin{align*}
        \EX[p_j]=\frac23\quad\text{and}\quad
        \EX[\min\{p_j,\xi\}]=\frac13\quad\text{and}\quad
        \EX[\min\{p_i,p_j\}]=\frac5{18}.
    \end{align*}

    Now let us fix a deterministic nonanticipative policy~$\alg$ resulting in the rates~$y^{\alg}$, completion times~$C_j^\alg$, and attained processing~$Y_j^\alg(t)=\int_0^t y^\alg_j(s)\dd s \leq p_j$.
    Our first goal is to show that
    \begin{align}
        \pr[C_j^\alg \leq t]=\EX[\Lambda_j(Y_j^\alg(t))].
        \label{eq:randomized-LB-completion-probability}
    \end{align}
    To see this, condition on all requirements~$p_{-j}$ other than~$p_j$, and let~$x$ be the processing given to~$j$ by time~$t$ under the given $p_{-j}$ in the case that it does not complete before time~$t$.
    If $p_j \le x$, then by nonclairvoyance it will be processed until completion before $t$, i.e., $C_j^\alg \le t$, and $Y_j^\alg(t) = p_j$.
    Otherwise, $Y_j^\alg(t) = x$.
    Thus, in general $Y_j^\alg(t)=\min\{p_j,x\}$.
    Since~$\Lambda_j$ is continuous and piecewise linear with~$\Lambda_j(0)=0$, we can write
    \begin{align*}
        \Lambda_j(\min\{p_j,x\})
        =\int_0^{\min\{p_j,x\}}\Lambda_j'(s)\dd s
        =\int_0^x\Lambda_j'(s)\ind[p_j>s]\dd s.
    \end{align*}
    Taking the conditional expectation over~$p_j$ and using independence gives
    \begin{align*}
        \EX[\Lambda_j(\min\{p_j,x\})\mid p_{-j}]
        &=\int_0^x\Lambda_j'(s)\EX[\ind[p_j>s]\mid p_{-j}]\dd s
        = \int_0^x \Lambda_j'(s) \pr[p_j > s] \dd s\\
        &= \int_0^x\Lambda_j'(s)e^{-\Lambda_j(s)}\dd s
        =1-e^{-\Lambda_j(x)}\\
        &= \pr[p_j\leq x\mid p_{-j}] =\pr[C_j^\alg \leq t\mid p_{-j}]\,.
    \end{align*}
    Now, taking the expectation over $p_{-j}$ yields $\pr[C_j^\alg \le t] = \EX\bigl[\Lambda_j(\min\{p_j,x\})\bigr]$.
    Since $Y_j^\alg(t) = \min\{p_j, x\}$, we thus obtain~\eqref{eq:randomized-LB-completion-probability}.

    Let~$d(t)$ count the jobs completed by time~$t$.
    By~\eqref{eq:randomized-LB-completion-probability} and linearity of expectation, we have
    \begin{equation}\label{eq:expected-number-finished-jobs}
        \EX[d(t)]
        =\sum_{j \in J} \pr[C_j^\alg \le t]
        =\sum_{j \in J} \EX\bigl[\Lambda_j(Y_j^\alg(t))\bigr]
        =\EX\Bigl[\sum_{j\in J}Y_j^\alg(t)+\sum_{j\in B}\min\{Y_j^\alg(t),\xi\}\Bigr].
    \end{equation}
    At each time, by the capacity constraint of vertex~$3$, the total rate that can be attained by jobs in $B \cup C$ is at most $1$.
    Therefore,
    \begin{equation}\label{eq:bound-B-C}
        \sum_{j \in B \cup C} Y_j^\alg(t) \le t.
    \end{equation}
    Similarly, $\sum_{j \in A} Y_j^\alg(t) + \sum_{j \in B} \min \{Y_j^\alg(t), \xi\} \le \sum_{j \in A \cup B} Y_j^\alg(t) \le t$ by the capacity constraint of vertex~$2$.
    We can also bound the left-hand side as follows: $\sum_{j\in A} Y_j^\alg(t) + \sum_{j\in B}\min\{Y_j^\alg(t), \xi\} \le \sum_{j \in A} p_j + \sum_{j \in B} \min\{p_j, \xi\}$.
    Together, this yields
    \begin{equation}\label{eq:bound-A-B}
        \sum_{j \in A} Y_j^\alg(t) + \sum_{j \in B} \min\{Y_j^\alg(t), \xi\} \le \min\Bigl\{t,\ \sum_{j \in A} p_j + \sum_{j \in B} \min\{p_j, \xi\}\Bigr\}.
    \end{equation}
    By summing the inequalities~\labelcref{eq:bound-B-C,eq:bound-A-B}, we obtain
    \[\sum_{j \in J} Y_j^\alg(t) + \sum_{j \in B} \min\{Y_j^\alg(t), \xi\} \le \min\Bigl\{2t,\ t+\sum_{j \in A} p_j + \sum_{j \in B} \min \{p_j, \xi\}\Bigr\}.\]
    Substituting this into~\eqref{eq:expected-number-finished-jobs} and using monotonicity and linearity of the expectation, we obtain
    \[\EX[d(t)] \le \min \Bigl\{2t,\ t + \EX\Bigl[\sum_{j \in A} p_j + \sum_{j \in B} \min\{p_j,\xi\}\Bigr]\Bigr\} = \min\Bigl\{2t,\, t + |A| + |B| \cdot \frac 1 3\Bigr\} = \min\{2t, t + 2N\}.\]
    Integrating the expected number of unfinished jobs $|U(t)| = |J| - d(t)$ yields
    \begin{align*}
        \EX\Bigl[\sum_j C_j^\alg\Bigr]
        &= \int_0^\infty \EX[|U(t)|] \dd t
        =\int_0^\infty(6N-\EX[d(t)])\dd t\\*
        &\geq\int_0^{2N}(6N-2t)\dd t+\int_{2N}^{4N}(4N-t)\dd t
        =10N^2.
    \end{align*}

    We now bound the optimal total completion time.
    Let~$y$ be the schedule that processes each group at rate~$\frac 1 2$ in nondecreasing order of processing requirements and leaves a group's allocation idle once the group is finished.
    For each group~$G\in\{A,B,C\}$, processing~$G$ alone at unit rate in this order would almost surely complete each~$j\in G$ at time~$\sum_{i\in G\colon p_i\leq p_j}p_i$, since the independent processing requirements have continuous distributions. Thus, the expected total completion time of~$G$ in~$y$ is
    \[2 \cdot \Bigl( \frac12 \sum_{i, j \in G\colon i \neq j} \EX[\min\{p_j,p_i\}] + \sum_{j \in G}\EX[p_j]\Bigr),\]
    where the factor~$2$ comes from the rate~$\frac12$.
    Using the formulas from above, we obtain for groups $G=A$ or $G=C$ the expected cost $|G|(|G|-1) \cdot \frac 1 2 + 2 |G| = \frac12|G|^2 + \frac32|G|$.
    For $G=B$, it is $\frac{5}{18}|G|^2 + \frac{19}{18}|G|$.
    Let~$Z$ denote the total completion time of~$y$. Plugging in the group sizes gives
    \begin{align*}
        \EX[Z]=\frac{N^2+(2N)^2}{2}+\frac5{18}(3N)^2+\frac{23}{3}N
        =5N^2+\frac{23}{3}N.
    \end{align*}
    Letting $N \to \infty$, this already yields a lower bound of $2$ for the given stochastic instance. We now slightly improve the schedule~$y$ in an event that occurs with large probability.

    Let~$W_G(\theta) \coloneqq \sum_{j\in G\colon p_j\leq \theta}p_j$ denote the work in group~$G$ contributed by jobs of size at most~$\theta$.
    Let~$\mathcal E_N$ be the event such that
    \begin{align}
        W_A(\infty)<1.02N,\quad W_C(1.7)<1.02N,\quad
        W_B(1.6)>1.09N,\quad\text{and} \quad W_C(\infty)>1.09N.
        \label{eq:randomized-LB-event}
    \end{align}
    Using~$\EX[p_j\ind[p_j\leq \theta]]=\int_0^\theta\pr[p_j>x]\dd x-\theta\pr[p_j>\theta]$, the expectations of these four sums are, respectively,
    \begin{align*}
        N,\qquad 2N(1-2.7e^{-1.7})<1.014N,\qquad
        N(2-2.6\sqrt3e^{-1.6})>1.0907N,\qquad \text{and} \quad 2N.
    \end{align*}
    Since~$\pr[p_j>x]\leq e^{-x}$, every summand in~$W_G(\theta)$ has second moment at most~$\EX[p_j^2]\leq2$.
    Using independence within each group and~$|G|\leq3N$, we therefore bound each sum's variance by~$6N$.
    The first two means lie below their thresholds by at least~$0.02N$ and~$0.006N$, while the last two lie above by at least~$0.0007N$ and~$0.91N$.
    Thus, violating any one inequality requires a deviation of at least~$0.0007N$ from the corresponding mean.
    Chebyshev's inequality and a union bound now give
    \begin{align*}
        \pr[\mathcal E_N^c]\leq4\cdot\frac{6N}{(0.0007N)^2}=O(1/N),
    \end{align*}
    so~$\pr[\mathcal E_N]=1-O(1/N)$.

     On~$\mathcal E_N$, throughout the time interval~$[T,T+2L)$ with~$T=2.04N$ and~$L=0.07N$, group~$A$ has finished, because its completion time in~$y$ is~$2W_A(\infty)<2.04N=T$.
    Meanwhile, $B$ and~$C$ are both active and process only jobs of size at most~$1.6$ and greater than~$1.7$, respectively, during this interval: by time~$T$, group~$C$ has received~$\frac12T=1.02N>W_C(1.7)$ units of processing, and by time~$T+2L$, groups~$B$ and~$C$ have each received~$\frac12(T+2L)=1.09N<\min\{W_B(1.6),W_C(\infty)\}$.
    As illustrated in \cref{fig:randomized-LB-block-exchange}, we replace this block by processing~$B$ alone at rate~$1$ for~$L$ time units, then~$C$ alone at rate~$1$ for~$L$ time units, preserving each group's scheduling order and the processing allocated to every job in the block.
    Since group~$A$ has already completed, this defines a feasible schedule~$\hat{y}$ that rejoins~$y$ at time~$T+2L$ and agrees with it outside the block.

    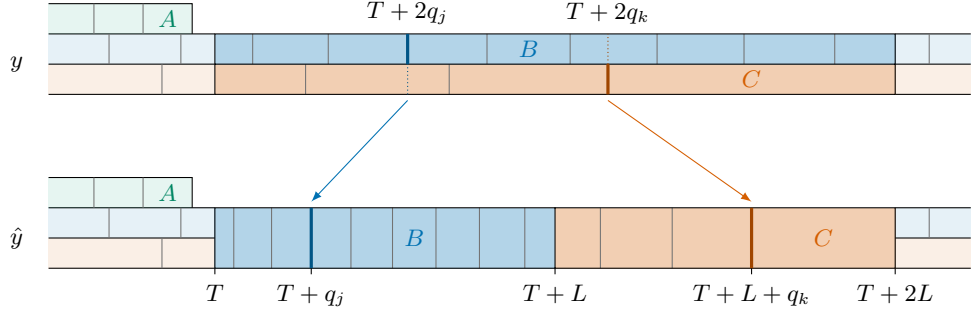
\begin{figure}[tb]
        \centering
        \begin{tikzpicture}[
            x=1cm, y=1cm, font=\footnotesize,
            Bfill/.style={fill=col6, fill opacity=0.3},
            Cfill/.style={fill=col5, fill opacity=0.3},
            Acont/.style={fill=col3, fill opacity=0.1},
            Bcont/.style={fill=col6, fill opacity=0.1},
            Ccont/.style={fill=col5, fill opacity=0.1},
            job/.style={black!55, thin},
            Bmark/.style={col6!75!black, line width=1.2pt},
            Cmark/.style={col5!75!black, line width=1.2pt},
            Bguide/.style={col6!75!black, densely dotted},
            Cguide/.style={col5!75!black, densely dotted},
            move/.style={-latex, thin},
        ]
            \def\rowh{0.8}
            \def\rowy{2.3}
            \def\Bjobs{.5,1.5,3.6,4.7,5.85,7,8.2}
            \def\Cjobs{1.2,3.1}

            \foreach \b in {0,\rowy} {
                \foreach \l/\r in {-2.2/0,9/10} {
                    \fill[Ccont] (\l,\b) rectangle (\r,\b+\rowh/2);
                    \fill[Bcont] (\l,\b+\rowh/2) rectangle (\r,\b+\rowh);
                    \draw[thin] (\l,\b) -- (\r,\b) (\l,\b+\rowh/2) -- (\r,\b+\rowh/2) (\l,\b+\rowh) -- (\r,\b+\rowh);
                }
                \foreach \x in {-1.4,-.45,9.45}
                    \draw[job] (\x,\b+\rowh/2) -- (\x,\b+\rowh);
                \draw[job] (-.7,\b) -- (-.7,\b+\rowh/2);
                \fill[Acont] (-2.2,\b+\rowh) rectangle (-.3,\b+3*\rowh/2);
                \draw[thin] (-2.2,\b+3*\rowh/2) -- (-.3,\b+3*\rowh/2) -- (-.3,\b+\rowh);
                \foreach \x in {-1.6,-.95}
                    \draw[job] (\x,\b+\rowh) -- (\x,\b+3*\rowh/2);
                \node[text=col3!85!black] at (-.625,\b+5*\rowh/4) {$A$};
            }
            \node[anchor=east] at (-2.4,\rowy+\rowh/2) {$y$};
            \node[anchor=east] at (-2.4,\rowh/2) {$\hat y$};

            \fill[Cfill] (0,\rowy) rectangle (9,\rowy+\rowh/2);
            \fill[Bfill] (0,\rowy+\rowh/2) rectangle (9,\rowy+\rowh);
            \foreach \x in \Bjobs
                \draw[job] (\x,\rowy+\rowh/2) -- (\x,\rowy+\rowh);
            \foreach \x in \Cjobs
                \draw[job] (\x,\rowy) -- (\x,\rowy+\rowh/2);
            \draw[thin] (0,\rowy) rectangle (9,\rowy+\rowh) (0,\rowy+\rowh/2) -- (9,\rowy+\rowh/2);
            \node[text=col6] at (4.15,\rowy+3*\rowh/4) {$B$};
            \node[text=col5] at (7.1,\rowy+\rowh/4) {$C$};
            \draw[Bguide] (2.55,\rowy) -- (2.55,\rowy+\rowh/2);
            \draw[Bmark] (2.55,\rowy+\rowh/2) -- (2.55,\rowy+\rowh);
            \draw[Cmark] (5.2,\rowy) -- (5.2,\rowy+\rowh/2);
            \draw[Cguide] (5.2,\rowy+\rowh/2) -- (5.2,\rowy+\rowh);
            \node[above] at (2.55,\rowy+\rowh) {$T+2q_j$};
            \node[above] at (5.2,\rowy+\rowh) {$T+2q_k$};

            \fill[Bfill] (0,0) rectangle (4.5,\rowh);
            \fill[Cfill] (4.5,0) rectangle (9,\rowh);
            \foreach \x in \Bjobs
                \draw[job] (\x/2,0) -- (\x/2,\rowh);
            \foreach \x in \Cjobs
                \draw[job] (4.5+\x/2,0) -- (4.5+\x/2,\rowh);
            \draw[thin] (0,0) rectangle (9,\rowh) (4.5,0) -- (4.5,\rowh);
            \node[text=col6] at (2.6375,\rowh/2) {$B$};
            \node[text=col5] at (8.05,\rowh/2) {$C$};
            \draw[Bmark] (1.275,0) -- (1.275,\rowh);
            \draw[Cmark] (7.1,0) -- (7.1,\rowh);

            \draw[move, col6] (2.55,\rowy-.08) -- (1.275,\rowh+.08);
            \draw[move, col5] (5.2,\rowy-.08) -- (7.1,\rowh+.08);

            \foreach \x/\t in {0/T,1.275/T+q_j,4.5/T+L,7.1/T+L+q_k,9/T+2L}
                \draw[thin] (\x,0) -- (\x,-.08) node[below] {$\t$};
        \end{tikzpicture}
        \caption{The block exchange on~$\mathcal E_N$.
        Group~$A$ finishes before~$T$, and the schedules~$y$ and~$\hat y$ coincide outside the block~$[T,T+2L)$.
        Inside the block, $\hat y$ processes the same work of~$B$ and then of~$C$ in the same order at rate~$1$.
        Hence, a job~$j\in B$ that completes at~$T+2q_j$ in~$y$ is advanced to~$T+q_j$, and a job~$k\in C$ that completes at~$T+2q_k$ is delayed to~$T+L+q_k$.
        Since the jobs of~$B$ in the block are shorter than those of~$C$, more completions move earlier than later.}
        \label{fig:randomized-LB-block-exchange}
    \end{figure}

    Let~$B'$ and~$C'$ be the jobs in the respective groups that complete during~$(T,T+2L]$ in~$y$, and write their completion times as~$T+2q_j$ with~$q_j\in(0,L]$.
    The exchange advances each~$B'$-completion by~$q_j$ and delays each~$C'$-completion by~$L-q_j$. All other completion times are unchanged.
    For every~$v\in[0,L]$, at least~$(L-v)/1.6-1$ jobs in~$B'$ have~$q_j\geq v$, while at most~$v/1.7+1$ jobs in~$C'$ have~$q_j\leq v$.
    Indeed, in~$y$, group~$B$ processes volume~$L-v$ during~$[T+2v,T+2L)$ in jobs of size at most~$1.6$, and group~$C$ processes volume~$v$ during~$(T,T+2v]$ in jobs of size greater than~$1.7$, with one partially processed boundary job in each group.
    Thus the total completion time~$\hat{Z}$ of~$\hat{y}$ satisfies
    \begin{align*}
        Z-\hat{Z}
        &=\sum_{j\in B'}q_j-\sum_{j\in C'}(L-q_j) \\
        &= \sum_{j\in B'}\int_0^L \ind[q_j \ge v] \dd v - \sum_{j \in C'} \int_0^L \ind[q_j \le v] \dd v \\
        &\geq\int_0^L\Bigl(\frac{L-v}{1.6}-\frac{v}{1.7}-2\Bigr)\dd v\\
        &=\frac{L^2}{2}\Bigl(\frac1{1.6}-\frac1{1.7}\Bigr)-2L
        \geq9\cdot10^{-5}N^2-0.14N.
    \end{align*}
    Outside~$\mathcal E_N$, we set~$\hat{y}=y$.
    Thus,
    \begin{align*}
        \hat{Z}\leq Z-\ind[\mathcal E_N]\bigl(9\cdot10^{-5}N^2-0.14N\bigr).
    \end{align*}
    Taking the expectation and using the unconditional baseline cost~$\EX[Z]=5N^2+\frac{23}{3}N$, we obtain
    \begin{align*}
        \EX[\OPT(p)]
        &\leq\EX[\hat{Z}]\\*
        &\leq5N^2+\frac{23}{3}N
            -\pr[\mathcal E_N]\bigl(9\cdot10^{-5}N^2-0.14N\bigr)\\
        &=(5-9\cdot10^{-5})N^2
            +9\cdot10^{-5}N^2\pr[\mathcal E_N^c]
            +\Bigl(\frac{23}{3}+0.14\pr[\mathcal E_N]\Bigr)N\\
        &=(5-9\cdot10^{-5})N^2+O(N),
    \end{align*}
    where we use~$\pr[\mathcal E_N^c]=O(1/N)$ in the last step.

    Finally, we apply the lower-bound direction of Yao's principle~\cite{Yao77}.
    Fix any~$c$-competitive randomized nonclairvoyant algorithm, and let~$R$
    denote its random bits, independent of~$p$.
    For every fixed~$R$, the resulting deterministic algorithm has expected
    cost at least~$10N^2$ under our input distribution.
    Averaging over~$R$ and applying the competitive guarantee for each fixed~$p$
    therefore gives
    \begin{align*}
        10N^2
        \leq \EX_{p,R}[\mathrm{ALG}(p;R)]
        \leq c\cdot\EX_p[\OPT(p)]
        \leq c\bigl((5-9\cdot10^{-5})N^2+O(N)\bigr).
    \end{align*}
    Dividing by~$N^2$ and letting~$N\to\infty$ gives
    $c\geq10/(5-9\cdot10^{-5})>2+10^{-5}$.
\end{proof}

\subsection{Dual Fitting for General PSP}
\label{subsec:general-psp-dual-fitting}

The previously best known competitive ratio for general PSP is~$27$, which is achieved by PF~\cite{JLM2026}.
Our concavity-based approach does not extend to general PSP, because even approximating~$\OPT$ by a concave function up to a constant factor is impossible (see \cref{prop:unbounded-concavity}).
Instead, we analyze PF directly via dual fitting against the standard mean-busy-time relaxation, as in~\cite{ImKM18,JLM2026}.
In contrast to these analyses, which split the unfinished jobs at a global quantile of their normalized rates, our dual solution discounts the credit of each job based on its own normalized rate.
This improves the guarantee to~$4$, even with arbitrary release dates.

\maintheoremgeneralPSP*

We introduce the relaxation and its dual in \cref{subsec:time-indexed-relaxation} and fit a dual solution to the PF schedule in \cref{subsec:general-dual-fitting}.
Throughout the remainder of this section, we assume that~$p_j > 0$ for every job~$j \in J$.
This is without loss of generality, because a job with~$p_j = 0$ completes at its release date~$r_j$ in every schedule (see \Cref{subsec:polytope-scheduling}).
In particular, it is never active, so it does not affect the rates chosen by PF, and it contributes exactly~$w_j r_j$ to both~$\PF$ and~$\OPT$.
Removing all such jobs therefore decreases both objective values by the same amount, so any bound~$\PF \leq c \cdot \OPT$ with~$c \geq 1$ for the remaining instance carries over to the original instance.
The empty remaining instance is immediate, so we also assume~$J\neq\emptyset$.

\subsubsection{A Continuous-Time Relaxation}
\label{subsec:time-indexed-relaxation}

We use the standard mean-busy-time relaxation of the weighted completion time, which charges each job its \emph{mean busy time} rather than its completion time.
The variables~$y_{jt} \geq 0$ describe the rate at which job~$j$ is processed at time~$t \geq r_j$, and the program is given by
\begin{alignat}{3}
    \LP\qquad \text{minimize}\quad
        & \sum_{j\in J} w_j \int_{r_j}^\infty t\cdot \frac{y_{jt}}{p_j} \dd t \notag\\*
    \text{subject to}\quad
        & \int_{r_j}^\infty \frac{y_{jt}}{p_j} \dd t \geq 1
        &&\quad \forall j\in J, \notag\\*
        & \sum_{j\in J \colon r_j\leq t} B_{dj} y_{jt} \leq 1
        &&\quad \forall t \geq 0,\ d\in [D], \notag\\*
        & y\geq0. && \notag
\end{alignat}
The dual program has a variable~$\alpha_j$ per job and a variable~$\beta_{dt}$ per resource and time:
\begin{alignat}{3}
    \DLP\qquad \text{maximize}\quad
        & \sum_{j\in J}\alpha_j-\sum_{d\in[D]}\int_0^\infty\beta_{dt}\dd t \notag\\
    \text{subject to}\quad
        & \alpha_j-w_jt\leq p_j\sum_{d\in[D]}B_{dj}\beta_{dt}
        &&\quad \forall j\in J,\ \forall t\geq r_j, \notag\\
        & \alpha,\beta\geq0. && \notag
\end{alignat}

Since the variables are indexed by continuous time, these programs are infinite-dimensional, and we verify the relaxation and weak duality directly.
\begin{lemma}
    \label{lem:mbt-relaxation}
    It holds that~$\LP \leq \OPT$, and every feasible solution of~$\DLP$ has objective value at most~$\LP$.
\end{lemma}
\begin{proof}
    Fix a feasible schedule for~$p$ with rates~$y^*$ and completion times~$C_j^*$.
    Setting~$y_{jt}\coloneqq y_j^*(t)$ for~$t\geq r_j$ gives a feasible solution of~$\LP$.
    Moreover,~$y_j^*(t)=0$ for~$t\notin [r_j,C_j^*]$ and~$\int_{r_j}^\infty y_j^*(t)/p_j\dd t = 1$, hence the objective value of~$y^*$ is
    \begin{align*}
        \sum_{j\in J} w_j \int_{r_j}^{C_j^*} t\cdot \frac{y_j^*(t)}{p_j} \dd t
        \leq \sum_{j\in J} w_j C_j^* \int_{r_j}^{C_j^*} \frac{y_j^*(t)}{p_j} \dd t
        = \sum_{j\in J} w_j C_j^*.
    \end{align*}
    Since the schedule is arbitrary, this proves that~$\LP\leq \OPT$.

    Next, let~$y$ and~$(\alpha,\beta)$ be feasible solutions for~$\LP$ and~$\DLP$, respectively.
    We obtain
    \begin{align*}
        \sum_{j \in J} w_j \int_{r_j}^\infty t\cdot \frac{y_{jt}}{p_j} \dd t
        &\geq \sum_{j\in J} \int_{r_j}^\infty \frac{y_{jt}}{p_j} \Bigl(\alpha_j - p_j \sum_{d\in[D]} B_{dj}\beta_{dt}\Bigr) \dd t \\
        &= \sum_{j\in J} \alpha_j \int_{r_j}^\infty \frac{y_{jt}}{p_j} \dd t
        - \sum_{d\in[D]} \int_0^\infty \beta_{dt} \sum_{j\in J \colon r_j \leq t} B_{dj} y_{jt} \dd t \\
        &\geq \sum_{j\in J} \alpha_j - \sum_{d\in[D]} \int_0^\infty \beta_{dt} \dd t.
    \end{align*}
    The first inequality is the dual constraint, and the second inequality uses both primal constraints together with~$\alpha,\beta\geq 0$.
    Since~$y$ was an arbitrary feasible solution of~$\LP$, the objective value of~$(\alpha,\beta)$ is at most~$\LP$.
\end{proof}

\subsubsection{Fitting a Dual Solution to Proportional Fairness}
\label{subsec:general-dual-fitting}

We construct a dual solution from the PF schedule that carries a free parameter~$\theta > 0$, which we optimize at the very end.
Let~$y^{\textsc{pf}}(t)$, $A(t)$, and~$C_j$ denote the rates, the set of active jobs, and the completion times of PF, and let~$\eta(t)\coloneqq\eta(A(t))$ be the Lagrange multipliers of~\eqref{psp-pf-cp} for~$A(t)$.
We define the dual variables as follows.
\begin{itemize}
    \item $\displaystyle \dualSa_j \coloneqq w_j \int_0^{C_j}
        \min\Bigl\{1,\ \frac{\theta}{s}\cdot\frac{p_j}{y_j^{\textsc{pf}}(s)}\Bigr\} \dd s$ for all~$j\in J$, and
    \item $\displaystyle \dualSb_{dt} \coloneqq
        \int_t^\infty \frac{\theta}{s}\cdot \eta_d(s) \dd s$ for all~$t>0$, $d\in [D]$.
\end{itemize}
By convention, the minimum equals~$1$ whenever~$s=0$ or~$y_j^{\textsc{pf}}(s) = 0$, so the interval~$[0,r_j)$ contributes~$w_j r_j$ to~$\dualSa_j$.
At~$t=0$, set~$\dualSb_{d0}\coloneqq\max_{j\in J}\dualSa_j/(p_j\sum_{e\in[D]}B_{ej})$ for every~$d\in[D]$. Every denominator is positive, so these prices are finite and satisfy the dual constraints at time zero. Their values at this single time do not affect the objective integral.

Intuitively, the value~$\dualSa_j$ of job~$j$ is a discounted version of its cost~$w_jC_j$ under PF. At every time, the discount factor is capped at~$1$, and it is small exactly when~$j$ is processed quickly relative to the elapsed time.
The price~$\dualSb_{dt}$ of resource~$d$ aggregates the future prices that PF itself attaches to that resource, scaled by~$\theta/s$.
Since~$\eta(s) = 0$ for~$s \geq \max_j C_j$, the quantity~$\dualSb_{dt}$ is finite for every~$t>0$.

\begin{lemma}
    \label{lem:general-dual-feasible}
    The pair~$(\dualSa,\dualSb)$ is feasible for~$\DLP$.
\end{lemma}
\begin{proof}
    Nonnegativity is clear, and the constraints at~$t=0$ hold by construction.
    Fix a job~$j$ and a time~$t \geq r_j$ with~$t>0$.
    Since the integrand defining~$\dualSa_j$ is bounded by~$1$, we have~$\dualSa_j \leq w_j C_j$, so the constraint holds trivially for~$t \geq C_j$.
    On the other hand, for~$t < C_j$, bounding the integrand by~$1$ on~$[0,t]$ gives
    \begin{align*}
        \dualSa_j - w_j t
        \leq w_j \int_t^{C_j} \min\Bigl\{1,\ \frac{\theta}{s}\cdot\frac{p_j}{y_j^{\textsc{pf}}(s)}\Bigr\} \dd s
        \leq p_j \int_t^{C_j} \frac{\theta}{s}\cdot \frac{w_j}{y_j^{\textsc{pf}}(s)} \dd s.
    \end{align*}
    Since $t \ge r_j$, job~$j$ is active at every time $s \in [t, C_j)$ and thus the KKT stationarity condition~\eqref{eq:kkt-stationarity} yields~$w_j/y_j^{\textsc{pf}}(s) = \sum_{d\in[D]} B_{dj}\eta_d(s)$ for every~$s \in [t, C_j)$.
    Hence,
    \[
        \dualSa_j - w_j t
        \leq p_j \int_t^{C_j} \frac{\theta}{s} \sum_{d\in[D]} B_{dj} \eta_d(s) \dd s
        \leq p_j \sum_{d\in[D]} B_{dj} \int_t^{\infty} \frac{\theta}{s} \eta_d(s) \dd s
        = p_j \sum_{d\in[D]} B_{dj} \dualSb_{dt}. \qedhere
    \]
\end{proof}

Next, we split the objective of~$\DLP$ into the~$\dualSa$-part and the~$\dualSb$-part and bound both separately.
\begin{lemma}
    \label{lem:general-dual-beta}
    It holds that~$\sum_{d\in[D]} \int_0^\infty \dualSb_{dt} \dd t \leq \theta \cdot \PF$.
\end{lemma}
\begin{proof}
    Exchanging the order of integration and using \cref{lem:pf-optimality}(i) yields
    \begin{align*}
        \sum_{d\in[D]} \int_0^\infty \dualSb_{dt} \dd t
        &= \sum_{d\in[D]} \int_0^\infty \int_0^s \frac{\theta}{s} \eta_d(s) \dd t \dd s
        = \theta \int_0^\infty \sum_{d\in[D]} \eta_d(s) \dd s \\
        &= \theta \int_0^\infty w(A(s)) \dd s
        = \theta \sum_{j\in J} w_j F_j.
    \end{align*}
    This is at most~$\theta \cdot \PF$ because~$F_j \leq C_j$ for every job~$j$.
\end{proof}

It remains to bound~$\sum_j \dualSa_j$ from below by a constant fraction of~$\PF$.
To do this, we first prove the following integral bound.
\begin{lemma}
    \label{lem:volume-general}
    Let~$\theta>0$ and~$C>0$, and let~$z\colon [0,C]\to\RR_{\geq0}$ be integrable with~$\int_0^C z(t) \dd t = 1$.
    Then
    \begin{align*}
        \int_0^{C} \min\Bigl\{1,\ \frac{\theta}{t\cdot z(t)}\Bigr\} \dd t \geq \frac{4\theta}{4\theta+1} C,
    \end{align*}
    where the minimum is~$1$ whenever~$t\cdot z(t)=0$.
\end{lemma}
\begin{proof}
    Abbreviate~$a\coloneqq 4\theta$ and~$\mu\coloneqq \bigl(\frac{a}{a+1}\bigr)^2 C$, so that~$0<\mu<C$.
    For any~$t\in (0,C]$ it holds that
    \begin{align*}
        \min\Bigl\{1,\ \frac{\theta}{t\cdot z(t)}\Bigr\} + \frac{\mu}{a} z(t)
        \geq \min\Bigl\{1,\ \sqrt{\frac{\mu}{t}} \Bigr\},
    \end{align*}
    because if~$t\cdot z(t)\leq \theta$, then the left side is at least~$1$. Otherwise, the minimum equals~$\frac{\theta}{t\cdot z(t)}$, and since the arithmetic mean of the values $\frac{\theta}{t\cdot z(t)}$ and $\frac{\mu}{a} z(t)$ is at least their geometric mean~$\sqrt{\theta\mu/(ta)}$, we conclude that $\frac{\theta}{t\cdot z(t)} + \frac{\mu}{a} z(t) \geq 2\sqrt{\theta\mu/(ta)} = \sqrt{\mu/t}$.

    We integrate this inequality over~$[0,C]$, use~$\int_0^C z(t)\dd t = 1$, and evaluate the resulting integral, whose integrand equals~$1$ on~$[0,\mu]$ and~$\sqrt{\mu/t}$ on~$[\mu,C]$.
    By the choice of~$\mu$, this gives
    \begin{align*}
        \int_0^C \min\Bigl\{1,\ \frac{\theta}{t\cdot z(t)}\Bigr\} \dd t
        &\geq \int_0^C \min\Bigl\{1,\ \sqrt{\tfrac{\mu}{t}}\Bigr\} \dd t - \frac{\mu}{a}
        = \mu + 2\sqrt{\mu}\bigl(\sqrt{C}-\sqrt{\mu}\bigr) - \frac{\mu}{a} \\
        &= 2\sqrt{\mu C} - \frac{a+1}{a}\,\mu
        = \frac{2a}{a+1}\,C - \frac{a}{a+1}\,C
        = \frac{4\theta}{4\theta+1}\,C. \qedhere
    \end{align*}
\end{proof}

We now prove \cref{thm:general-psp-completion} by bounding the objective value of~$(\dualSa,\dualSb)$ from below in terms of~$\PF$.
\begin{proof}[Proof of \cref{thm:general-psp-completion}]
    For each job~$j$, we apply \cref{lem:volume-general} to~$C = C_j>0$ and~$z(t) = y_j^{\textsc{pf}}(t)/p_j$, which vanishes on~$[0,r_j)$ and integrates to~$1$ over~$[0,C_j]$.
    With this choice, the integral in \cref{lem:volume-general} equals~$\dualSa_j/w_j$, so summing over all jobs gives
    \begin{align*}
        \sum_{j\in J} \dualSa_j
        \geq \frac{4\theta}{4\theta+1} \sum_{j \in J} w_j C_j
        = \frac{4\theta}{4\theta+1}\cdot \PF.
    \end{align*}
    Combining this with \cref{lem:general-dual-beta,lem:general-dual-feasible,lem:mbt-relaxation} yields
    \begin{align*}
        \OPT \geq \LP \geq \sum_{j\in J}\dualSa_j - \sum_{d\in[D]}\int_0^\infty \dualSb_{dt}\dd t
        \geq \Bigl(\frac{4\theta}{4\theta+1} - \theta\Bigr)\cdot \PF.
    \end{align*}
    The function~$\theta \mapsto \frac{4\theta}{4\theta+1} - \theta$ has derivative~$\frac{4}{(4\theta+1)^2} - 1$, which vanishes at~$\theta = \frac14$, where the function attains its maximum value~$\frac12-\frac14 = \frac14$.
    Hence,~$\PF \leq 4\cdot\OPT$, so~$\PF$ is~$4$-competitive on general polytopes with arbitrary release dates.
\end{proof}

\section{Conclusion}
\label{sec:conclusion}

Support submodularity characterizes concavity of the optimal weighted completion time for all job weights and yields improved guarantees for Proportional Fairness.

\paragraph{Weighted completion time.}
PF is~$2$-competitive on support-submodular PSP, where the crucial ingredient of our analysis is the concavity of the optimal-value function.
This result is optimal, since no nonclairvoyant algorithm can do better than~$2$ even on a single machine~\cite{MotwaniPT94}.
Thus, unrelated machines and a single machine admit the same nonclairvoyant guarantee despite their different offline complexities~\cite{smith1956various,Sitters05,Sitters17}.

For arbitrary release dates, we show that PF is~$3$-competitive on support-submodular PSP.
In contrast to the case of uniform release dates, PF is not better than~$2.19$-competitive even on a single machine~\cite{JLM2026}.
PF is therefore not optimal in this setting, as $2$-competitive nonclairvoyant algorithms exist for a single machine with arbitrary release dates~\cite{JaegerSSW24}.

For general PSP, our dual-fitting argument improves PF's competitive ratio from~$27$~\cite{JLM2026} to~$4$.
Moreover, we show that every randomized nonclairvoyant algorithm has a competitive ratio of at least~$2+10^{-5}$, even with unit weights and uniform release dates.
Hence, general PSP is strictly harder than a single machine under nonclairvoyance.

\paragraph{Weighted flow time.}
Even on a single machine, speed augmentation is necessary: without it, no nonclairvoyant algorithm is $O(1)$-competitive, even for unit weights~\cite{MotwaniPT94}, and no deterministic clairvoyant algorithm is $O(1)$-competitive for weighted flow time~\cite{BansalC09}.
For support-submodular PSP, we show that PF with $(2+\eps)$-speed is~$O(1/\eps)$-competitive via a potential-function argument that leverages the concavity of the optimal-value function.
Our analysis is tight for PF~\cite{Edmonds00}.
In particular, PF is $O(1)$-speed $O(1)$-competitive on unrelated machines, contrary to the belief stated in~\cite{ChenIP25}.
This improves the previous~$(e+\eps)$-speed~$O(1/\eps^2)$ guarantee~\cite{ImKM18} on the intersection with PF-monotone PSP, including polymatroids, generalized flow scheduling, and RS-S with separable utilities.

For graph scheduling, PF is~$(4+\eps)$-speed~$O(1/\eps)$-competitive, giving the first~$O(1)$-speed~$O(1)$-competitive nonclairvoyant algorithm.
Furthermore, for every downward-closed polytope between a binary packing polytope with at most~$c$ ones per column and its integer hull, our potential-function argument shows that PF with $(2c+\eps)$-speed is~$O(c/\eps)$-competitive.
For general PSP, no deterministic nonclairvoyant algorithm can be $O(1)$-competitive with speed~$o(\sqrt{\log n})$~\cite{ImKM18}.

\newpage
\begin{appendices}
\crefalias{section}{appendix}
\crefalias{subsection}{subappendix}

\section{Comparison with CIP25}
\label{sec:comparison-chen-im-petety}

Chen, Im, and Petety~\cite{ChenIP25} obtain scalable clairvoyant algorithms by gradient descent on the \emph{residual optimum}, the optimal cost of scheduling the remaining processing requirements.
They assume either that the residual optimum is supermodular in the set of jobs~\cite[Theorem~3.2]{ChenIP25} or that the rate polytope satisfies a substitutes condition~\cite[Theorem~3.3]{ChenIP25}.
We compare the latter condition with support submodularity, and their potential function with ours.

\paragraph{Substitutes and support submodularity.}
Let~$\cP\subseteq\RR^n_{\geq0}$ be a downward-closed rate polytope contained in a box~$\mathcal{B}=\prod_{j\in J}[0,L_j]$.
The hypothesis of~\cite[Theorem~3.3]{ChenIP25} is that, for every $c\in\RR^n_{\geq0}$, the valuation
\begin{equation*}
    v_c(x)\coloneqq\max\bigl\{\langle c,z\rangle\bigm| z\in\cP,\ z\leq x\bigr\}
\end{equation*}
for every $x \in \mathcal{B}$
is a linear-substitute valuation~\cite[Definition~2.11]{ChenIP25}.

\begin{proposition}
    \label{prop:linear-substitutes-imply-lattice}
    If~$v_c$ is a linear-substitute valuation for every~$c\in\RR^n_{\geq0}$, then~$h_\cP$ satisfies~\eqref{eq:lattice-submodular}. Consequently,~$\cP$ is support submodular and PF-monotone for every positive weight vector.
\end{proposition}
\begin{proof}
    Fix~$c\in\RR^n_{\geq0}$. By~\cite[Lemma~2.13]{ChenIP25}, the dual profit~$\pi_c(q)\coloneqq\max_{x\in \mathcal{B}}\{v_c(x)-\langle q,x\rangle\}$ is submodular (see also \cite{MilgromS09}).
    For~$q\in\RR^n_{\geq0}$, we have~$\pi_c(q)=h_\cP(c-q)$, because~$z\leq x$ gives~$\langle c,z\rangle-\langle q,x\rangle\leq\langle c-q,z\rangle$, with equality for~$x=z\in\cP\subseteq \mathcal{B}$.

    Now let~$a,b\in\RR^n_{\geq0}$ and~$c\coloneqq a\vee b$.
    The price vectors~$c-a$ and~$c-b$ are nonnegative, and their componentwise minimum and maximum are~$c-(a\vee b)$ and~$c-(a\wedge b)$.
    Hence, submodularity of~$\pi_c$ gives~$h_\cP(a)+h_\cP(b)\geq h_\cP(a\vee b)+h_\cP(a\wedge b)$, which is~\eqref{eq:lattice-submodular}.
    The remaining claims follow from \cref{lem:lattice-submodular}.
\end{proof}

Support submodularity is strictly weaker than this hypothesis.
Unrelated machines are support submodular (\Cref{prop:unrel-machines-support-submodular}) but in general not PF-monotone~\cite[Proposition~4.13]{JLM2026}, so by \cref{prop:linear-substitutes-imply-lattice}, they do not satisfy the hypothesis in general.
Chen, Im, and Petety treat unrelated machines by separate arguments~\cite[Sections~5.5--5.7]{ChenIP25}.

\paragraph{Potential functions.}
The potential of~\cite{ChenIP25} combines the residual optimum of the algorithm's remaining jobs with that of a joint instance containing one copy of each job for the algorithm and one for the optimal schedule.
Supermodularity of the residual optimum ensures that this potential does not increase when a job arrives~\cite[Lemma~4.4]{ChenIP25}, and the gradient-descent direction controls its drift.
Our potential instead evaluates~$\OPT_0$ on the lag between PF and an arbitrary fixed feasible schedule (\Cref{subsec:overview-flow}).
An arriving job has zero lag, so arrivals do not change the potential, and the drift is controlled by the concavity of~$\OPT_0$ and its supporting prices from \cref{thm:support-submodular-psp-concavity}.
PF itself never computes a residual optimum and only depends on the set of active jobs, their weights, and their columns~$B_j$.

\section{Unbounded Concavity for General PSP}
\label{sec:unbounded-concavity}

Throughout this appendix, all release dates are~$0$. For fixed weights and rate polytope, we write~$\OPT(p)$ for the optimal weighted completion time as a function of the processing requirements~$p$.

For support-submodular rate polytopes, the map~$p\mapsto \OPT(p)$ is concave (\Cref{thm:support-submodular-psp-concavity}), and this concavity is the only property of the polytope used in the competitive analysis of PF (cf.~\Cref{thm:main-completion}).
It is therefore natural to ask whether for general PSP there is a concave~$c$-approximation of~$\OPT$, because such an approximation yields~$2c$-competitiveness for PF.
\begin{definition}[$c$-concavity]
    \label{def:gamma-concavity}
    Let~$c\geq 1$.
    A PSP instance is \emph{$c$-concave} if there exists a concave positively homogeneous function~$G\colon \RR_{\geq0}^n\to\RR_{\geq0}$ such that for all~$p\in \RR_{\geq0}^n$ we have
    \begin{align*}
        \OPT(p) \leq G(p) \leq c \cdot \OPT(p).
    \end{align*}
\end{definition}

As mentioned, $c$-concavity implies that PF is~$2c$-competitive for weighted completion time when all jobs are released at time~$0$.
Indeed, \Cref{lem:subinstance-of-structured-instance}, the bound~$\OPT\leq G$, and the concavity and positive homogeneity of~$G$ (as in \cref{subsec:main-proof-completion}) give
\begin{align*}
    \PF(p)
    = \int_0^\infty w(A(t)) \dd t
    \leq 2 \int_0^\infty \OPT(y^{\textsc{pf}}(t)) \dd t
    \leq 2 \int_0^\infty G(y^{\textsc{pf}}(t)) \dd t
    \leq 2\cdot G(p)
    \leq 2c \cdot \OPT(p).
\end{align*}

\begin{figure}[tb]
\centering
\begin{tikzpicture}[
    scale=1.0,
    vtx/.style={circle, draw, inner sep=0pt, minimum size=4.5pt, fill=white},
    qvtx/.style={vtx, fill=qcol},
    edge/.style={draw, thick},
    lev/.style={font=\footnotesize, anchor=east},
    plab/.style={font=\footnotesize, anchor=west},
]
    \colorlet{c0}{myred}
    \colorlet{c1}{mygreen}
    \colorlet{c2}{myblue}
    \colorlet{qcol}{myorange}

    \draw[qcol, line width=4pt, opacity=.4, line cap=round]
        (4.5,2.8) -- (4.5,1.4) -- (3.5,0);

    \foreach \x in {1,4.5,8} { \draw[edge] (4.5,2.8) -- (\x,1.4); }
    \foreach \c/\a/\b/\d in {1/0/1/2, 4.5/3.5/4.5/5.5, 8/7/8/9} {
        \foreach \x in {\a,\b,\d} { \draw[edge] (\c,1.4) -- (\x,0); }
    }

    \node[qvtx] at (4.5,2.8) {};
    \node[qvtx] at (4.5,1.4) {};
    \node[qvtx] at (3.5,0) {};
    \foreach \x in {1,8} { \node[vtx] at (\x,1.4) {}; }
    \foreach \x in {0,1,2,4.5,5.5,7,8,9} { \node[vtx] at (\x,0) {}; }

    \node[lev] at (-0.6,2.8) {$L_0$};
    \node[lev] at (-0.6,1.4) {$L_1$};
    \node[lev] at (-0.6,0)   {$L_2$};
    \node[plab] at (9.6,2.8) {$p_v = m^{2}$};
    \node[plab] at (9.6,1.4) {$p_v = m^{1}$};
    \node[plab] at (9.6,0)   {$p_v = m^{0}$};

    \node[font=\footnotesize, color=qcol, anchor=east] at (3.3,0.75) {$Q$};
\end{tikzpicture}
\caption{The instance of \cref{prop:unbounded-concavity} for $m = 3$ and $k = 2$.
    Each root-leaf path $Q$ carries one packing constraint and contributes the part~$p^{(Q)} = \bm{1}_Q$. A vertex at depth~$d$ lies on exactly~$m^{k-d} = p_v$ such paths.}
\label{fig:unbounded-superadditivity}
\end{figure}
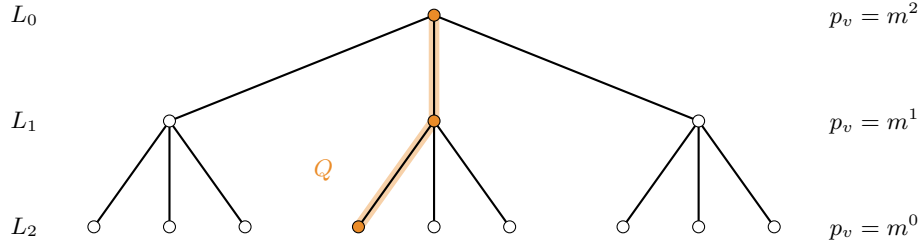

We now show that, for every constant~$c$, there exist PSP instances that are not~$c$-concave.
Hence, the~$c$-concavity framework cannot give PF a constant competitive ratio for general PSP.
\begin{proposition}
    \label{prop:unbounded-concavity}
    For every~$c\geq 1$, there is a PSP instance with unit weights that is not~$c$-concave.
\end{proposition}
\begin{proof}
    Fix $k \geq 1$ and $m \geq 2$.
    Let $T$ be the complete $m$-ary rooted tree of depth $k$ (see \cref{fig:unbounded-superadditivity}), let the job set equal the vertex set of~$T$, and let $\mathcal{Q}$ be the set of the $m^k$ root-leaf paths of $T$, each identified with its vertex set, so $|Q| = k+1$ for all $Q \in \mathcal{Q}$.
    All weights are $1$, a vertex $v$ at depth $\mathrm{lev}(v)$ has $p_v = m^{k - \mathrm{lev}(v)}$, and the rate polytope has one packing constraint per root-leaf path, that is,
    \begin{align*}
        \cP_T
        = \Bigl\{ y \in \RR^n_{\geq 0} \Bigm| \sum_{v \in Q} y_v \leq 1 \ \forall Q \in \mathcal{Q} \Bigr\}.
    \end{align*}
    We decompose $p$ into the $m^k$ parts $p^{(Q)} = \bm{1}_Q$ for $Q \in \mathcal{Q}$.
    This is a decomposition of~$p$, because the subtree below a vertex~$v$ at depth~$d$ has~$m^{k-d}$ leaves, so~$v$ lies on exactly~$m^{k-d} = p_v$ paths of~$\mathcal{Q}$.

    Each part is expensive.
    Since any $Q' \in \mathcal{Q}$ meets $Q$ in a prefix of $Q$, for vectors supported on~$Q$, the constraint of~$Q$ implies all other constraints, so $(\cP_T)_Q = \{ y \ge 0 \mid \supp(y) \subseteq Q,\ \sum_{v \in Q} y_v \leq 1\}$.
    Thus~$p^{(Q)}$ is a single unit-speed machine with $k+1$ unit jobs, and $\OPT(p^{(Q)}) = 1 + 2 + \dots + (k+1) = \tfrac{(k+1)(k+2)}{2}$.
    Summing over the $m^k$ parts yields
    \begin{align*}
        \sum_{Q \in \mathcal{Q}} \OPT\bigl(p^{(Q)}\bigr)
        = m^k \cdot \frac{(k+1)(k+2)}{2}.
    \end{align*}

    The combined instance is cheap.
    Every root-leaf path contains exactly one vertex of each depth, so the indicator $\bm{1}_{L_d}$ of the $d$-th level $L_d$ lies in~$\cP_T$.
    Running $\bm{1}_{L_k}, \bm{1}_{L_{k-1}}, \dots, \bm{1}_{L_0}$ in this order processes level $d$ in time $m^{k-d}$ and completes all~$|L_d| = m^d$ of its jobs at time $\sum_{e \ge d} m^{k-e} = \sum_{i=0}^{k-d} m^i$.
    Hence, counting pairs $(d,i)$ with $d + i = s$ gives
    \begin{align*}
        \OPT(p)
        \leq \sum_{d=0}^{k} m^d \sum_{i=0}^{k-d} m^i
        = \sum_{s=0}^{k} (s+1) m^s
        \le (k+1) m^k \sum_{j \ge 0} m^{-j}
        = (k+1) m^k \cdot \frac{m}{m-1}.
    \end{align*}

    Now suppose the instance is~$c$-concave witnessed by~$G$.
    A concave positively homogeneous function satisfies~$G(a)+G(b) = 2\left(\frac12 G(a)+\frac12 G(b) \right) \leq 2 G\left(\frac12 (a+b)\right) = G(a+b)$ for all~$a,b\in \RR_{\geq0}^n$.
    Hence, for~$p = \sum_{Q\in\mathcal{Q}} p^{(Q)}$ we have
    \begin{align*}
        m^k\cdot \frac{(k+1)(k+2)}{2}
        = \sum_{Q\in \mathcal{Q}} \OPT(p^{(Q)})
        \leq \sum_{Q\in \mathcal{Q}} G(p^{(Q)})
        \leq G(p) \leq c\cdot \OPT(p)
        \leq c (k+1)m^k \cdot \frac{m}{m-1}.
    \end{align*}
    This forces~$c \geq \frac{m-1}{m}\cdot \frac{k+2}{2}$.
    For~$k = \lceil 2c \rceil$ and an integer~$m>c+1$, however, $\frac{m-1}{m}\cdot\frac{k+2}{2} \geq \frac{m-1}{m}(c+1) > c$, a contradiction.
\end{proof}

\section{Non-Support-Submodular Rate Sets}
\label{sec:non-support-submodular-PSP}
In this section, we exhibit scheduling problems whose rate sets are not support submodular.
We first show that multicommodity flow scheduling and all-or-nothing scheduling realize the $m$-ary rooted tree instance of \cref{prop:unbounded-concavity}.
Hence, for every constant~$c\geq1$, both problems contain instances that are not~$c$-concave. In particular, taking~$c=1$ shows that neither problem is support submodular.
Next, in \cref{subsec:graph-scheduling}, we study graph scheduling and characterize exactly the graphs for which the matching polytope is support submodular.
Finally, we turn to resource scheduling with substitutes (\Cref{subsec:multidim-ssPSP}), which we have shown is support submodular when~$\rho_j = \rho_j' = 1$ for all~$j\in J$, and prove that this is sharp in the sense that a single job with~$\rho_j<1$ can already break support submodularity.

\subsection{Multicommodity Flow Scheduling}
In \emph{multicommodity flow scheduling}, we are given a directed graph~$G = (V,E)$ with arc capacities~$c_e > 0$.
Each job~$j\in J$ must route~$p_j$ units of flow from a source~$s_j \in V$ to a sink~$t_j \in V$, and the flows of all jobs must jointly respect the arc capacities.
Writing~$P_j$ for the set of~$(s_j,t_j)$-paths in~$G$ and~$x_{jP}$ for the flow that job~$j$ sends along~$P\in P_j$, the rate of job~$j$ is the total flow it sends, and the feasible rate polytope is
\begin{align*}
    \cP = \Bigl\{ y \in \RR_{\geq0}^n \Bigm| y_j = \sum_{P \in P_j} x_{jP} \ \forall j \in J,\ \sum_{j\in J}\sum_{P \in P_j\colon e \in P} x_{jP} \leq c_e \ \forall e \in E,\ x \geq 0 \Bigr\}.
\end{align*}
Each arc acts as a shared capacity constraint among all jobs whose flow crosses it.

\begin{proposition}
    \label{prop:mcf-not-gamma-concave}
    For every~$c\geq1$, multicommodity flow scheduling contains a unit-weight instance that is not~$c$-concave.
    In particular, multicommodity flow scheduling is not support submodular.
\end{proposition}
\begin{proof}
    Let~$T$, $\mathcal{Q}$, and~$\cP_T$ be the tree instance of \cref{prop:unbounded-concavity}, and index the root-leaf paths as~$\mathcal{Q} = \{Q_1,\dots,Q_{m^k}\}$ according to the left-to-right order of the leaves of~$T$.
    The leaves below a vertex~$v\in V(T)$ form a contiguous block, so~$I_v \coloneqq \{i\mid v\in Q_i\}$ is an interval of~$[m^k]$.

    Take~$G$ to be the directed path with arcs~$e_1,\dots,e_{m^k}$ in this order, all of unit capacity, and let job~$v\in V(T)$ be the commodity whose source and sink are the first and the last node of the subpath formed by~$\{e_i\mid i\in I_v\}$.
    Since~$G$ is a path, this subpath is the unique~$(s_v,t_v)$-path, hence~$y_v$ is precisely the flow that commodity~$v$ sends along it, and the capacity constraint of~$e_i$ reads
    \begin{align*}
        \sum_{v\in V(T)\colon i\in I_v} y_v
        = \sum_{v\in Q_i} y_v
        \leq 1.
    \end{align*}
    Ranging over~$i\in[m^k]$ gives exactly the constraints of~$\cP_T$, so this instance realizes the tree instance.
    By \cref{prop:unbounded-concavity} it is not~$c$-concave, and since support submodularity implies concavity of~$\OPT$ by \cref{thm:support-submodular-psp-concavity}, it follows that the polytope is not support submodular.
\end{proof}

\subsection{All-or-Nothing Scheduling}
In \emph{all-or-nothing scheduling}, we are given a collection~$\mathcal{S}$ of subsets of jobs.
A single unit of resource is split among the sets. If we choose~$x_S \geq 0$ for each~$S \in \mathcal{S}$ with~$\sum_{S} x_S \leq 1$, then each set~$S$ contributes rate~$x_S$ to every job in~$S$ simultaneously.
The rate of job~$j$ is thus~$y_j = \sum_{S \ni j} x_S$.
Taking the downward closure yields the rate polytope
\begin{align*}
    \cP = \Bigl\{ y \geq 0 \Bigm| y_j \leq \sum_{S \in \mathcal{S}\colon j \in S} x_S \ \forall j \in J,\ \sum_{S \in \mathcal{S}} x_S \leq 1,\ x \geq 0 \Bigr\}.
\end{align*}
Equivalently,~$\cP$ is the downward closure of the convex hull of the origin and the indicator vectors~$\bm{1}_S$ of the sets~$S \in \mathcal{S}$.

\begin{proposition}
    \label{prop:aon-not-gamma-concave}
    For every~$c\geq1$, all-or-nothing scheduling contains a unit-weight instance that is not~$c$-concave.
    In particular, all-or-nothing scheduling is not support submodular.
\end{proposition}
\begin{proof}
    Let~$T$, $\mathcal{Q}$, and~$\cP_T$ be the instance of \cref{prop:unbounded-concavity}, with the root-leaf paths indexed as~$\mathcal{Q} = \{Q_1,\dots,Q_{m^k}\}$ according to the left-to-right order of the leaves.
    Take~$\mathcal{S}$ to be the family of antichains of~$T$, that is,
    \begin{align*}
        \mathcal{S}
        \coloneqq \bigl\{ S\subseteq V(T) \bigm| |S\cap Q|\leq1 \ \forall Q\in\mathcal{Q} \bigr\}.
    \end{align*}

    We claim that the resulting polytope is~$\cP_T$.
    Every~$\bm{1}_S$ with~$S\in\mathcal{S}$ satisfies the constraints of~$\cP_T$ by definition of~$\mathcal{S}$.
    As~$\cP_T$ is convex and downward-closed, it contains the downward closure of~$\mathrm{conv}\{\bm{1}_S\mid S\in\mathcal{S}\}$, which is the polytope of this instance.
    For the reverse inclusion, note that the leaves below a vertex~$v$ form a contiguous block, so the set~$I_v\coloneqq\{i\mid v\in Q_i\}$ of rows in which the column of~$v$ has a one is an interval of~$[m^k]$.
    The constraint matrix of~$\cP_T$ is therefore an interval matrix and hence totally unimodular, so the polytope~$\cP_T = \{y\geq0 \mid By\leq\bm{1}\}$ is integral.
    Every vertex of~$T$ lies on some root-leaf path, so~$y_v\leq1$ for all~$y\in\cP_T$, hence the integral points of~$\cP_T$ are precisely the vectors~$\bm{1}_S$ with~$S\in\mathcal{S}$.
    Integrality implies that $\cP_T$ is the convex hull of these points.
    Thus all-or-nothing scheduling realizes the tree instance, which by \cref{prop:unbounded-concavity} is not~$c$-concave.
    Since support submodularity implies concavity of~$\OPT$ by \cref{thm:support-submodular-psp-concavity}, the polytope is not support submodular.
\end{proof}

\subsection{Graph Scheduling}
\label{subsec:graph-scheduling}

Let~$G=(V,E)$ be a loopless multigraph, and recall from \cref{subsec:non-support-submodular-problems} the matching polytope~$\cP_G$ of graph scheduling.
We now characterize exactly when graph scheduling is support submodular.
We call a connected component of~$G$ a \emph{star} if it has a vertex that is incident to all of its edges, and a \emph{triangle} if its underlying simple graph is a triangle.
A component is a star or a triangle if and only if any two of its edges share an endpoint.

\begin{proposition}
    \label{prop:graph-scheduling-stars}
    Let~$G=(V,E)$ be a loopless multigraph.
    The matching polytope~$\cP_G$ is support submodular if and only if every connected component of~$G$ is a star or a triangle.
\end{proposition}
\begin{proof}
    Let~$E_1,\dots,E_m$ be the edge sets of the connected components of~$G$ with at least one edge, and define
    \begin{align*}
        \cS \coloneqq \Bigl\{ y \in \RR^E_{\geq 0} \Bigm| \sum_{e \in E_i} y_e \leq 1 \quad \forall i \in [m] \Bigr\}.
    \end{align*}
    This is the polymatroid of the partition matroid rank function~$f(F) = |\{ i \in [m] \mid F \cap E_i \neq \emptyset \}|$.
    Hence,~$\cS$ is support submodular by \cref{prop:polymatroid-support-submodular}.
    We first show that~$\cP_G = \cS$ if every component of~$G$ is a star or a triangle.

    Suppose that any two edges in the same component share an endpoint.
    Then a matching contains at most one edge of each~$E_i$, so~$\cP_G \subseteq \cS$.
    Conversely, the vertices of~$\cS$ are the incidence vectors of edge sets with at most one edge of each~$E_i$, and these edge sets are matchings, since edges of different components are disjoint.
    Thus,~$\cP_G = \cS$.

    For the other direction, take~$\alpha = \bm 1$, so that~$H_\alpha(U)$ is the maximum of~$\sum_{e \in U} y_e$ over~$y \in (\cP_G)_U$.
    Recall that every~$y \in \cP_G$ satisfies the degree constraints~$\sum_{e \in \delta(v)} y_e \leq 1$, and that~$\cP_G$ contains the incidence vector of every matching of~$G$.
    Now suppose that some component contains two disjoint edges~$f$ and~$f'$.
    Let~$u_0 u_1 \dots u_k$ be a shortest path from an endpoint~$u_0$ of~$f$ to an endpoint~$u_k$ of~$f'$, and let~$x$ and~$z$ be the other endpoints of~$f$ and~$f'$, respectively.
    Then~$k \geq 1$ and~$x \neq z$ as~$f \cap f' = \emptyset$, and by minimality of~$k$ neither~$x$ nor~$z$ lies on the path.
    Hence~$x u_0 u_1 \dots u_k z$ is a path with at least three edges, and we denote its first three edges by~$e_1 = v_0v_1$, $e_2 = v_1v_2$, and~$e_3 = v_2v_3$.
    We choose~$U = \{e_1, e_2\}$ and~$W = \{e_2, e_3\}$.
    The degree constraints of~$v_1$ and~$v_2$ give~$y_{e_1} + y_{e_2} \leq 1$ and~$y_{e_2} + y_{e_3} \leq 1$ for all~$y \in \cP_G$, and~$\bm 1_{\{e_2\}}$ and~$\bm 1_{\{e_1,e_3\}}$ are incidence vectors of matchings.
    Thus,~$H_\alpha(U) = H_\alpha(W) = H_\alpha(U \cap W) = 1$ and~$H_\alpha(U \cup W) \geq 2$, so
    \begin{align*}
        H_\alpha(U) + H_\alpha(W)
        = 2
        < 3
        \leq H_\alpha(U \cup W) + H_\alpha(U \cap W),
    \end{align*}
    and~$\cP_G$ is not support submodular.
\end{proof}

Combining \cref{prop:graph-scheduling-stars} with \cref{thm:main-concave} characterizes the graphs for which the optimal weighted completion time of graph scheduling is concave for all weights.

\begin{corollary}
    \label{cor:graph-scheduling-concavity}
    Let~$G=(V,E)$ be a loopless multigraph.
    Every connected component of~$G$ is a star or a triangle if and only if, for every weight vector~$w\in\RR^E_{>0}$, the optimal-value function~$p\mapsto\OPT(p)$ of graph scheduling on~$G$ with weights~$w$ is concave on~$\RR^E_{\geq0}$.
\end{corollary}
\begin{proof}
    By \cref{prop:graph-scheduling-stars}, every connected component of~$G$ is a star or a triangle if and only if the matching polytope~$\cP_G$ is support submodular.
    Since~$\cP_G$ is downward-closed, \Cref{thm:main-concave} shows that this holds if and only if~$p\mapsto\OPT(p)$ is concave on~$\RR^E_{\geq0}$ for every weight vector~$w\in\RR^E_{>0}$.
\end{proof}

If some component of~$G$ is neither a star nor a triangle, then \cref{cor:graph-scheduling-concavity} yields a weight vector for which~$\OPT$ is not concave.
The following proposition shows that unit weights suffice already on a path with three edges, and that the matching polytope of this path is moreover not PF-monotone.
\begin{proposition}
    \label{prop:graph-scheduling-path}
    Let~$G$ be the path~$v_0v_1v_2v_3$ with edges~$e_b = v_{b-1}v_b$ for~$b\in[3]$ and unit weights.
    Then~$\cP_G$ is not PF-monotone, and the optimal-value function~$p\mapsto\OPT(p)$ is not concave.
\end{proposition}
\begin{proof}
    The rate polytope is~$\cP_G = \{y\in\RR^3_{\geq0} \mid y_1+y_2\leq 1,\ y_2+y_3\leq1\}$.
    For~$S = \{e_1,e_2,e_3\}$, both constraints are tight in the optimal solution of~\eqref{psp-pf-cp}, since otherwise~$y_1$ or~$y_3$ could be increased.
    Substituting~$y_1 = y_3 = 1-y_2$, the objective becomes~$2\log(1-y_2)+\log(y_2)$, which is maximized at~$y_2 = \frac13$.
    Hence~$y(S) = (\frac23,\frac13,\frac23)$.
    For~$S\setminus\{e_1\} = \{e_2,e_3\}$, the constraint~$y_2+y_3\leq1$ is tight and the objective~$\log(y_2)+\log(y_3)$ is maximized at~$y_2 = y_3 = \frac12$.
    Thus, removing~$e_1$ decreases the rate of~$e_3$ from~$\frac23$ to~$\frac12$, so~$\cP_G$ is not PF-monotone.

    For the second claim, recall that a concave positively homogeneous function is superadditive (cf.\ the proof of \cref{prop:unbounded-concavity}).
    The jobs~$e_1$ and~$e_2$ share the vertex~$v_1$, so their total rate is at most~$1$ at every time, and hence the first of them completes no earlier than at time~$1$ and the second no earlier than at time~$2$.
    Thus~$\OPT(1,1,0) = 3$, and symmetrically~$\OPT(0,1,1) = 3$.
    For~$p = (1,2,1)$, however, processing~$e_1$ and~$e_3$ simultaneously at rate~$1$ during~$[0,1]$ and then~$e_2$ at rate~$1$ during~$[1,3]$ is feasible and has weighted completion time~$1+1+3 = 5$.
    Therefore,~$\OPT(1,1,0)+\OPT(0,1,1) = 6 > 5 \geq \OPT(1,2,1)$, so~$\OPT$ is not superadditive and hence not concave.
\end{proof}

\subsection{RS-S Is Not Support Submodular for \texorpdfstring{$\rho<1$}{rho<1}}
\label{subsec:rho-leq-1}

In \cref{subsec:multidim-ssPSP} we showed that resource scheduling with substitutes (RS-S) is support submodular when~$\rho_j=\rho_j'=1$ for all~$j\in J$.
We now show that this fails for every smaller parameter~$\rho\in(0,1)$ by constructing an RS-S instance whose support function violates submodularity.
The construction uses two resources, two linear jobs whose slopes depend on~$\rho$, and a single job whose utility is the constant-elasticity-of-substitution (CES) aggregate of its two resource allocations.

\begin{proposition}
    \label{prop:ras-rho-not-support-submodular}
    For every~$\rho\in(0,1)$, there is a resource scheduling with substitutes instance whose feasible rate set is not support submodular.
\end{proposition}
\begin{proof}
    Let~$c \coloneqq 3\cdot 2^{1/\rho-1}$, and consider two resources~$A$ and~$B$ and three jobs with utilities
    \begin{align*}
        u_1(x_1) \coloneqq c\, x_{1A}, \qquad
        u_2(x_2) \coloneqq c\, x_{2B}, \qquad
        u_3(x_3) \coloneqq \left(x_{3A}^\rho + x_{3B}^\rho \right)^{1/\rho}.
    \end{align*}
    Jobs~$1$ and~$2$ have separable utilities, so only job~$3$ has~$\rho_3=\rho<1$.
    Formally, RS-S requires every~$f_{jd}$ to be increasing, which we obtain by adding~$\eta\, x_{1B}$ to~$u_1$ and~$\eta\, x_{2A}$ to~$u_2$ for some~$\eta>0$.
    All support function values below depend continuously on~$\eta$, and the violation of submodularity that we derive is strict, so it persists for sufficiently small~$\eta>0$.
    For readability, we present the case~$\eta=0$.
    Let~$\alpha=\bm{1}$, so that~$h_U(\alpha)$ is the maximum of~$\sum_{j\in U} u_j(x_j)$ over all resource allocations~$x\geq0$ with~$\sum_{j\in J} x_{jA}\leq 1$ and~$\sum_{j\in J} x_{jB}\leq 1$, where~$u_j(x_j)$ denotes the rate of job~$j$ under its allocation~$x_j=(x_{jA},x_{jB})$.
    We disprove submodularity of~$U\mapsto h_U(\alpha)$ by considering the sets~$U = \{1,3\}$,~$V = \{2,3\}$ and their union and intersection.

    For~$U\cap V=\{3\}$, we have
    \begin{align*}
        h_{\{3\}}(\alpha)
        = \max\{(x_{3A}^{\rho} + x_{3B}^{\rho})^{1/\rho} \mid 0\leq x_{3A}, x_{3B}\leq 1\}
        = (1+1)^{1/\rho}
        = 2^{1/\rho}.
    \end{align*}

    For~$U=\{1,3\}$ we maximize~$c\cdot x_{1A} + \left(x_{3A}^\rho + x_{3B}^\rho\right)^{1/\rho}$ subject to~$x_{1A}+x_{3A}\le 1$ and~$x_{3B}\le 1$.
    Job~$1$ does not use resource~$B$, so~$x_{3B}=1$ at the optimum.
    Substituting~$x_{1A}=1-x_{3A}$ and writing~$x=x_{3A}$ reduces the problem to maximizing
    \begin{align*}
        F(x) = c\cdot (1-x) + (1+x^\rho)^{1/\rho}, \qquad x\in[0,1].
    \end{align*}
    Its derivative is~$F'(x) = -c + (1+x^\rho)^{1/\rho-1}x^{\rho-1}$.
    Since~$\rho<1$ we have~$x^{\rho-1}\to\infty$ as~$x\to0^+$, hence~$F'(0^+) = \infty$, while~$F'(1) = -c + 2^{1/\rho-1} = -2\cdot 2^{1/\rho-1} < 0$.
    Hence, the maximizer~$x^\ast$ is interior,~$x^\ast\in(0,1)$, and satisfies~$F'(x^\ast)=0$, i.e.,~$(1+(x^\ast)^\rho)^{1/\rho-1} = c\cdot (x^\ast)^{1-\rho}$.
    Substituting this into~$F$ yields the optimal value,
    \begin{align*}
        F(x^\ast)
        &= c\cdot (1-x^\ast) + (1+(x^\ast)^\rho)^{1/\rho}
        = c - c\cdot x^\ast + (1+(x^\ast)^\rho) (1+(x^\ast)^\rho)^{1/\rho - 1} \\
        &= c - c\cdot x^\ast + (1+(x^\ast)^\rho)\cdot c\cdot (x^\ast)^{1-\rho}
        = c - c\cdot x^\ast + c \cdot (x^\ast)^{1-\rho} + c\cdot x^\ast \\
        &= c + c\cdot (x^\ast)^{1-\rho}
        = c + (1+(x^\ast)^\rho)^{1/\rho-1}.
    \end{align*}
    Since~$x^\ast<1$, we have~$1+(x^\ast)^\rho<2$, and because~$\tfrac1\rho-1>0$ the map~$t\mapsto t^{1/\rho-1}$ is increasing, so
    \begin{align*}
        F(x^\ast)
        < c + 2^{1/\rho-1}
        = 3\cdot 2^{1/\rho-1} + 2^{1/\rho-1}
        = 2^{1/\rho+1}.
    \end{align*}
    Thus,~$h_{\{1,3\}}(\alpha) < 2^{1/\rho+1}$, and by symmetry~$h_{\{2,3\}}(\alpha) < 2^{1/\rho+1}$.

    For the set~$U\cup V = \{1,2,3\}$ we maximize~$c\cdot x_{1A} + c\cdot x_{2B} + \left(x_{3A}^\rho + x_{3B}^\rho\right)^{1/\rho}$ subject to~$x_{1A}+x_{3A}\le 1$ and~$x_{2B}+x_{3B}\le 1$.
    Substituting~$x_{1A}=1-x_{3A}$ and~$x_{2B}=1-x_{3B}$ gives
    \begin{align*}
        F(x_{3A},x_{3B})
        = 2c - c\cdot (x_{3A}+x_{3B}) + (x_{3A}^\rho + x_{3B}^\rho)^{1/\rho}.
    \end{align*}
    For~$0<\rho<1$, the generalized~$\rho$-mean is at most the arithmetic mean, that is,~$(x_{3A}^\rho + x_{3B}^\rho)^{1/\rho} \le 2^{1/\rho-1}(x_{3A}+x_{3B})$, hence
    \begin{align*}
        F(x_{3A},x_{3B})
        \leq 2c - \bigl(c - 2^{1/\rho-1}\bigr)(x_{3A}+x_{3B}).
    \end{align*}
    Since~$c = 3\cdot 2^{1/\rho-1} > 2^{1/\rho-1}$, the coefficient~$c - 2^{1/\rho-1} = 2\cdot 2^{1/\rho-1}$ is positive, so the right-hand side is maximized at~$x_{3A}=x_{3B}=0$, where it equals~$2c$, and the original objective also equals~$2c$ there.
    Thus,
    \begin{align*}
        h_{\{1,2,3\}}(\alpha)
        = 2c
        = 3\cdot 2^{1/\rho},
    \end{align*}
    attained at~$x_{1A}=x_{2B}=1$ and~$x_3=0$.
    It follows that submodularity of the support function is violated, because
    \[
        h_{\{1,3\}}(\alpha) + h_{\{2,3\}}(\alpha)
        < 2^{1/\rho+1} + 2^{1/\rho+1}
        = 2^{1/\rho+2}
        = 3 \cdot 2^{1/\rho} + 2^{1/\rho}
        = h_{\{1,2,3\}}(\alpha) + h_{\{3\}}(\alpha). \qedhere
    \]
\end{proof}

\section{Polynomial-Time Implementation of Proportional Fairness}
\label{sec:pf-polynomial-time}

PF recomputes its rates by solving~\eqref{psp-pf-cp} for the current set of active jobs whenever a job is released or completes (see \cref{subsec:proportional-fairness}).
We show that approximate solutions, which can be computed in polynomial time, suffice for all our guarantees up to a factor of~$1+\eps$.

\paragraph{Computational model.}
Our computational claims assume rational input and an explicit matrix~$B$, or a well-described rational rate polytope with a polynomial-time separation oracle.
For unrelated machines and generalized flow scheduling, the compact extended formulations in \cref{subsec:unrel-ssPSP,subsec:lattice-ssPSP} supply the required representation. Graph scheduling admits polynomial-time separation for its matching polytope~\cite[Theorem~12.21]{KorteVygen}.

\paragraph{Approximate rates.}
Since each job is released once and completes once, PF solves at most~$2n$ instances of~\eqref{psp-pf-cp}.
However, PF cannot always be executed exactly, because for some rational inputs the optimal solution of~\eqref{psp-pf-cp} is irrational~\cite[Example~48]{JainV10}.
For a nonempty active set~$S$ and~$\delta>0$, the ellipsoid method yields feasible rates~$\widetilde y(S)$ with~$\abs{\widetilde y_j(S)-y_j(S)}\leq\delta$ for every~$j\in S$ in time polynomial in the input encoding length and~$\log(1/\delta)$~\cite{JLM2026}.
The maximum rate~$y_j^{\max}$ equals~$1/\max_d B_{dj}$ for explicit~$B$ and is computable by linear optimization in the oracle model.
By \cref{lem:pf-optimality}, we have~$y_j(S)\geq w_jy_j^{\max}/w(S)$ for every~$j\in S$.
Thus, for~$\rho\in(0,1)$, choosing
\begin{align*}
    \delta\coloneqq\frac{\rho}{1+\rho}\min_{j\in S}\frac{w_jy_j^{\max}}{w(S)}
\end{align*}
ensures that the feasible approximate rates satisfy~$\widetilde y_j(S)\geq y_j(S)/(1+\rho)$ for every~$j\in S$.
The minimum has polynomial encoding length, so the running time is polynomial in the input length and~$\log(1/\rho)$.

\paragraph{Effect on the guarantees.}
The \emph{approximate schedule} uses at every time~$t$ the rates~$\widetilde y(A(t))$, where~$A(t)$ is its own set of active jobs.
Applying \cref{lem:subinstance-of-structured-instance} along the approximate schedule's own active sets, and using monotonicity and homogeneity of the optimal weighted completion time~$\OPT_0$ with all release dates set to~$0$, gives completion-time ratios~$2(1+\rho)$ and~$3+2\rho$ in the proof of \cref{thm:main-completion}.
Likewise, the proof of \cref{thm:c-approximate-flow-time} uses effective speed~$\kappa/(1+\rho)$ and gives ratio~$\kappa/[\kappa-2c(1+\rho)]$ provided~$\kappa>2c(1+\rho)$; choosing sufficiently small~$\rho=O(\eps/c)$ preserves the stated flow-time bounds.
The proof of \cref{thm:general-psp-completion} uses the KKT conditions only through~\eqref{eq:kkt-stationarity} and \cref{lem:pf-optimality}(i).
For the approximate schedule, we use the exact multipliers~$\eta(A(t))$ of its own active sets~$A(t)$. Since~$\widetilde y_j(S)\geq y_j(S)/(1+\rho)$, the identity~\eqref{eq:kkt-stationarity} then yields~$w_j/\widetilde y_j(A(t))\leq(1+\rho)\sum_{d\in[D]}B_{dj}\eta_d(A(t))$ for every~$j\in A(t)$.
Multiplying the resource prices in \cref{subsec:general-dual-fitting} by~$1+\rho$ therefore restores dual feasibility and multiplies the bound of \cref{lem:general-dual-beta} by~$1+\rho$, whereas the lower bound on the job values, which rests on \cref{lem:volume-general}, does not involve the KKT conditions and holds verbatim.
With~$\theta=\frac14$ as in that proof, this gives the ratio~$4/(1-\rho)$, which is at most~$4+\eps$ for~$\rho\leq\eps/(4+\eps)$.

\end{appendices}

\section*{Acknowledgments}

SJ and AL were supported by the Deutsche Forschungsgemeinschaft (DFG, German Research Foundation) under Germany’s Excellence Strategy -- The Berlin Mathematics Research Center MATH$^+$ (EXC-2046/1, EXC-2046/2, project ID: 390685689).
BZ was supported by the Deutsche Forschungsgemeinschaft (DFG, German Research Foundation), project ID 547924951.

\section*{AI Methodology Statement}
We used large language models in the preparation of this paper as follows.
The constructions for \Cref{thm:concavity-implies-support-submodularity,thm:randomized-LB} were found with the assistance of ChatGPT-6 Astra.
The initial hint that we can use our linear certificates for the weighted completion time also for analyzing weighted flow time was given by ChatGPT-5.5.
The evaluation of the integral in the proof of \Cref{lem:volume-general} was carried out with the assistance of Claude Opus 4.8.
We used Claude Opus 5.5 to work out the connection of support submodularity, the condition of \cite{KT03}, and the substitutes condition of \cite{ChenIP25}.
All AI-assisted contributions were verified by the authors, who take full responsibility for their correctness.
Beyond this, AI tools were used only for minor language editing.
All other results, proofs, and ideas in this paper are our own.

\printbibliography

\end{document}